# Where does a molecule begin? Threshold masses of three-particle Coulomb systems and a predictive theory of their Born–Oppenheimer offsets

Roberto Peverati

*Department of Chemistry and Chemical Engineering, Florida Institute of Technology,*
*150 W University Boulevard, Melbourne, Florida 32901, USA*
*E-mail: rpeverati@fit.edu*

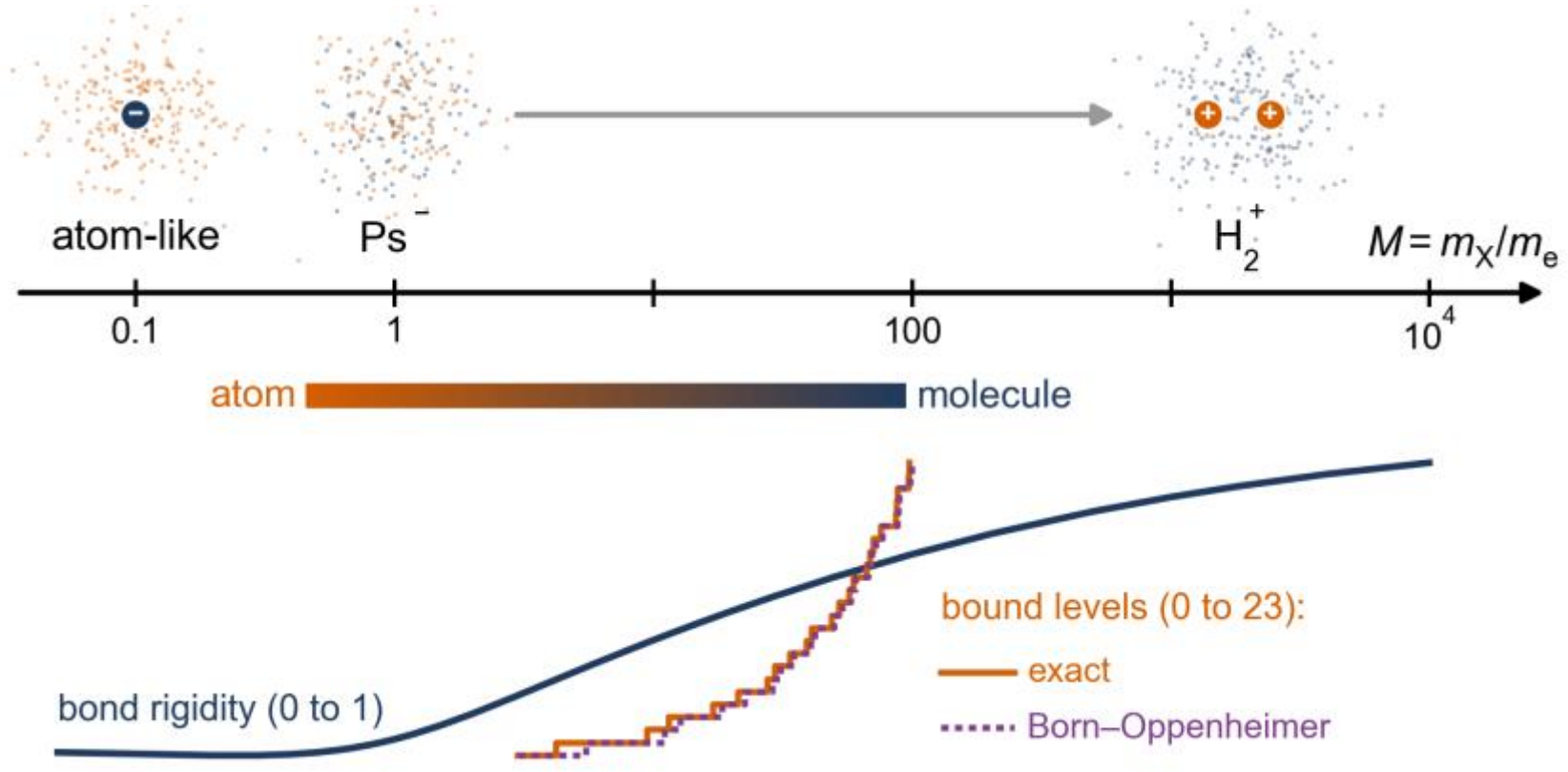

**Abstract.** The Born–Oppenheimer (BO) approximation serves as the foundational framework for the contemporary concept of molecular structure, which is central to chemistry and the molecular sciences. Its principal assumption is the treatment of nuclei as infinitely heavy particles. In this work, we test the boundaries of the BO approximation by examining systems with variable nuclear masses to investigate the emergence of molecular structure. We examine the three-particle family $\{X^+,X^+,e^-\}$ as a function of the mass ratio $M = m_X/m_e$, using deterministic variational calculations and the exact BO approximation. We accurately compute the threshold masses of all 23 excited levels bound below $M \approx 100$ with certified upper bounds. The BO approximation places every one of them too late by a nearly constant amount, 1.2 to 1.6 (0.23 to 0.54 with the adiabatic correction). We explain this constancy with a parameter-free theory, and test it with predictions frozen before the exact calculations: 18 of 18 symmetric levels pass (typical miss 0.003). For unequal nuclear masses none of 7 levels falls within the band, but the theory still captures over 90% of the offset. Where does a molecule begin? The ground state changes character gradually, over a crossover 2.3 decades wide in $M$. The sharp landmarks are the threshold masses: the first rotational level binds at $M$ = 4.208, in the ortho-type nuclear-spin species, and the first vibrational level at $M$ = 9.526; BO places them at 5.54 and 11.08. Thus, the BO picture orders the spectrum correctly at finite mass and shifts it by an amount that can be computed.

*Supplementary Information: derivations, numerical methods, validation, the complete threshold map with certified bounds, the comparison with earlier estimates, the pre-specified prediction log and additional results are provided in the Supplementary Information accompanying this preprint.*

## 1. Introduction

The Born–Oppenheimer (BO) approximation[1] is the usual starting point for discussion of molecular structure. By separating the motion of the electrons from that of the nuclei, it provides a potential energy surface, which defines an equilibrium geometry and on which the nuclei vibrate and rotate. The approximation is derived as an expansion in the fourth root of the electron-to-nuclear mass ratio,[1,2] $\kappa = (m_e/M)^{1/4}$, and it becomes exact as the nuclei become infinitely heavy. The exact molecular Hamiltonian, however, treats both nuclei and electrons quantum mechanically, and the Coulomb interactions among them depend only on their distances, so the Hamiltonian has no preferred orientation and no built-in geometry. How does a bond of definite length emerge from it, and at which nuclear mass does this happen? These are old questions in the foundations of chemistry.[3,4] Woolley pointed out that the exact eigenstates of the molecular Coulomb Hamiltonian are invariant under rotations, inversion, and the permutation of identical particles, so that a definite molecular structure cannot be a property of such a state; it appears only in time-dependent states or once the nuclei are treated classically.[5,6] Sutcliffe and Woolley later examined what the spectral theory of the full Hamiltonian does and does not imply, and concluded that the potential energy surface is a useful construct added to molecular quantum mechanics, rather than derived from it.[7,8] Jecko replied with a summary of the rigorous mathematical results, which justify the BO approximation as an asymptotic statement for heavy nuclei, without saying how heavy is heavy enough.[9] Frolov's calculations on this very family showed that the discrete spectrum survives when one mass grows without limit, but that the limit in which both nuclear masses grow is singular, a failure he called adiabatic divergence.[8] Our work sits between these positions: rather than asking whether the BO picture can be derived, we ask at which finite mass it becomes accurate enough to tell which levels exist.

Three-particle Coulomb systems of the type $\{X^+,X^+,e^-\}$ offer the simplest setting for these questions. The nuclear mass can be varied continuously, from the positronium-like case ($M = m_X/m_e = 1$) to $H_2^+$ ($M = 1836$) and beyond, and the problem remains tractable at high precision. The change of character of the ground state along this family has been examined with several descriptors. Mátyus and co-workers followed the density about the centre of mass and found the change for $0.4 < M < 0.8$;[10] King, Longford, and Cox located an energetic marker near 0.49;[11] Salas and co-workers, using a marginal distribution, placed it near 1;[12] and Kais and Shi examined the stability boundary in the charge ratio.[13] Exactly solvable models of Müller-Herold showed sharp, mass-driven transitions in the density and in the angular correlation,[14,15] and Ludeña and co-workers showed that criteria based on the density depend on the reference point from which the particle positions are measured.[16] All of these studies concern the ground state and smooth descriptors. Thus, they locate a change of character, but they do not say when the system acquires the rotation and vibration that make it a molecule.

Excited levels offer sharper landmarks. Each rotational or vibrational level of the family is bound only above a critical mass, and below that mass it dissolves into the continuum. Frolov and co-workers first located the lowest of these threshold masses between 1989 and 1994,[17–21] with finite exponential bases of 600 to 1500 functions, three to five digits, and no error bars; their vibrational thresholds are 3 to 7% too high. What is still missing is (i) converged critical-mass values with certified bounds over a wider range of levels, (ii) a comparison with the thresholds of the BO approximation, which is the approximation that defines the molecular picture, and (iii) an explanation of the difference between the two. The last point allows a direct test of the BO picture: does it fail early or late, by how much, and why?

In the present work, we compute the threshold masses of every excited level of the ground electronic manifold that becomes bound below $M \approx 100$ (23 levels; the first binds at $M = 4.208$) using a deterministic perimetric-Laguerre variational solver. We set the upper limit at $M \approx 100$ to cover the muonic molecular ions and the whole ground-state crossover; above it, levels appear at increasingly short intervals, and the cost of the calculations grows quickly with J. We then compare the results with the thresholds of our own exact BO and BO-plus-adiabatic calculations. We find that the BO approximation places every threshold too late by a nearly constant amount, and we explain why. The explanation is then tested with predictions that were frozen before the exact calculations were run, for the symmetric family and for unequal nuclear masses. We limit ourselves to isolated eigenstates of three particles. These have a bond length but no shape, and they do not localise; shape, and the role of an environment, are left for future work.

Our main contributions are the following.

(i) *Threshold masses.* Exact thresholds of 23 levels between $M = 4.2$ and 98.3, with certified upper bounds and, for most levels, a relative precision of $10^{-6}$–$10^{-9}$; the threshold laws and the size of the states near the boundary; and a check against the muonic ions pp$\mu$, dd$\mu$, and tt$\mu$.

(ii) *A theory of the BO offsets.* The zero-energy BO equation is a linear eigenvalue problem in $M$, so that every correction of order $1/M$ shifts a threshold mass by a constant. The adiabatic part of the shift follows at first order, and the non-adiabatic part from the vibrational and rotational mass functions, with no fitted parameter. Pre-specified predictions pass for all 18 levels tested in the symmetric family. For unequal nuclear masses, none of the 7 levels tested falls within the band we set in advance; even there, however, the theory accounts for 91 to 99.6% of the offset, and its misses are 4 to 40 times smaller than those of simple null models, so that it appears incomplete rather than wrong.

(iii) *The ground-state crossover.* The ground state changes from atom-like to molecule-like over 2.3 decades in $M$, with no sharp point. The descriptors proposed in the literature cluster at the onset of this crossover, and the thresholds are the only sharp landmarks.

(iv) *The BO limit and the entanglement.* A numerical verification of the asymptotic BO expansion, with the adiabatic correction at order $\kappa^4$ and the non-adiabatic one at order $\kappa^6$, and the residual electron–nuclear entanglement of the ground state, which in the BO limit is purely orientational.

The paper is organised as follows. The next section defines the three-particle family and its thresholds. It also describes the exact and BO calculations, the theory of the offsets, and the pre-specified tests. The results are then presented in the next section in the order validation, theory, tests, thresholds, and ground-state crossover, followed by the BO limit and the entanglement. The conclusions answer the question of the title.

## 2. Theory and methods

### 2.1 The three-particle family and its thresholds

We consider $\{X^+,X^+,e^-\}$ systems with two particles of charge +1 and mass $m_X$, and one electron, in the centre-of-mass frame and in Hartree atomic units. The only parameter is $M = m_X/m_e$. With $\boldsymbol{R}$ the vector joining the two $X^+$ and $\boldsymbol{\rho}$ the position of the electron relative to their midpoint (Jacobi coordinates), the internal Hamiltonian is:

$$\hat{H}(M) = -\frac{1}{M}\nabla_{\boldsymbol{R}}^2 - \frac{2M+1}{4M}\nabla_{\boldsymbol{\rho}}^2 + \frac{1}{R} - \frac{1}{|\boldsymbol{\rho}+\boldsymbol{R}/2|} - \frac{1}{|\boldsymbol{\rho}-\boldsymbol{R}/2|}\,. \quad (1)$$

Both coordinates are independent of the mass, so that $M$ enters only through the kinetic energy, and there is no mass-polarisation (cross) term. The lowest dissociation threshold is that of a hydrogen-like atom plus a distant $X^+$, $E_{th} = -M/[2(M+1)]$, and a level is bound if its energy lies below $E_{th}$. We label the levels $(J,v)$, with $J$ the total orbital angular momentum (spin is not part of the calculation) and $v$ the vibrational index. The critical mass $M_c(J,v)$ of a level is the mass at which its energy crosses $E_{th}$. The most recent CODATA 2022 masses obtained from NIST[22] are used whenever a physical system is named.

Note that the exchange of the two $X^+$ particles gives the rotational tower of the ground manifold the symmetry $(-1)^J$. The $J = 1$ level is thus antisymmetric, as for ortho-type species with fermionic X, and $J = 2$ is symmetric. This will matter when we compare the first rotation with the first vibration. All levels considered here have natural parity, $P = (-1)^J$, and belong to the ground ($1s\sigma_g$) manifold. We did not compute the antisymmetric $J = 0$ and symmetric $J = 1$ states, which in BO terms belong to the $2p\sigma_u$ manifold and bind only at much larger mass (BO thresholds 331 and 816), the states of unnatural parity, or the levels that bind above $M \approx 100$.

### 2.2 Exact three-body calculations

The exact calculations were performed with an in-house variational solver in a perimetric basis of Laguerre functions, in the tradition of Pekeris and as used by King, Cox, and co-workers.[11,23,24] The integrals are analytic and the matrices do not depend on the masses, so that a mass scan needs only new diagonalisations. Near-threshold states require a diffuse tail, which we provide with a multi-block basis (a compact set plus diffuse sets). The solver is extended to $J > 0$ with bipolar harmonics (tested for $J = 1$ to 7), and the energies are converged by extrapolating the geometric tail of the basis sequence.

Two numerical details deserve mention. First, an orthogonalisation cutoff of $10^{-12}$ proved necessary, since a cutoff of $10^{-14}$ produced a spurious root. For the highest angular momenta, $J = 6$ and 7, the matrix elements are long sums of nearly cancelling terms, and double precision was not enough; we therefore assembled the overlap and Hamiltonian matrices in double-double (about 32-digit) arithmetic, for which a cutoff of $10^{-14}$ is safe, and two further checks guard against spurious roots. Second, a threshold is certified when a variational energy lies at least $10^{-11}$ $E_h$ below $E_{th}$ at a given mass, well above the rounding floor of $10^{-14}$ $E_h$. This certificate relies on the margin being much larger than the rounding errors, rather than on interval arithmetic, and it proves binding at the stated mass, not at every larger mass.

### 2.3 Born–Oppenheimer and adiabatic thresholds

The BO baseline uses the exact $1s\sigma_g$ potential curve, computed in prolate spheroidal coordinates to about $10^{-12}$ $E_h$. For an electron of reduced mass $\mu_e$, the clamped-nuclei energy is exactly $\mu_e U(\mu_e R)$, with $U(r)$ the unit-mass curve, so that a single curve serves the whole family. The vibration–rotation levels follow from the nuclear radial equation:

$$-\frac{1}{M}\frac{d^2\chi}{dR^2} + \frac{J(J+1)}{MR^2}\chi + [\mu_e U(\mu_e R) + c\,E_{ad}(R)]\chi = E\,\chi\,, \quad (2)$$

where $E_{ad} = \mu_e^2 D(\mu_e R)/M$ is the diagonal adiabatic correction, with $c = 0$ for BO and $c = 1$ for BO plus adiabatic (BO+ad). Each model's levels are measured from its own dissociation limit, $-\mu_e/2$ for BO and $-\mu_e/2 + \mu_e^2/(4M)$ for BO+ad, which differ from the exact $E_{th}$ by terms of order $1/M$. The thresholds $M_c^{BO}$ and $M_c^{BO+ad}$ are the masses at which a level reaches that limit. They are therefore defined in the same way as the exact ones, and their differences from the exact $M_c$ are the offsets studied below. We obtain them as eigenvalues of the zero-energy problem of eqn (3), with an independent node-count bisection of the radial equation as a check (agreement to $10^{-9}$). The derivations and the numerical methods are given in the ESI (Sections S1.2–S1.4 and S2.6).

## 2.4 Why the offsets are nearly constant

The near-constancy of the offsets follows from the structure of the zero-energy radial problem. At threshold, we scale the internuclear distance as $r = \mu_e R$, with $\mu_e = 2M/(2M+1)$ the reduced mass of the electron, and multiply the BO radial equation by $M/\mu_e^2$. The equation then becomes, exactly, a linear eigenvalue problem in the mass,

$$-\frac{d^2u}{dr^2} + \left[\frac{L(L+1)}{r^2} + c\,\Delta D\right] u = -\Lambda\,\Delta U\,u\,, \qquad (3)$$

where $\Lambda = M + 1/2$, $\Delta U(r)$ is the BO potential and $\Delta D(r)$ the adiabatic correction, both for unit mass and measured from their asymptotes, and $c = 0$ for BO and $c = 1$ for BO+ad. The mass enters only through $\Lambda$, so that a level appears when $\Lambda$ reaches an eigenvalue $\Lambda_c$ of this mass-independent problem, and $M_c = \Lambda_c - 1/2$. A shooting solution of eqn (3) reproduces all our BO threshold masses to $10^{-9}$. Every correction of order $1/M$ to the Hamiltonian then enters eqn (3) as a mass-independent perturbation, and shifts $M_c$ by a constant. Since zero-energy states sample about the same region of $R$, the shift is nearly independent of the level; we regard this last step, a WKB argument, as interpretation and not as a derivation.

The adiabatic correction is of first order and can be evaluated without any fit. To first order, a perturbation of eqn (3) shifts $\Lambda_c$ by its expectation value in the zero-energy solution $u$, divided by $\langle u|-\Delta U|u\rangle$. Adding the adiabatic correction therefore lowers the threshold by:

$$M_c^{\mathrm{BO}} - M_c^{\mathrm{BO+ad}} \simeq \frac{\langle u|-\Delta D|u\rangle}{\langle u|-\Delta U|u\rangle}\,, \qquad (4)$$

which gives 1.016, 1.021, and 1.030 for $v$ = 1–3, against 1.018, 1.022, and 1.031 from the difference between the BO and BO+ad thresholds (agreement to 0.2%). For unequal nuclear masses, $m_A$ and $m_B = q m_A$, the same argument gives:

$$M_c = \frac{(1+q)\,\Lambda_c}{2q} - \frac{1}{1+q}\,, \qquad (5)$$

where the two lowest electronic states, $1s\sigma_g$ (gerade, g) and $2p\sigma_u$ (ungerade, u), become coupled and degenerate at dissociation, so the same argument requires a two-channel (g/u) problem, with the exact threshold built in (derivation in the ESI, Section S1.9).

## 2.5 Non-adiabatic mass functions

The remaining, non-adiabatic part is described by the $R$-dependent vibrational and rotational masses $m_v(R)$ and $m_r(R)$ of the Bunker–Moss effective Hamiltonian.[25] In the mass-scaled convention used here, they are:

$$m_v(R) = 4\langle \partial_R \phi_0 | \hat{Q}(\hat{h} - E_0)^{-1}\hat{Q} | \partial_R \phi_0 \rangle\,, \qquad (6)$$

and

$$m_r(R) = 4\left\langle \frac{\hat{L}_y \phi_0}{R} \right| (\hat{h} - E_0)^{-1} \left| \frac{\hat{L}_y \phi_0}{R} \right\rangle, \qquad (7)$$

where $\varphi_0$ is the clamped-nuclei $1s\sigma_g$ function, $\hat{h}$ and $E_0$ are its electronic Hamiltonian and energy, $\hat{Q} = 1 - |\varphi_0\rangle\langle\varphi_0|$, and the derivative is taken at a fixed position of the electron relative to the midpoint; only $\Sigma_g$ states contribute to $m_v$ and only $\Pi_g$ states to $m_r$. These functions are known for $H_2^+$ from the work of Bunker and Moss, Kutzelnigg, Schwenke, and Mátyus and co-workers.[25–29] We recomputed them from the exact BO wave function on 2597 $R$ points, converged to $3 \times 10^{-11}$, so that one mass-scaled pair serves the whole family (Fig. 1). At the equilibrium distance, they give $m_v(R_e)$ = 0.34655 and $m_r(R_e)$ = 0.07542, which match the values inferred from the exact energies (0.3465 and 0.0754) without any tuning. As a code check, they reproduce 421 of the 423 entries of Table 4 of Moss (1996) within 0.004 $cm^{-1}$.[30] Both functions tend to 1/2 as $R^{-3}$. Our BO and BO+ad calculations use the bare nuclear reduced mass, M/2, for the nuclear motion and the electron reduced mass $\mu_e$ for the electronic problem, and the mass functions are defined relative to this choice, $M \rightarrow M + m(R)$. Their common limit, $m(\infty) = 1/2$, then restores the exact reduced mass of the ion and the atom at dissociation, $\mu = M(M+1)/(2M+1)$. With a different convention, for example atomic instead of nuclear masses, part of the non-adiabatic offset would move into the BO+ad thresholds; the theory is unchanged, but the split between the adiabatic and non-adiabatic parts depends on this choice. The mass functions themselves are not new.[27–29] What is new here is that, in the $\mu_e$-scaled convention, a single pair serves the whole family, and that we use it with no fitted parameter to predict the offsets of the threshold masses.

We include the mass functions by replacing $M$ with $M + m_v(R)$ in the radial kinetic energy and with $M + m_r(R)$ in the centrifugal term, which turns eqn (3), with $c = 1$, into:

$$-\frac{d}{dr}\left(q_v \frac{du}{dr}\right) + \left[q_r \frac{L(L+1)}{r^2} + \Delta D\right] u = -\Lambda\,\Delta U\,u\,, \qquad (8)$$

with

$$q_{v,r}(r) = \frac{\Lambda}{\Lambda + m_{v,r}(r)}\,. \qquad (9)$$

Its eigenvalues give the predicted thresholds, with no fitted parameter. To first order in the mass functions, the non-adiabatic offset, $\Delta M_c^{nad} = M_c^{BO+ad} - M_c$, is a weighted average,

$$\Delta M_c^{nad} \simeq \frac{\int [m_v (du/dr)^2 + m_r L(L+1) u^2/r^2] dr}{\Lambda_c \int (-\Delta U)\, u^2\, dr}, \qquad (10)$$

so that the offsets measure $m_v$ and $m_r$ mostly between 3 and 8 bohr, where the zero-energy states have their weight. The two routes differ by at most 0.0025 in $M_c$, and we quote the solutions of eqn (8).

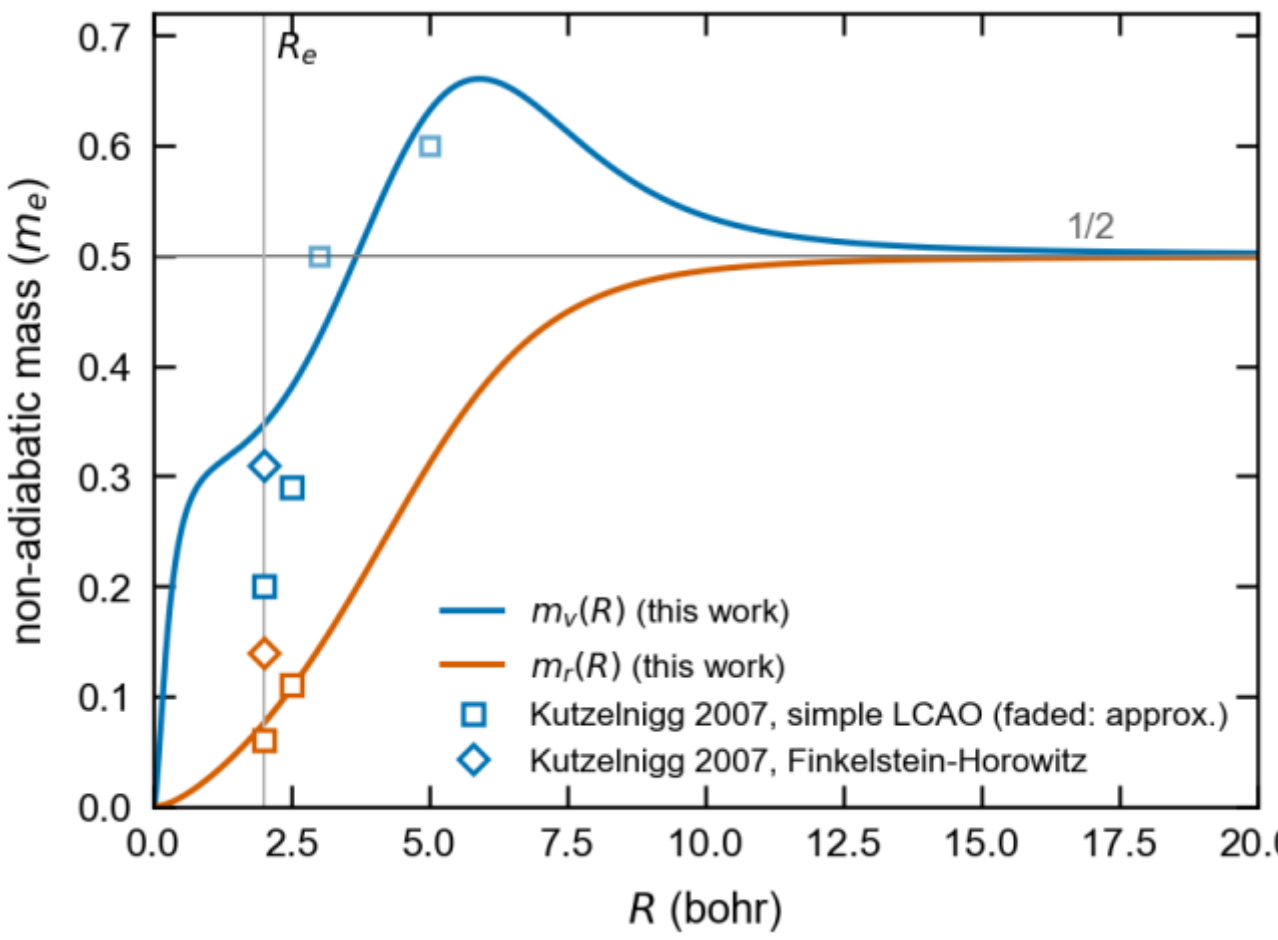

**Fig. 1** Non-adiabatic vibrational and rotational mass functions, $m_v(R)$ and $m_r(R)$, of the $\{X^+,X^+,e^-\}$ family in the $\mu_e$-scaled convention, computed from the exact Born–Oppenheimer wave function (lines). Open symbols: estimates of Kutzelnigg[26] (squares, simple LCAO wave function, faded where approximate; diamonds, Finkelstein–Horowitz function). Vertical line, the equilibrium distance $R_e$; horizontal line, the common limit 1/2.

### 2.6 Pre-specified predictions

To test the theory, we froze its predictions before running the exact calculations. Each prediction, with a stated band, was written to a read-only file, hashed (SHA-256), and recorded in a dated local commit. There were three freezes for the symmetric family: the (0,4) level; the (1,1), $J = 3$, and $J = 4$ levels; and a map of 14 levels. A separate freeze covered the unequal-mass family. A prediction passes if the exact threshold lies within the band plus the numerical error. The bands were 0.0095 for $v = 4$, 0.0295 for the others (the largest residual of the earlier $J > 0$ predictions), and 0.03 for unequal masses.

Since the freezes are held by the author and not time-stamped by a third party, we call these predictions pre-specified and not blind. For completeness, we list what was known at each freeze. No exact three-body calculation of a predicted level was run before its freeze, with one exception: a scale optimisation of the (1,1) basis, run 4 min before the second freeze, which computed no threshold and showed only that the level is bound at its BO+ad mass, as already implied by the earlier levels. Before the third freeze, we had seen the earlier exact thresholds, the BO and BO+ad map and the estimates of Frolov for (2,1), (1,2) and (2,2), and a non-frozen run of the same prediction code had preceded the freeze by 9 min. After the first three freezes, short code tests produced approximate values of some of the predicted thresholds; the frozen files were not changed, and their hashes were verified after the exact calculations. A detailed log is given in the ESI.

### 2.7 Code generation and use of generative AI tools

All solvers and analysis scripts of this work were written with the assistance of a generative AI coding agent (Claude Code, Anthropic; the models are listed in the Acknowledgements), working in our project repository under our direction. We specified each task in natural-language prompts; the agent wrote code and tests, ran the calculations we approved, and reported results, which we checked. No numerical result in this paper was generated by the language model itself. Every number comes from deterministic code that is archived with the paper and covered by test suites, and each was reproduced or spot-checked independently of the run that produced it.

## 3. Results and discussion

### 3.1 Validation

Before turning to the thresholds, we validated the three-body and BO solvers against published values.

**Three-body solver.** For the ground states of He and $H^-$ (infinite nuclear mass)[31] and $Ps^-$,[32] our energies agree with the literature to $10^{-10}$ $E_h$ or better. For $H_2^+$, with levels labelled $(J,v)$ as in the rest of this work, the levels (0,0), (0,1), (1,0), (1,1), (2,0) and (2,1) agree with Tang et al. to $10^{-12}$–$10^{-13}$ $E_h$ at their proton mass, and (3,0) agrees to $1.2 \times 10^{-11}$ $E_h$.[33] Tang et al. give no value for (4,0). Our energy lies $3 \times 10^{-9}$ $E_h$ below a Dunham-type extrapolation, $E_0 + Bx - Dx^2 + Hx^3$ with $x = J(J+1)$, fitted to their $J = 0$–3 levels with $v = 0$, which matches the size of the neglected $x^4$ term. The $2^1P$ and $2^3P$ levels of He agree with Drake and Bondy to $10^{-10}$ $E_h$.[34] To validate the solver across the whole mass family, we compared it with the variational energies of King et al.,[11] who studied the same three-particle systems, recomputing every entry with their masses (CODATA 2010[35]). All 43 entries of their Table III agree with ours within one unit of the last digit they consider converged (worst case 0.64 unit). Their 2856-term energies of $Mu_2^+$, $H_2^+$ (−0.597139063079 $E_h$) and $\mu D_2^+$ agree with ours to all 12 printed digits. Where our energy is lower ($D_2^+$ by 2.7 × $10^{-10}$ $E_h$; $H^-$, $Ps^-$ and $Mu^-$ by 1–4 × $10^{-11}$ $E_h$), it is the better upper bound. At the CODATA 2022 mass used in the rest of this work, the $H_2^+$ ground state is −0.5971390630805 $E_h$.

BO solver. Our clamped-nuclei energies reproduce the benchmark value of Madsen and Peek,[36] $U(2\ a_0)$ = −0.6026342144949 $E_h$, to 4 × $10^{-13}$ $E_h$, and the equilibrium distance of Scott et al.,[37] $R_e$ = 1.99719332 $a_0$. They also agree with the 13-figure tabulation of the 1s$\sigma_g$ and 2p$\sigma_u$ energies by Madsen and Peek[36] to 3 × $10^{-12}$ $E_h$ for $1 \leq R \leq 90\ a_0$, and with the tables of Wind[38] and Scott et al.[37] to the precision given there. The BO thresholds obtained from eqn (3) agree to $10^{-9}$ with a second, independent calculation in which we count the nodes of the threshold solution and bisect in M.

Both solvers are therefore accurate well beyond the precision that the threshold masses require. Full validation details are given in the ESI (Section S3 and Table S3).

## 3.2 The Born–Oppenheimer offsets of the threshold masses

We located the threshold masses of the excited levels in three independent ways. The exact values $M_c$ come from the three-body solver: for each level, we follow its energy as a function of M and find where it crosses the dissociation threshold, and the crossing gives a certified upper bound on $M_c$. The BO and BO+ad values come from the BO solver, as eigenvalues of the zero-energy problem of eqn (3), each measured from its own dissociation limit. Table 1 lists the three sets for the nine lowest excited levels, together with the certified upper bounds.

The first rotational level, the exchange-antisymmetric $J$ = 1, binds at $M_c(1,0)$ = 4.2082827(5), and the first vibrational level at $M_c(0,1)$ = 9.525634(4). The BO approximation places them at 5.5375 and 11.0815, overestimating them by 1.33 and 1.56; even with the adiabatic correction they are still overestimated by 0.36 and 0.54. The same holds for every level we computed: across the 23 levels of the map discussed below, the BO threshold exceeds the exact one by 1.19 to 1.56, and the BO+ad threshold by 0.23 to 0.54.

**Table 1** Threshold masses of the lowest excited levels of {$X^+$,$X^+$,$e^-$}: exact value, certified upper bound, BO and BO+ad values, and the non-adiabatic offset $\Delta_{na} = M_c^{BO+ad} - M_c$, exact and predicted without fitted parameters

| **(*J*,*v*)** | $M_c$ | **Bound** | $M_c^{BO}$ | $M_c^{BO+ad}$ | $\Delta_{na}$ | $\Delta_{na}^{pred}$ |
|---|---|---|---|---|---|---|
| (0,1) | 9.525634(4) | 9.52586 | 11.0815 | 10.0632 | 0.5375 | 0.5281 |
| (0,2) | 29.63939(1) | 29.64016 | 31.1678 | 30.1456 | 0.5062 | 0.5124 |
| (0,3) | 59.31866(2) | 59.32056 | 60.8467 | 59.8157 | 0.4970 | 0.5017 |
| (0,4)[a] | 98.26550(11) | 98.26962 | 99.7974 | 98.7585 | 0.4930 | 0.4958 |
| (1,0) | 4.2082827(5) | 4.2082830 | 5.5375 | 4.5706 | 0.3623 | 0.3328 |
| (2,0) | 11.4894217(3) | 11.4894218 | 12.7397 | 11.7786 | 0.2892 | 0.2778 |
| (1,1)[a] | 17.094282(1) | 17.0942819 | 18.4945 | 17.5236 | 0.4294 | 0.4165 |
| (3,0)[a] | 21.376199(1) | 21.3762034 | 22.5996 | 21.6381 | 0.2619 | 0.2562 |
| (4,0)[a] | 33.914349(2) | 33.9143596 | 35.1246 | 34.1616 | 0.2472 | 0.2441 |

Masses in units of $m_e$. $M_c$: extrapolated (threshold-law fits for $J$ = 0, 1 and 2, basis extrapolation otherwise); the digits in parentheses are the uncertainty in the last digits, and the lower side is not rigorous. Bound: certified upper bound (a variational root at least $10^{-11}$ $E_h$ below threshold at that mass). Odd $J$: exchange-antisymmetric; even $J$: symmetric. [a] Pre-specified (Table 2). The lowest symmetric $J$ = 1 level (2p$\sigma_u$-like) binds near $M$ = 813.54(7) (exploratory; BO 815.56).

The offsets are strikingly regular (Fig. 2a). Along the vibrational series, BO minus exact is 1.5558, 1.5284, 1.5280 and 1.5319 for $v$ = 1–4, and BO+ad minus exact is 0.5375, 0.5062, 0.4970 and 0.4930, although the threshold mass itself grows tenfold, from 9.5 to 98.3. Along the rotational series the offsets are smaller but equally steady: 1.33, 1.25, 1.22 and 1.21 for $J$ = 1–4 (BO), and 0.36, 0.29, 0.26 and 0.25 (BO+ad). Thus, the offset depends on the angular momentum and, weakly, on $v$, but not on the size of $M_c$. Note also that the order in which the levels appear is the same in the exact, BO and BO+ad calculations: the BO picture gets the sequence of the spectrum right, and only shifts it.

The theory of eqns (3)–(10) accounts for both parts of the offset. The adiabatic part, about 1.02 of the BO offset for $J$ = 0, follows at first order from eqn (4): it predicts 1.016, 1.021 and 1.030 for $v$ = 1–3, against 1.018, 1.022 and 1.031 from the two solved models, and 0.966 and 0.960 for $J$ = 1 and 2, against 0.967 and 0.961. The agreement is at the 0.2% level, without any fit. The adiabatic part is large because, in the region sampled by the zero-energy state, the adiabatic correction measured from its asymptote (about 0.075 near $R_e$) is comparable to the depth of the well (about 0.10).

The non-adiabatic part, the remaining 0.23 to 0.54, comes from the mass functions of Fig. 1, which enter through eqn (8). Without any adjustable parameter, they give 0.528, 0.512 and 0.502 for $v$ = 1–3, against the exact 0.5375, 0.5062 and 0.4970, and 0.333 and 0.278 for $J$ = 1 and 2, against 0.3623 and 0.2892 (Table 1). The residuals in the offset, +0.0095, −0.0062 and −0.0046 for $v$ = 1–3 and +0.030 and +0.011 for $J$ = 1 and 2, are of the size of the higher-order terms that the theory neglects. The largest is for $J$ = 1, which has the smallest threshold mass, where these terms should be largest (an expectation, not a computed result). Note that the zero-energy states sample mainly 3 to 8 bohr, where $m_v(R)$ has its maximum (0.661 at 5.9 bohr), so that the weighted vibrational mass, 0.48–0.50, is close to the asymptotic value 1/2 even though $m_v(R_e)$ is only 0.347. The non-adiabatic offset is smaller for $J > 0$ because 77–83% of the kinetic weight at threshold is centrifugal, and the rotational mass $m_r$ that governs it is small at short distance.

The same pair of functions also fixes the energy at large mass. As we show below, the exact energy approaches the BO+ad energy as $c_6\kappa^6$, and $m_v(R_e) = 0.34655$ reproduces the vibrational mass inferred from $c_6$ (0.3465 ± 0.0003), while $m_r(R_e)$ = 0.07542 reproduces the value inferred from the exact rotational spacings (0.0754). One pair of functions thus explains the offsets of the thresholds, where $M$ is between 4 and 100, and the energies at large $M$, without any tuning. Therefore, the near-constancy of the BO offsets is not an accident of the levels we happened to compute: it is a structural consequence of the zero-energy problem, and its size is set by known non-adiabatic physics.

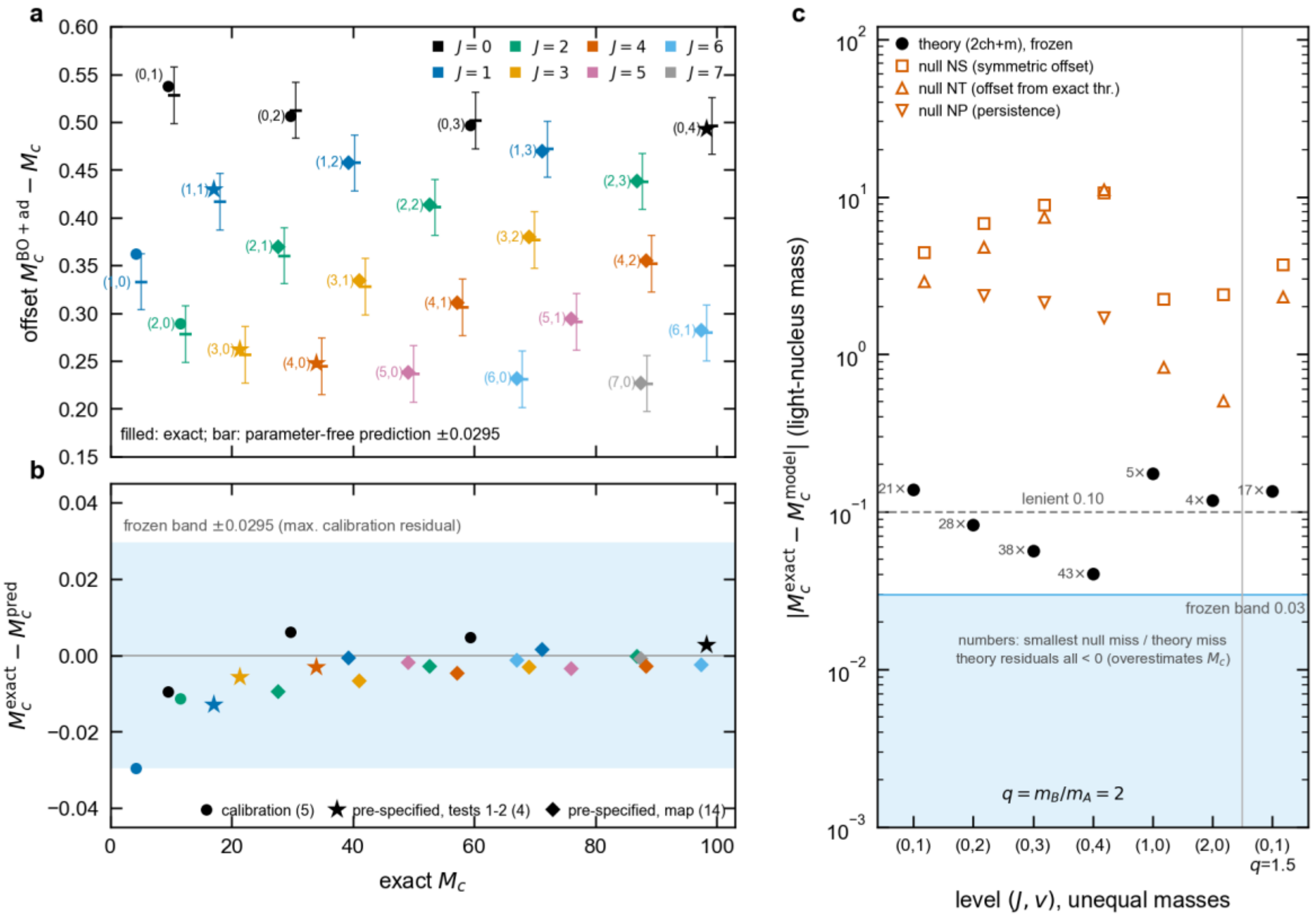

**Fig. 2** Born–Oppenheimer offsets of the threshold masses. (a) Non-adiabatic offset, $M_c^{BO+ad} - M_c$, against the exact threshold mass for the 23 levels of the symmetric family: filled markers, exact; short bars, parameter-free prediction with the frozen band of ±0.0295; colour, $J$. (b) Exact minus predicted threshold mass, with the frozen band: circles, the five levels known before any prediction; stars, the four levels of the first two pre-specified tests; diamonds, the 14 levels of the map. (c) Unequal-mass family ($q$ = 2, and 1.5 for the last level): |exact − model| for the two-channel theory (filled circles) and for the three null models (open symbols), on a logarithmic scale; the numbers give the ratio of the smallest null-model miss to the theory miss.

### 3.3 Pre-specified tests in the symmetric family

We tested the theory on 18 levels whose exact thresholds had not been computed when the predictions were frozen (see Theory and methods); Table 2 and Fig. 2b collect the results. The first freeze covered $v$ = 4 (band ±0.0095), the second the (1,1), (3,0) and (4,0) levels, and the third the 14 remaining levels of the map below $M \approx 100$ (band ±0.0295 for both), which extend the test to $J$ = 5–7 and to the second and third vibrational levels of $J > 0$. All 18 levels lie within their bands. For $v$ = 4, the prediction 98.2627 misses the exact 98.26550(11) by +0.0028, where a miss is the exact minus the predicted threshold. The (1,1), (3,0) and (4,0) predictions miss by −0.0129, −0.0056 and −0.0032. The 14 levels of the map have a median |miss| of 0.0027 and the largest miss of −0.0095, for (2,1). The last of them, (7,0), needed the extended-precision assembly; its prediction, 87.5223, misses the exact 87.521468(1) by only −0.0008. The two $J$ = 6 levels, recomputed with the extended-precision assembly, miss by −0.0012 and −0.0025.

**Table 2** Pre-specified tests of the offset theory in the symmetric family: BO+ad, predicted and exact threshold masses of the 18 levels tested

| $(J,v)$ | $M_c^{\mathrm{BO+ad}}$ | Predicted | Exact $M_c$ | Exact − pred. |
|---|---|---|---|---|
| *Test 1* | | | | |
| (0,4)[a] | 98.7585 | 98.2627 | 98.26550(11) | +0.0028 |
| *Test 2* | | | | |
| (1,1) | 17.5236 | 17.1072 | 17.094282(1) | −0.0129 |
| (3,0) | 21.6381 | 21.3818 | 21.376199(1) | −0.0056 |
| (4,0) | 34.1616 | 33.9175 | 33.914349(2) | −0.0032 |
| *Test 3* | | | | |
| (2,1) | 28.0669 | 27.7068 | 27.6972471 | −0.0095 |
| (1,2) | 39.7665 | 39.3093 | 39.3086240 | −0.0007 |
| (3,1) | 41.4085 | 41.0810 | 41.0743473 | −0.0067 |
| (5,0) | 49.3534 | 49.1173 | 49.115375(5) | −0.0019 |
| (2,2) | 53.0027 | 52.5920 | 52.5891645 | −0.0029 |
| (4,1) | 57.4757 | 57.1697 | 57.1649498(5) | −0.0047 |
| (6,0)[b] | 67.2154 | 66.9849 | 66.983717(1) | −0.0012 |
| (3,2) | 69.3939 | 69.0171 | 69.0138886 | −0.0032 |
| (1,3) | 71.6612 | 71.1896 | 71.1912322 | +0.0016 |
| (5,1) | 76.2429 | 75.9521 | 75.9486530(8) | −0.0034 |
| (2,3) | 87.2254 | 86.7875 | 86.7873305 | −0.0002 |
| (7,0) | 87.7486 | 87.5223 | 87.521468(1) | −0.0008 |
| (4,2) | 88.7090 | 88.3571 | 88.3541941(3) | −0.0029 |
| (6,1)[b] | 97.6986 | 97.4193 | 97.416732(2) | −0.0025 |

Predictions frozen before any exact calculation of the level (test 1: 14 November 2026; test 2: 15 November; test 3: 16 November; author-held, SHA-256 hash). Band: ±0.0295, except [a] ±0.0095. All 18 levels lie within their bands (median |exact − pred.| of test 3: 0.0027). [b] $J$ = 6: recomputed with the extended-precision assembly used for (7,0). Exact values without parentheses have uncertainties of $10^{-9}$–$3 \times 10^{-8}$.

These tests show that the offset of every computed level, over $J$ = 0–7, $v$ = 0–4 and $M_c$ = 4.2–98.3, is predicted without a fitted parameter to a few $10^{-3}$, about 1% of the offset. They confirm the constancy of the offset more than they discriminate between theories, however. A persistence null model, which assumes that the offset of $v$ = 4 equals that of $v$ = 3, predicts 98.2615 and also passes (miss +0.0040). For (1,1), (3,0) and (4,0), persistence estimates chosen after the fact are 4.6 to 8.4 times worse than the theory, but since they were not frozen, the comparison is only suggestive; we did not evaluate null models for the map. The discriminating evidence for the mass-function theory is therefore the parameter-free value of the offsets and the untuned agreement of $m_v(R_e)$ and $m_r(R_e)$ described above.

Note also that 13 of the 14 residuals of the map, and all earlier residuals for $J > 0$, are negative: the predicted offset is systematically a little too small. A band of ±0.01 would have been passed by all 14 levels of the map, so the band set in advance was generous, and the common sign points to a small missing term that we have not identified. Finally, the freezes were held by the author (a read-only file, a SHA-256 hash and a dated local commit) and not time-stamped by a third party, and some small-basis values were seen after two of the freezes (see Theory and methods); we therefore call the predictions pre-specified rather than blind.

### 3.4 Unequal masses: the limits of the theory

Table 3 and Fig. 2c show the test for unequal nuclear masses, $\{A^+(M), B^+(qM), e^-\}$, with $M$ the mass of the lighter nucleus and $q$ = 2 ($HD^+$-like) and 1.5. Here the theory becomes a two-channel (g/u) zero-energy problem, with the exact threshold built in, exact $g/u$ couplings and second-order 2 × 2 mass matrices. Seven predictions were frozen with a band of ±0.03 (0.0368 for $J$ = 1), a lenient band of 0.10 and three null models. None of the seven lies within the band. For $q$ = 2, the misses (exact minus theory) are −0.1396, −0.0834, −0.0567 and −0.0404 for $v$ = 1–4, and −0.1760 and −0.1180 for $J$ = 1 and 2; for $q$ = 1.5 and $v$ = 1, the miss is −0.1362. Three of the six levels at $q$ = 2 ($v$ = 2–4) are within the lenient band.

**Table 3** Test of the two-channel offset theory for unequal nuclear masses, $\{A^+(M),B^+(qM),e^-\}$: threshold mass $M$ of the lighter nucleus

| $q$, $(J,v)$ | Exact $M_c$ | Theory | Exact − theory | Null/theory[a] |
|---|---|---|---|---|
| 2, (0,1) | 11.573761(8) | 11.7133 | −0.1396 | 21 |
| 2, (0,2) | 29.005219(29) | 29.0886 | −0.0834 | 28 |
| 2, (0,3) | 53.41741(10) | 53.4741 | −0.0567 | 38 |
| 2, (0,4) | 84.3454(4) | 84.3858 | −0.0404 | 43 |
| 2, (1,0) | 5.405970 | 5.5819 | −0.1760 | 5 |
| 2, (2,0) | 11.029758(1) | 11.1478 | −0.1180 | 4 |
| 1.5, (0,1) | 11.675521(7) | 11.8117 | −0.1362 | 17 |

Theory frozen on 16 November 2026, before the first exact run, with a band of ±0.03 (±0.0368 for (1,0)) and a lenient band of 0.10: 0 of 7 levels lie within the band, and 3 of 7 within 0.10. [a] Ratio of the smallest miss of the three null models to the miss of the theory.

Nevertheless, the theory captures most of the offset. Along the $J$ = 0 series at $q$ = 2, the exact thresholds differ from the BO+ad ones by 4.02, 6.40, 8.56 and 10.28, so that the theory misses only 3.5%, 1.3%, 0.7% and 0.4% of the offset, and it predicts the change of the offset from one level to the next to within 0.02–0.06. Its misses are 4 to 43 times smaller than those of the closest null model (4 to 5 times for $J$ = 1 and 2, and 28 to 43 times against persistence for $v$ = 2–4). All misses have the same sign, so that the theory places every threshold too late, and they shrink with $v$. Thus, the theory describes most of the physics of the offset for unequal masses as well, but its stated uncertainty was too small, by a factor of up to six.

The failure is not a numerical artefact. At $q$ = 1, the unequal-mass code reproduces the symmetric threshold, 9.525634, to $1.5 \times 10^{-5}$ once the closed channel receives its own diffuse functions, and enlarging the basis at $q$ = 2 and 1.5 moves the $J$ = 0 thresholds by less than $1.2 \times 10^{-6}$, far below the misses. The cause, however, is not identified. The miss at $q$ = 1.5 is almost the same as that at $q$ = 2 (−0.1362 against −0.1396), although the square of the mass asymmetry differs by a factor of almost three, so the missing term does not scale with the square of the asymmetry, as we had expected. Candidates are the splitting of the second dissociation channel, $B^+$ + (Ae), which lies an amount of order $1/M$ above the open one; $g/u$ mixing beyond the two-channel truncation ($2p\pi_u$ and higher $\Sigma$ states); and terms of order $1/M$ that are odd in the mass asymmetry and therefore vanish in the equal-mass family. None of them has been tested. We conclude that the theory is established for equal nuclear masses and incomplete beyond them.

### 3.5 Threshold masses and the bound-state map

We computed the threshold masses of all 23 bound levels of the $1s\sigma_g$ manifold with $M_c \lesssim 100$, from the first rotational level, (1,0) at $M = 4.208$, to (0,4) at $M = 98.27$ (Tables 1 and 2; the complete map is shown in Fig. 3). Fourteen of these thresholds are reported here for the first time: the lowest levels of $J = 3$ to 7, (3,1), (4,1), (5,1), (6,1), (3,2), (4,2), (1,3), (2,3) and (0,4). For $J = 3$ and 4, only a model formula was available,[20] and it differs from our values by 8% and 17%. The other nine thresholds, (1,0), (2,0), (0,1), (1,1), (0,2), (0,3), (2,1), (1,2) and (2,2), were first estimated by Frolov and co-workers with finite basis sets,[17,18,20,21,39] to three to five significant figures and without error bars. With two exceptions, those estimates lie 0.06% to 7% above our converged values, as expected for variational finite-basis calculations: for (2,0) the earlier bracket contains our value, and the (1,1) value of Frolov and Wardlaw[39] lies only 0.0007% above ours. Our thresholds are given to seven to nine significant figures, each with a proven upper bound (see Exact three-body calculations), and can serve as benchmarks for other methods. For the two $J = 6$ levels, the double-precision basis sequences did not converge reliably, and we recomputed them with the extended-precision assembly used for $J = 7$. This gives $M_c(6,0) = 66.983717(1)$ and $M_c(6,1) = 97.416732(2)$; for their largest bases, one of our two guards against spurious roots flags the overlap matrix at the level of rounding errors (ESI, Table S5).

In detail, the earlier estimates for $J > 0$ and for (1,1) agree with ours to 0.3% or better:[17,18,20,21,39] 4.2133 and 4.219 against 4.2082827 for (1,0); the bracket 11.489–11.490, which contains our 11.4894217, for (2,0); and 17.125 and 17.0944 against 17.094282 for (1,1). Their vibrational thresholds, 10.204 and 10.045 for (0,1), 30.70 for (0,2) and 61.135 for (0,3), are 3% to 7% too high, as expected for strongly diffuse near-threshold states described with 600 to 700 functions, and their estimates for (2,1), (1,2) and (2,2) are 0.06%, 0.17% and 1.3% high. All earlier values, with their sources, are listed in ESI Table S6.

The exact thresholds also fix the order in which the levels appear. Rotational and vibrational levels interleave from the start: (1,0) at 4.21, (0,1) at 9.53, (2,0) at 11.49, (1,1) at 17.09, (3,0) at 21.38, (2,1) at 27.70, (0,2) at 29.64 and (4,0) at 33.91. There is therefore no range of M in which a complete rotational ladder exists before vibration begins. The order of appearance is the same in the exact, BO and BO+ad calculations. For the S, P and D levels it agrees with the order first reported by Frolov.[20]

Read naively, the sequence suggests that the system rotates before it vibrates, since (1,0) binds at 4.21 and (0,1) only at 9.53, as Frolov's ordering[20] also implies. This picture is incomplete, because it ignores the exchange symmetry of the two $X^+$ ions. Exchange gives the rotational levels of the ground manifold the symmetry $(-1)^J$, so that, once nuclear spin is included, even- and odd-$J$ levels belong to different nuclear-spin species,[5] para and ortho for spin-½ fermions, which do not interconvert in the isolated system. The (1,0) level is therefore not a rotational excitation of the ground state: it is the ground state of the other species. For spinless bosonic X, odd-$J$ levels do not exist at all. Within either species, the first excitation is vibrational: (0,1) at 9.53 precedes (2,0) at 11.49, and (1,1) at 17.09 precedes (3,0) at 21.38. Between 4.21 and 9.53, each species has a single bound level, and therefore no spectrum at all. To our knowledge, it has not been pointed out that the published sequence combines two nuclear-spin species.

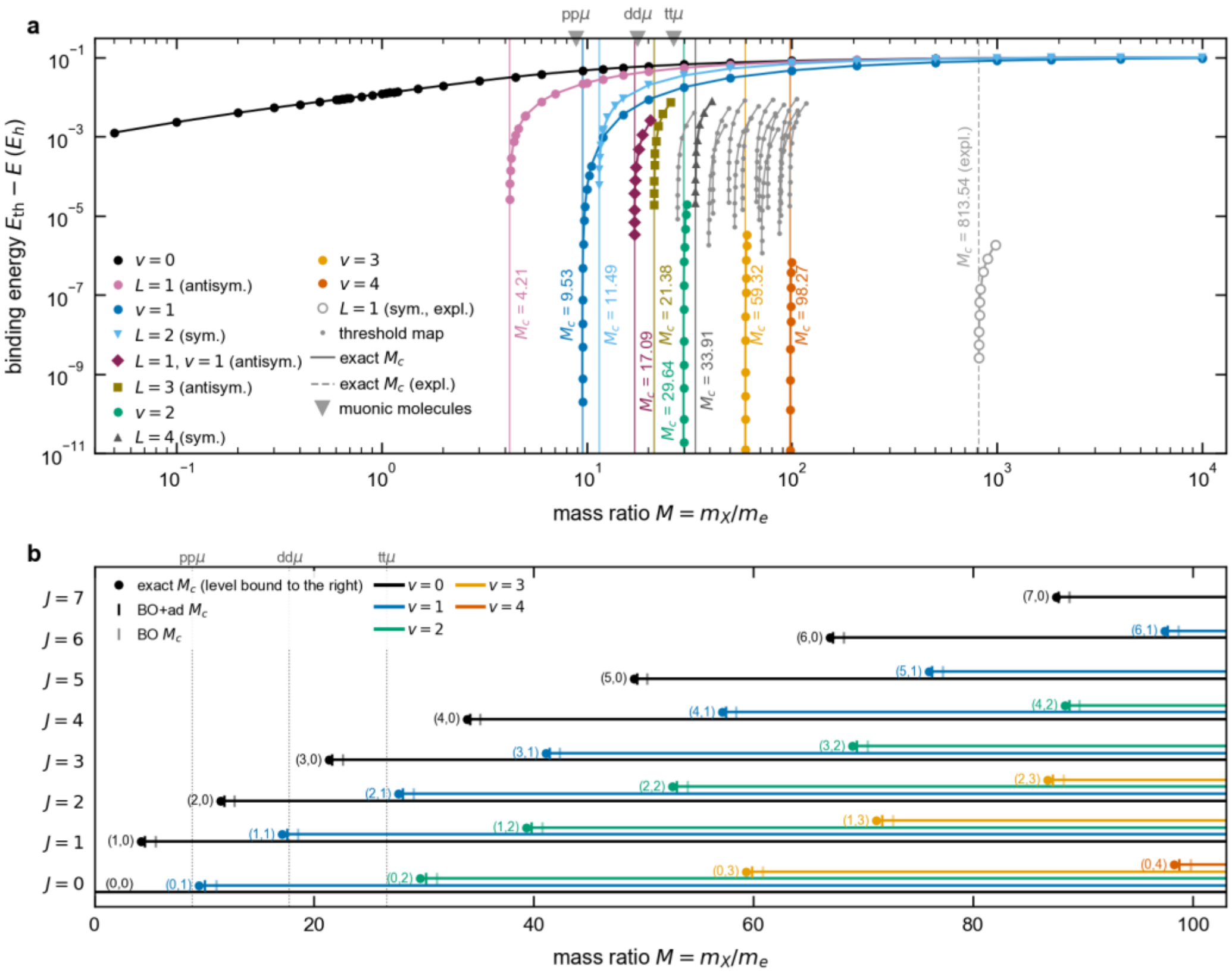

**Fig. 3** Bound levels of $\{X^+,X^+,e^-\}$ near their thresholds. (a) Binding energy $E_{th} - E$ against $M$ (both logarithmic) for $v = 0$–4, the lowest levels of $J = 1$–4, the (1,1) level, the 14 levels of the map (grey dots, $M/M_c - 1 \leq 0.2$) and the exploratory symmetric $J = 1$ level near $M = 813.5$; vertical lines, exact threshold masses; triangles, the mass ratios of pp$\mu$, dd$\mu$ and tt$\mu$. (b) Map of the bound states up to $M \approx 100$: each level $(J,v)$ is a line that starts at its exact threshold mass (filled circle; colour, $v$), with the BO+ad and BO thresholds marked by ticks; dotted lines, the muonic ions.

If we count the bound levels at fixed M, including the ground state, we find 1 at $M = 2$, 3 at 10, 8 at 30, 15 at 60 and 24 at 100. BO undercounts the levels by one at $M = 5$, 10, 30, 40, 50, 60 and 70, that is, at 7 of the 12 masses we sampled. The adiabatic correction removes the undercount at 5, 40, 50, 60 and 70, but not at 10 and 30. Near the bottom of the mass range, therefore, the BO spectrum is often missing a whole level, and BO+ad does not fully repair it.

Near threshold, the binding energy $E_b = E_{th} - E$ follows

$$E_b \simeq A\,(M - M_c)^2 \quad (J = 0)\,, \tag{11}$$

and

$$E_b \simeq B\,(M - M_c) \quad (J > 0)\,, \tag{12}$$

For $J = 0$, the quadratic law of eqn (11) holds (Fig. 4a): a fit $E_b = A(M - M_c)^p$ with the power $p$ left free gives $p = 2.001$, 2.004, 2.006 and 2.011 for $v = 1$–4. For $J > 0$, the ratio $E_b/(M - M_c)$ is constant near threshold (Fig. 4b), as eqn (12) requires, with slopes $B = 2.50 \times 10^{-3}$, $1.712 \times 10^{-3}$ and $1.211 \times 10^{-3}$ $E_h$ per unit of $M$ for $J = 2$, 3 and 4. The $J = 1$ levels, (1,0) and (1,1), are linear only asymptotically: for (1,1), the ratio $E_b/(M - M_c)$ still changes over the range of Fig. 4b, from $3.9 \times 10^{-4}$ to $7.3 \times 10^{-4}$ $E_h$, so we do not quote a single slope, and the thresholds of both levels rest on converged crossings rather than on a fit. The quadratic law is the expected behaviour of an ion bound in an s wave to a neutral polarisable atom: the long-range interaction is the polarisation tail $-C_4/R^4$, with no dipole term, so that the binding is of short-range type. The reduction of the near-threshold problem to this two-body picture is due to Frolov and Bishop.[18,19] For $J > 0$, the centrifugal barrier keeps the state compact at threshold and the law becomes linear, which corrects the quadratic law stated earlier for the $P$ and $D$ states.[18] These laws are not new. We report them because the central values of $M_c$ for $J = 0$ are obtained by extrapolating the binding energy to zero with the quadratic form, whereas the upper bounds do not depend on any fit.

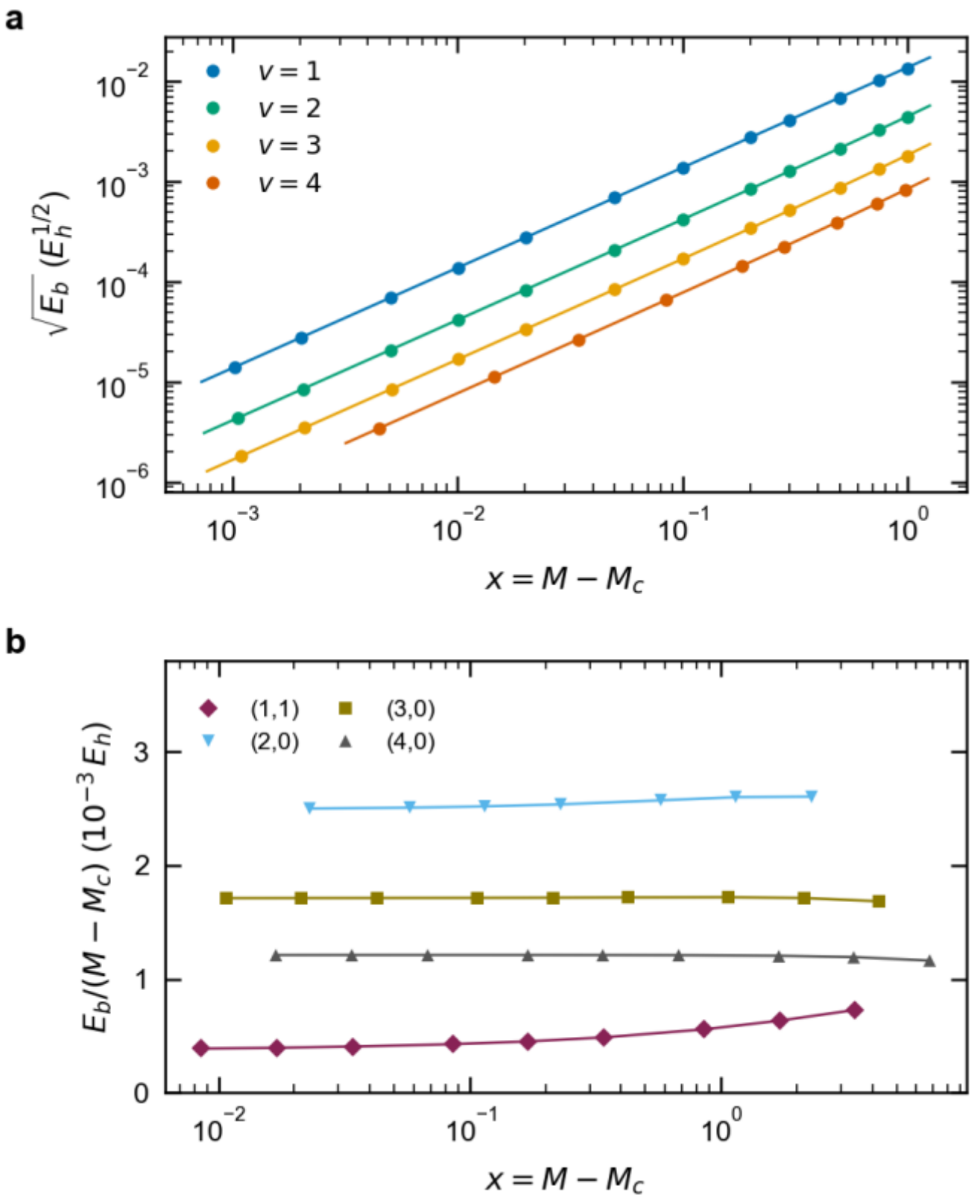

**Fig. 4** Threshold laws. (a) $(E_b)^{1/2}$ against $x = M - M_c$ for the $J = 0$ levels $v$ = 1–4 (logarithmic axes), with fits $c1x + c2x^2$ ($x \leq 0.3$, $M_c$ fixed); a slope of one is the quadratic law of eqn (11). (b) $E_b/(M - M_c)$ against $x$ for the (1,1), (2,0), (3,0) and (4,0) levels; a flat line is the linear law of eqn (12). The (1,1) ratio is still changing at $x \approx 10^{-2}$.

Near threshold, the $J = 0$ states are also enormous. The mean X–X distance approaches the universal halo value, $\langle r_{XX} \rangle \to 1/(2k_b)$, with $k_b = (2\mu E_b)^{1/2}$ and $\mu$ the ion–atom reduced mass (Fig. 5). The ratio $2k_b\langle r_{XX} \rangle$, which equals 1 for an ideal halo, stays within 2% of 1 for $M - M_c \lesssim 10^{-2}$ and rises above 1 as the state becomes more compact. For $v = 1$ it reaches about 11 000 bohr at $M - M_c \approx 0.001$, with one $X^+$ in the atom and the other far away. At its threshold, a new vibrational level is thus an atom plus a distant ion, not a compact molecule. This is one more reason why the threshold mass, and not the shape of the wave function, is the sharp marker.

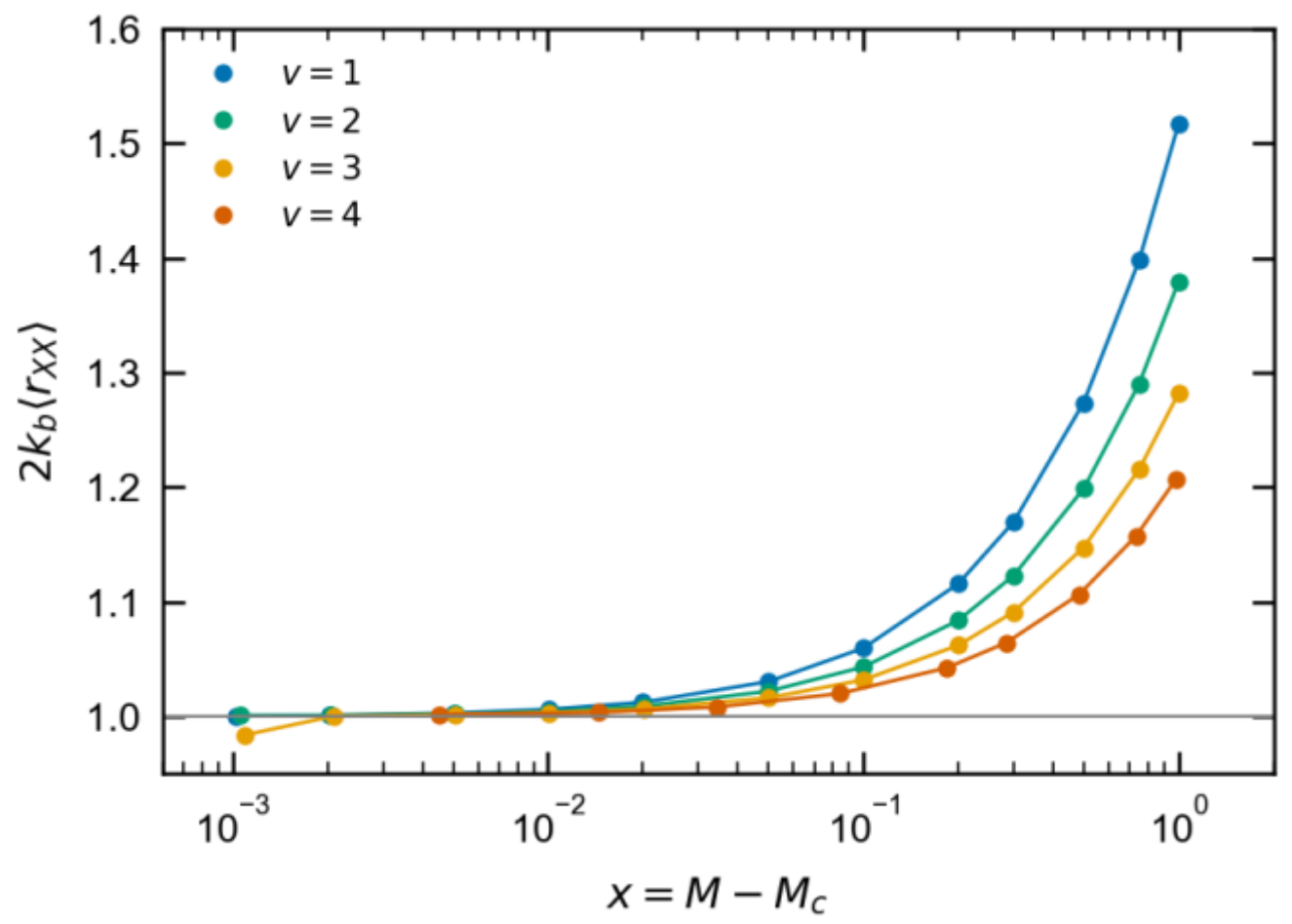

**Fig. 5** Size of the $J = 0$ levels near threshold: $2k_b\langle r_{XX} \rangle$ against $x = M - M_c$ for $v$ = 1–4 (logarithmic $x$ axis), with $k_b = (2\mu E_b)^{1/2}$ and $\mu$ the ion–atom reduced mass. The ratio tends to 1, the zero-range halo limit $\langle r_{XX} \rangle \to 1/(2k_b)$, as $x \to 0$.

### 3.6 Muonic molecular ions

Replacing the electron by a negative muon, which is 206.77 times heavier, gives the muonic molecular ions ppμ, ddμ and ttμ: two protons, deuterons or tritons bound by a $\mu^-$. Because only the mass ratio enters the scaled Hamiltonian, these ions are members of our family with $M = m_a/m_\mu$ (a = $p$, d, t), that is, $M$ = 8.880, 17.752 and 26.585, well inside the map. Muonic molecular ions of this kind, and above all dtμ, are the intermediates of muon-catalysed fusion,[40] and their bound spectra are known accurately,[39] so they provide an independent check of the thresholds.

Our thresholds predict the levels (0,0) and (1,0) for ppμ; in addition (0,1), (1,1) and (2,0) for ddμ; and in addition (3,0) for tt$\mu$, with no (0,2), (2,1) or (4,0) level in any of them. Fig. 6 compares, level by level, our predictions with the complete bound spectra of Frolov and Wardlaw (2, 5 and 6 bound states):[39] apart from the ground level, which is always bound, our thresholds predict 24 cases, 10 bound and 14 not bound, and all 24 agree. The closest cases (ESI, Table S11) are informative: ddμ lies only 3.8% above $M_c(1,1)$ = 17.094, consistent with the very weak binding of that level, and ttμ lies 4.0% below $M_c(2,1)$ = 27.697, which confirms that its sixth bound state is (3,0) and not (2,1).

Direct solutions at the mass ratios used by Frolov also reproduce his binding energies of the ddμ (1,1) and (2,0) levels and of the ttμ (2,0) level, $3.510170 \times 10^{-4}$, 1.5372612 $\times 10^{-2}$ and $3.0694483 \times 10^{-2}$ muon atomic units (1 muon atomic unit = 206.77 $E_h \approx$ 5626 eV), to within one unit in the last of the seven digits he gives.[41] For ttμ (1,1) we obtain $8.034498 \times 10^{-3}$ (45.21 eV), as in the earlier calculations.[39,42] These are consistency checks with independent calculations rather than tests of the thresholds themselves; they show, however, that the map can be read directly as the spectrum of real molecules.

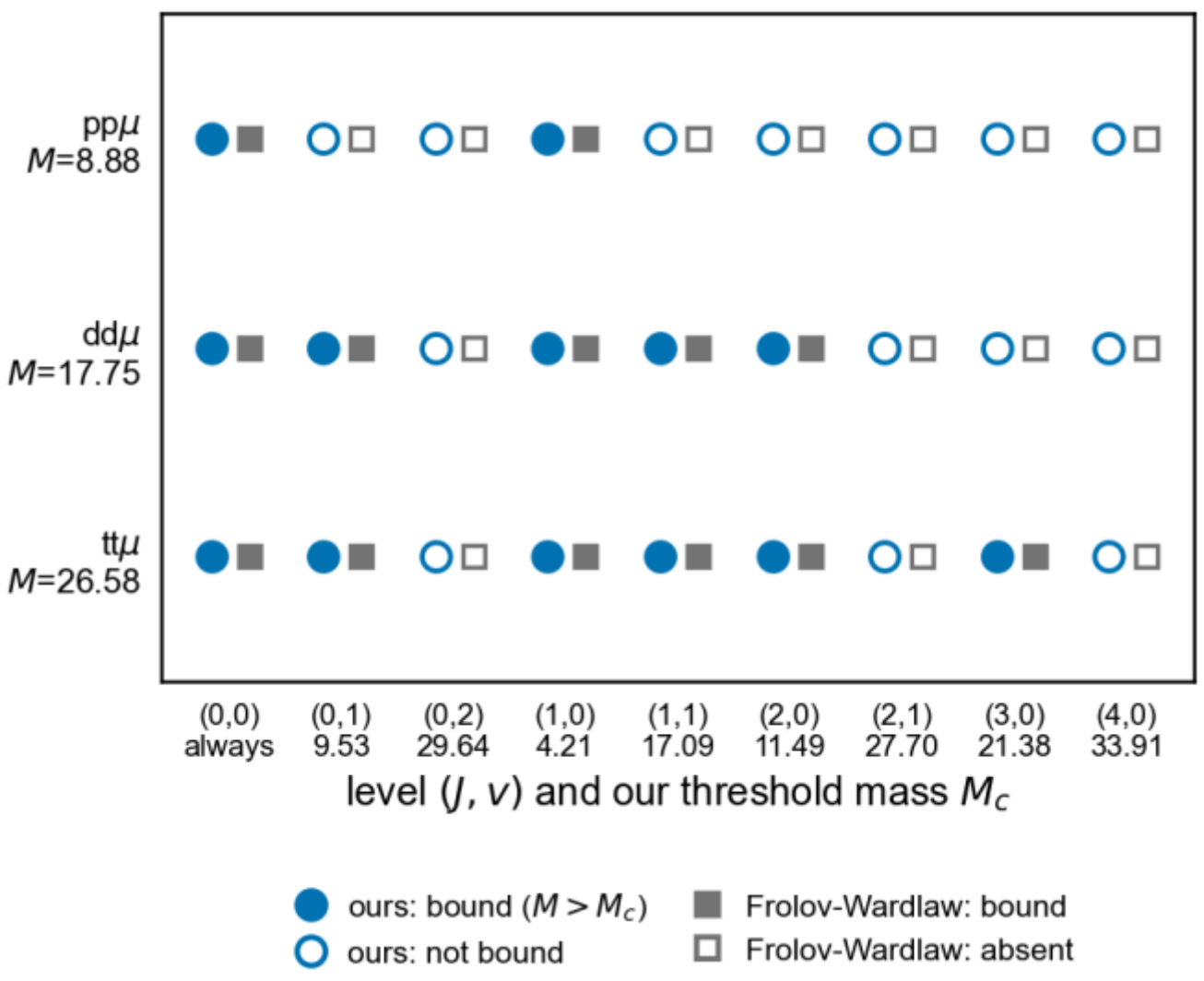

**Fig. 6** Bound levels of pp$\mu$, dd$\mu$ and tt$\mu$. Each column is a level ($J$,$v$) with a computed threshold mass; circles, our prediction from the thresholds; squares, the spectra of Frolov and Wardlaw; filled, bound; open, not bound. All 24 predictions agree.

### 3.7 The ground-state crossover

We now turn to the ground state, which is bound at every M. Its energy is an isolated eigenvalue of a Hamiltonian that depends analytically on 1/M. By the analytic perturbation theory of Kato,[43] the ground state is then a real-analytic function of M: the energy and the wave function have convergent Taylor expansions about every M. No sharp transition can therefore occur at any finite mass, and we see none in a scan of 30 masses from $M = 0.05$ to $10^4$, nor in the fidelity susceptibilities discussed below. The change of character of the ground state, from a floppy three-body system to a molecule with a bond length, is therefore a smooth crossover, and the question is not where it happens but how wide it is. We measure the width with the local exponent of the Lindemann ratio $\Delta R/\langle R\rangle$, the relative width of the distribution of the X–X distance. This ratio is the classic melting criterion for solids and clusters,[44] and here it measures how far the system is from a rigid bond:

$$f(M) = -4\,\frac{d\ln(\Delta R/\langle R\rangle)}{d\ln M}, \qquad (13)$$

which is zero at the maximum of the ratio and tends to 1 in the BO limit, where the ratio falls as $\kappa = M^{-1/4}$. It passes 0.1, 0.5 and 0.9 at $M = 0.455$, 1.52 and 95 (Fig. 7a), a width of 5.3 in ln $M$, or 2.3 decades. A second definition, the range over which the Lindemann ratio falls from 90% to 10% of its maximum, gives 4.3 decades. Either way, the crossover is broad.

Table 4 collects the markers proposed for the atom-to-molecule change, together with those we found, all evaluated on the same exact wave functions. In order of mass, they are the maximum of the Lindemann ratio ($M \approx$ 0.2–0.3); a band of extrema of the electron–nuclear entanglement measures (0.36–0.48); the minimum of the energetic marker of King and co-workers, which we locate without a fit at 0.4924222 (their fit gives 0.491 ± 0.003);[11] the maximum of $E/\mu_{Xe}$ (about 0.5); the minimum of the charge-stability boundary of Kais and Shi (0.543, converted by us);[13] the change of the X density about the centre of mass from a maximum to a hole (0.635 ± 0.01, inside the interval 0.4–0.8 found by Mátyus and co-workers with the same criterion);[10] the sign change of the marginal-distribution criterion of Salas and co-workers (about 0.98, interpolated by us);[12] the maximum of the nuclear kinetic energy (1.11); and the steepest rise of the dilation-corrected fidelity susceptibility (1.39). None of them is sharp, and each depends on the descriptor and, for the density, on the reference point: about the midpoint of the two $X^+$, the density shows no such change at all.[16]

**Table 4** Markers of the atom-to-molecule change of the ground state, and the first threshold masses, in order of $M$

| $M$ | Marker | Source |
|---|---|---|
| 0.2–0.3 | Maximum of the Lindemann ratio | This work |
| 0.36–0.48 | Extrema of the electron–nuclear entanglement measures | This work |
| 0.4–0.8 | X density about the centre of mass: maximum to hole | Mátyus *et al.*[10] |
| 0.491(3) | Minimum of $g = (E_0 - E_{th})/E_{th}$ (fitted) | King *et al.*[11] |
| 0.4924222(1) | Minimum of $g$, without fit | This work |
| ≈ 0.5 | Maximum of $E/\mu_{Xe}$ | This work |
| 0.543[a] | Minimum of the charge-stability boundary | Kais and Shi[13] |
| 0.635(10) | X density about the centre of mass: maximum to hole | This work |
| ≈ 0.98[a] | Sign change of the marginal-distribution criterion | Salas *et al.*[12] |
| 1.11 | Maximum of the nuclear kinetic energy | This work |
| 1.39 | Steepest rise of the dilation-corrected fidelity susceptibility | This work |
| 4.2082827(5) | First rotational level binds ($J = 1$, antisymmetric) | This work |
| 9.525634(4) | First vibrational level binds | This work |
| 11.4894217(3) | First symmetric rotational level binds ($J = 2$) | This work |

The first eleven entries are smooth markers of the ground state; the last three are sharp thresholds. Width of the crossover: $f(M)$ of eqn (13) passes 0.1, 0.5 and 0.9 at $M = 0.455$, 1.52 and 95 (2.3 decades); the ground-state markers span 0.32–0.36 of this width. [a] Converted (Kais and Shi) or interpolated (Salas *et al.*) by us from the published parameters.

However, the markers are not scattered across the crossover. Together they span 0.32 of its width (0.36 with the fidelity marker), and all of them lie below its midpoint, at $M = 1.5$. Our list is not exhaustive, and several of its entries were found during this work rather than proposed in advance (Table 4 distinguishes the two). The clustering does not depend on them, however: the markers proposed in the literature span only 0.17 of the width. The ground-state descriptors are thus onset indicators of a broad change, not competing definitions of a point. This region is also out of reach of the BO picture: plain BO has no bound state below $M = 0.965$, and BO+ad is qualitatively wrong below $M \approx 3$, where it overestimates the Lindemann ratio by up to 66%.

The threshold masses computed in this work behave differently. From 4.21 to 98.27, they span 1.16 widths and cover the rest of the crossover (Fig. 7b), so the descriptors agree with each other much better than with the thresholds. To our knowledge, the thresholds have not been used before as markers of the emergence of molecular structure: earlier work on this question relied on ground-state descriptors,[10–13] and the earlier threshold calculations addressed the stability of the ions rather than their molecular character.[18–20] Unlike the ground-state descriptors, each threshold marks a sharp event, the appearance of a new vibrational or rotational level, at a well-defined mass, and together they follow the build-up of the molecular spectrum across the rest of the crossover.

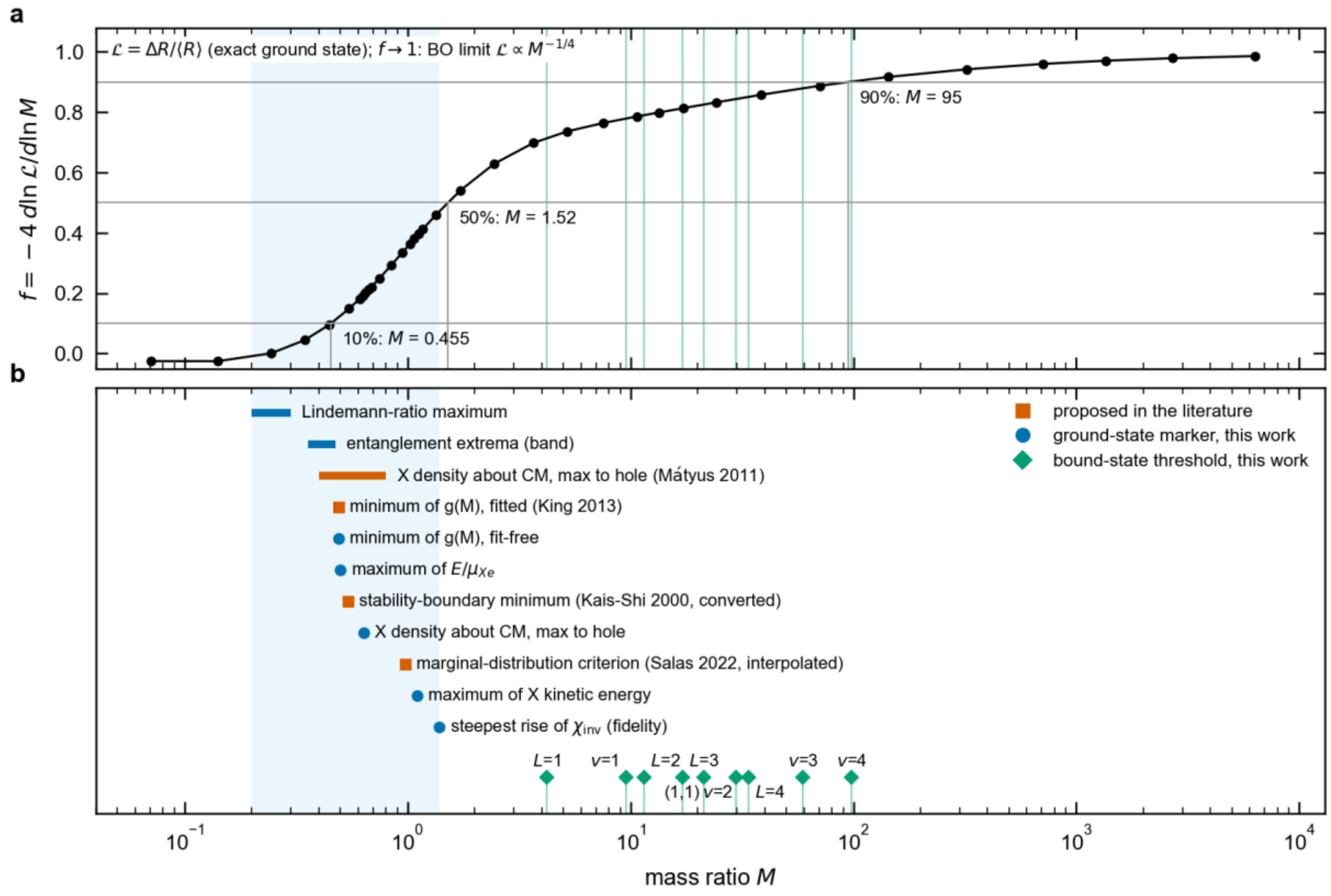

**Fig. 7** The ground-state crossover. (a) Local exponent $f(M)$ of the Lindemann ratio, eqn (13), with its 10%, 50% and 90% crossings at $M = 0.455$, 1.52 and 95. (b) The markers of Table 4, one per row, coloured by provenance (proposed in the literature or found in this work), and the first threshold masses; the shaded band spans the ground-state markers.

As a further test for a sharp change, we computed the fidelity susceptibility $\chi_F$, which measures how quickly the ground-state wave function changes with ln M and peaks at quantum phase transitions in many-body systems.[45,46] Most of $\chi_F$ comes from a trivial effect: as M grows, the wave function contracts as a whole, and this dilation accounts for 98% of $\chi_F$ at $M = 1$. With the dilation removed ($\chi_F^{inv}$; definitions in ESI, Section S1.13), the susceptibility rises smoothly near $M \approx 1.4$ and has only a broad maximum near $M = 6.9$, 2.6% above its background, before it tends to 1/32 at large M, the value set by the $M^{-1/4}$ scaling of the vibrational width (ESI, Section S3.12). No sharp feature appears, which confirms the picture above. The limit is not by itself a signature of structure: in exactly solvable harmonic models (ESI, Section S3.12), a structureless system also reaches a plateau, at 3/64, while a model with a bond reaches 1/32.

### 3.8 Approach to the Born–Oppenheimer limit

As M grows, the exact energy approaches the BO limit as an asymptotic series in $\kappa = M^{-1/4}$. Measured from plain BO, the leading term is of fourth order,

$$E - E_{\mathrm{BO}} \simeq a_4\kappa^4\,, \tag{14}$$

and measured from BO with the adiabatic correction, it is of sixth order,

$$E - E_{\mathrm{BO+ad}} \simeq c_6\kappa^6 + c_8\kappa^8\,. \tag{15}$$

The difference $E - E_{BO}$ is positive wherever BO is bound. A fit $E - E_{BO} = a\kappa^p$ with the power p left free gives $p = 3.995(2)$, and $M(E - E_{BO})$ tends to $a_4 = 0.1746$, which equals the adiabatic correction at $R_e$ (0.1748). The leading failure of plain BO is therefore the diagonal adiabatic correction, at order $\kappa^4$, and not the non-adiabatic coupling.

Once the adiabatic correction is included, $E - E_{BO+ad}$ is negative at every M and falls as $\kappa^6$ (Fig. 8a): the same free fit gives $p = 6.07(1)$. Because free fits are biased by the next order, we quote fits with fixed exponents, which give $c_6 = -0.03931(3)$ and $c_8 = -0.074(2)$; the value of $c_6$ still depends slightly on the fit window. It is reproduced by the vibrational mass function at the equilibrium distance, $c_6 = -m_v(R_e)(2k)^{1/2}/4$, where k is the curvature of the BO potential at $R_e$. The first non-adiabatic correction to the energy is thus a vibrational-mass effect.

The structural descriptors approach the BO limit with the powers of κ that the BO expansion predicts: 1 for the Lindemann ratio, 2 for $\langle R \rangle - R_e$, for $-dE/d \ln M$ and for the vibrational spacing, and 4 for the rotational spacing $d_1$ (ESI, Table S12). The fitted exponents, 0.980, 2.036, 2.11, 1.905 and 3.952, differ from these values by at most 0.11, and the local slopes at the largest mass, 0.986, 2.025, 2.08, 1.94 and 3.97, are closer still. The deviations therefore come from the next order of the series, not from a different law. With the adiabatic correction, BO overestimates the spacings by only about 0.15/M (vibration) and 0.075/M ($d_1$) in relative terms, whereas the errors of plain BO change sign at intermediate mass. For $H_2^+$, the exact spacings are 2191.10 $cm^{-1}$ (vibration) and 58.232 $cm^{-1}$ ($d_1$).

The rotational spacings approach the rigid-rotor limit from below (Fig. 8b). For a rigid rotor, the ratio of the first two spacings is $d_2/d_1 = 3$. The exact ratio is 2.475 at $M = 20$, 2.932 for $Mu_2^+$ ($\mu^+\mu^+e^-$, $M = 206.77$), 2.992 for $H_2^+$ and 2.99853(10) at $M = 10^4$ (ESI, Table S13). BO reproduces it to $10^{-4}$ or better for $M \geq 100$, so the deficit is centrifugal distortion, which BO already contains, and not a non-adiabatic effect. At small M, in contrast, rotation and vibration are not separated at all: at $M = 12$, the vibrational spacing is only 2.2 times $d_1$.

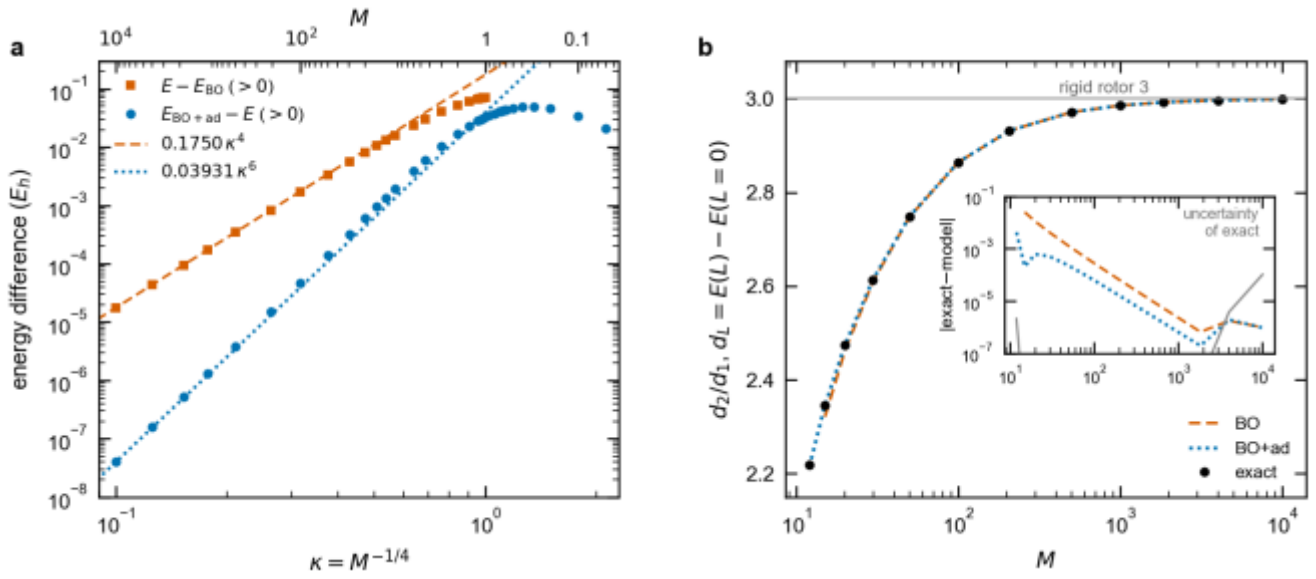

**Fig. 8** Approach to the Born–Oppenheimer limit. (a) $E - E_{BO}$ (squares) and $E_{BO+ad} - E$ (circles) against $\kappa = M^{-1/4}$ (logarithmic axes), with the laws $0.1750\kappa^4$ (dashed) and $0.03931\kappa^6$ (dotted). (b) Ratio of the first two rotational spacings, $d_2/d_1$, against $M$: exact (circles), BO (dashed) and BO+ad (dotted), with the rigid-rotor value 3; the inset shows |exact − model| against the uncertainty of the exact ratio.

### 3.9 Electron–nuclear entanglement

We also quantify the entanglement between the electron and the nuclei across the family. Electron–nuclear entanglement in molecular eigenstates is well known, also in BO states,[47–50] but, to our knowledge, its exact decomposition into orientational and radial parts in the laboratory frame, and its behaviour from $Ps^-$ to the BO limit, have not been evaluated before. We partition the ground state ($J = 0$) between the vector R joining the two $X^+$ and the electron coordinate ρ relative to their midpoint. Since $J = 0$, the reduced state of **R** is block-diagonal in the partial wave $l$, only even $l$ occur, and its von Neumann entropy splits exactly into an orientational and a radial part,

$$S = S_{\text{orient}} + S_{\text{rad}} \,, \tag{16}$$

with

$$S_{\text{orient}} = -\sum_l w_l \ln \frac{w_l}{2l+1}, \qquad S_{\text{rad}} = \sum_l w_l S_l \,, \tag{17}$$

where $w_l$ is the weight of partial wave $l$ and $S_l$ the entropy of the normalised radial block $l$. We evaluate $S$ from the exact wave function by projecting it onto partial waves up to $l$ = 64 and diagonalising each radial block on a quadrature grid (ESI, Sections S1.12 and S2.9); the neglected partial waves contribute less than $10^{-5}$. The entropy is nonzero at every M (Fig. 9): 1.495 nats at $M = 0.05$, a maximum of 1.604 near $M \approx 0.41$, 1.516 for $Ps^-$, 0.299 for $H_2^+$ and 0.272 at $M = 10^4$, converged to about $10^{-5}$. Its smooth extrema, and those of two radial-only measures, the logarithmic negativity and the radial mutual information (ESI, Section S1.12), fall in a narrow band, $M$ = 0.36–0.48, inside the onset region of the crossover but coincident with none of the other markers.

In the BO limit, the radial part vanishes, as $\kappa^2 \ln(1/\kappa)$, but the entropy does not: it tends to 0.249338 nats, all of it orientational. The electron keeps a small content of $l = 2, 4, \ldots$ ($w_0 = 0.95865$, $w_2 = 0.03926$) that is locked to the direction of the molecular axis, and eqn (17) turns these weights into a finite orientational entropy. This is the entanglement of the electron with the orientation of the molecular frame, which any $J = 0$ state carries in a laboratory-frame partition; it is neither a failure of the BO approximation nor a result about decoherence. The orientational coherence of the axis, once the electron is traced out, saturates at 0.94, so that the electron alone does not make the molecular axis classical; for that, an environment is needed.

As an independent check, we evaluated the body-frame partition of Sanz-Vicario and co-workers, in which the orientational part is absent by construction, with our exact $H_2^+$ wave function. It gives 0.047954 bits against their 0.0479102 (0.1%), even though their energy is $6.6 \times 10^{-5}$ $E_h$ above the exact value.[49] The BO product state, the clamped-nuclei electronic state times the BO nuclear state, is by contrast qualitatively wrong at small mass (2.42 against 1.17 nats at $M = 3$), and its error at large M appears to fall as 1/M (three points only).

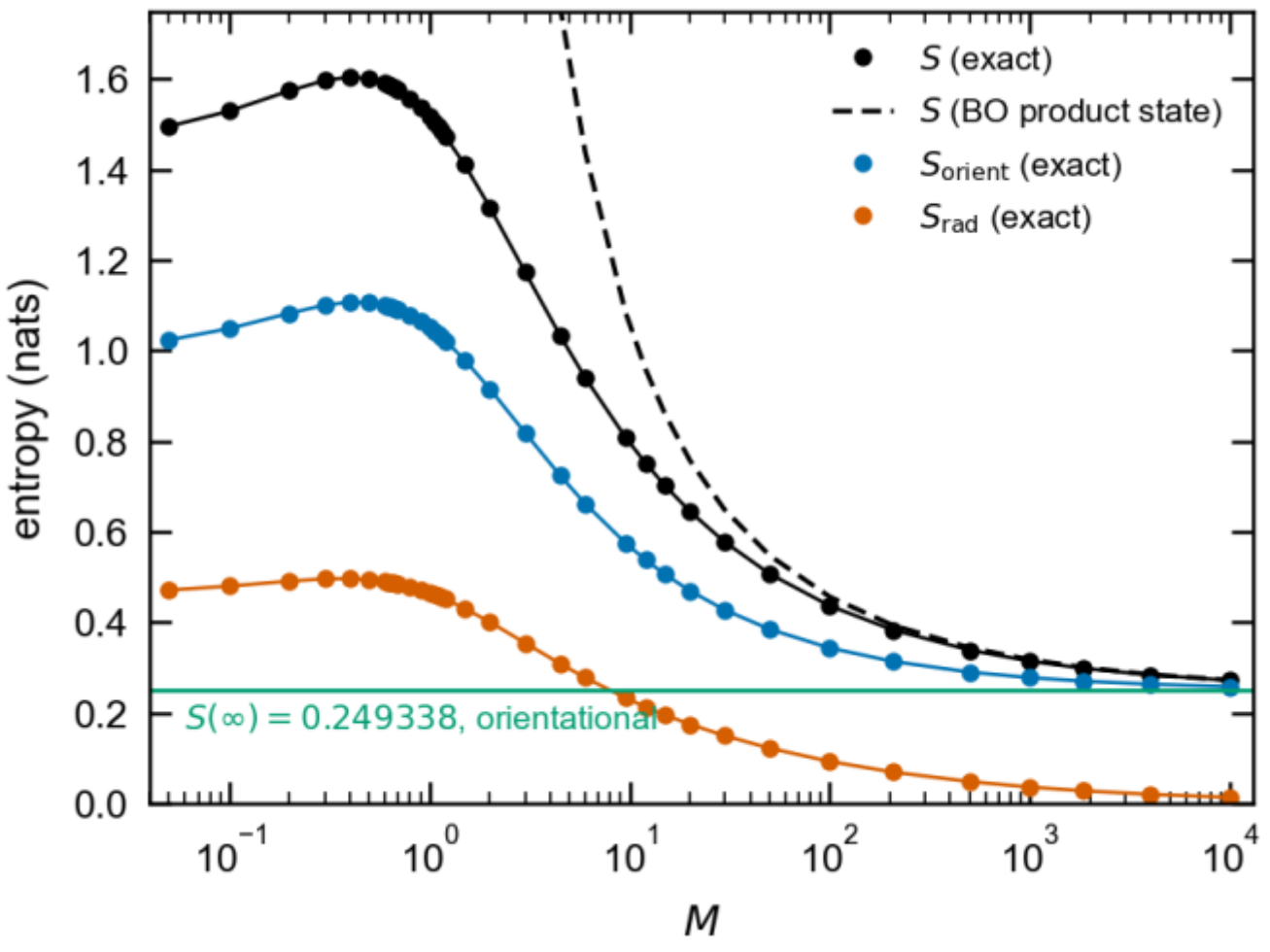

**Fig. 9** Electron–nuclear entanglement of the exact ground state against $M$: total entropy $S$ and its orientational and radial parts, eqns (16) and (17); dashed line, $S$ of the BO product state ($M \geq 3$); horizontal line, the BO limit, 0.249338 nats, which is purely orientational.

## 4. Conclusions

In this work, we have asked where a molecule begins in the three-particle family $\{X^+, X^+, e^-\}$. The answer is gradually. The ground state changes character over a crossover 2.3 decades wide in $M$, and the descriptors proposed to mark the change, which we evaluated on the same exact wave functions, cluster at its onset without singling out a point. The sharp landmarks are the threshold masses: the first rotational level binds at $M = 4.208$, in the ortho-type nuclear-spin species, and the first vibrational level at $M =$ 9.526. Beyond them, the levels keep appearing one at a time: we have determined the threshold masses of all 23 levels of the $1s\sigma_g$ manifold below $M \approx 100$, with a relative precision of $10^{-6}$–$10^{-9}$ for most levels and certified upper bounds, which, to our knowledge, makes them the most precise reported for these systems. The muonic ions pp$\mu$, dd$\mu$ and tt$\mu$ confirm the map, level by level.

The BO approximation places every one of these thresholds too late, by a nearly constant amount, 1.2 to 1.6 (0.23 to 0.54 with the adiabatic correction). We have shown that this constancy is structural: the zero-energy BO equation is a linear eigenvalue problem in $M$, so that every correction of order $1/M$ shifts a threshold mass by a constant. The adiabatic part of the shift follows at first order, to 0.2%, and the non-adiabatic part follows from the vibrational and rotational mass functions, with no fitted parameter; the same functions reproduce the $\kappa^6$ term of the energy at large mass. Pre-specified predictions of the theory passed for all 18 levels tested in the symmetric family, with a typical miss of 0.003 against a band of 0.03.

The theory has a clear boundary. It is established for equal nuclear masses; for unequal masses it still captures 91 to 99.6% of the offset and outperforms simple null models by a factor of 4 to 40, but it misses its band, which points to a missing term, most likely from the second dissociation channel. The central threshold values rest on extrapolation, whereas the upper bounds are certified, and our results concern isolated three-particle eigenstates, which have a bond length but no shape and, even in the BO limit, no classical orientation.

These limits also define the next steps. The unequal-mass discrepancy is the most informative open problem, and the same exact-plus-BO approach extends naturally to $v \geq 5$, to heteronuclear and positronic systems and, most interestingly, to molecular shape in systems of three heavy particles, which we leave to future work.

More broadly, our results give a practical way to read the BO picture at finite mass. BO gets the order of the spectrum right, from the first bound level onward, and it misplaces each threshold by an amount that, in the symmetric family, a parameter-free theory predicts in advance to a few thousandths. The threshold map provides benchmark values for any method that treats nuclei and electrons on an equal footing, and the offset theory turns the question of whether the BO approximation holds into a quantitative one: by how much it fails, and why. In this sense, a molecule does not begin at a single mass. It begins level by level, at masses that we can now compute exactly and predict from the BO picture itself.

## Author contributions

R.P.: conceptualisation, methodology, software, investigation, writing.

## Conflicts of interest

There are no conflicts to declare.

## Data availability

The code, the numerical results (including the frozen pre-specified predictions with their SHA-256 hashes) and the scripts that regenerate every figure and table are openly available at https://github.com/peverati/Ethreshold_BOoffset and archived on Zenodo (https://doi.org/10.5281/zenodo.24398482), under MIT (code) and CC BY 4.0 (data) licences.

## Acknowledgements

This paper is dedicated to Martin Head-Gordon, who was my mentor and postdoc advisor between 2012 and 2016. His long-standing work on energy decomposition analysis and on the nature of the chemical bond inspired this project. In his paper with Levine on the origin of the covalent bond in $H_2^+$ and $H_2$,[51] the emergence of the chemical bond is traced back to the lowering of the electronic kinetic energy by delocalisation over both centres, an elegant explanation obtained inside the Born–Oppenheimer picture. The present paper asks the question that comes one step earlier, at which mass ratio that picture, and with it a bond of definite length, emerges from the full three-body problem, and I like to think of it as a small footnote to his work.

During the preparation of this study and manuscript, the author used Claude Code (Anthropic) with the models Opus 5.5 (82%), Sonnet 5.5 (14%), Fable 5.2 (3.5%), and Haiku (0.2%) for the purposes of writing and testing the solver and analysis code, running and post-processing calculations under the author's direction, searching, checking, and summarising the literature, generating figures from computed data, and drafting working versions of the text and the ESI. The author rewrote the text, has reviewed and edited all other output, and takes full responsibility for the content of this publication. The literature search was assisted by the same tool; every cited work was checked against the publisher's record and read in full by the Author. No numerical result in this paper was generated by the language model itself. Every number comes from deterministic code that is archived with the paper and covered by test suites, and each was reproduced or spot-checked independently of the run that produced it. Representative prompts, paraphrased, were: write a variational three-body solver in a perimetric Laguerre basis for arbitrary masses, with tests against the hydrogen, helium, and $H_2^+$ limits; locate the mass at which the first vibrational level becomes bound, with a certified upper bound and an extrapolated value, and report the convergence honestly; compute the BO and adiabatic thresholds and explain the offset from the exact values; freeze the prediction in a read-only, hashed file before running the exact calculation; set up the long calculation so that the author can launch it; read this paper in full and compare its numbers with ours; check the literature for prior work on these thresholds.

# Electronic Supplementary Information

## Where does a molecule begin? Threshold masses of three-particle Coulomb systems and a predictive theory of their Born–Oppenheimer offsets

Roberto Peverati

## Contents

All quantities are in Hartree atomic units ($m_e = \hbar = e = 1$) unless stated otherwise; $M = m_X/m_e$ is the mass ratio, $\kappa = M^{-1/4}$ the Born–Oppenheimer (BO) expansion parameter, and $\Lambda = M + 1/2$ the coupling of the zero-energy BO problem. Levels are labelled $(J, v)$, with $J$ the total orbital angular momentum and $v$ the vibrational index; spin is not part of the calculation.

# S1 Theory and derivations

## S1.1 Hamiltonian in Jacobi coordinates

For two particles X⁺ of mass $M$ at $\boldsymbol{x}_0$, $\boldsymbol{x}_1$ and an electron of mass 1 at $\boldsymbol{x}_2$, we use the Jacobi coordinates

$$\boldsymbol{R} = \boldsymbol{x}_1 - \boldsymbol{x}_0, \qquad \boldsymbol{\rho} = \boldsymbol{x}_2 - 1/2\,(\boldsymbol{x}_0 + \boldsymbol{x}_1), \qquad \boldsymbol{X}_{\mathrm{cm}} = \frac{M(\boldsymbol{x}_0 + \boldsymbol{x}_1) + \boldsymbol{x}_2}{2M + 1}. \tag{S1}$$

The chain rule gives

$$\nabla_0 = M/M_t\,\nabla_{\mathrm{cm}} - \nabla_R - 1/2\,\nabla_\rho, \quad \nabla_1 = M/M_t\,\nabla_{\mathrm{cm}} + \nabla_R - 1/2\,\nabla_\rho, \quad \nabla_2 = 1/M_t\,\nabla_{\mathrm{cm}} + \nabla_\rho, \tag{S2}$$

with $M_t = 2M + 1$. Inserting these into the kinetic energy, all cross terms cancel:

$$\frac{\nabla_0^2 + \nabla_1^2}{2M} + \frac{\nabla_2^2}{2} = \frac{\nabla_{\mathrm{cm}}^2}{2M_t} + \frac{1}{M}\nabla_R^2 + \frac{2M+1}{4M}\nabla_\rho^2. \tag{S3}$$

After removal of the centre of mass, the internal Hamiltonian is eqn (1) of the main text,

$$\hat{H}(M) = -\frac{1}{M}\nabla_R^2 - \frac{2M+1}{4M}\nabla_\rho^2 + \frac{1}{R} - \frac{1}{|\boldsymbol{\rho} + \boldsymbol{R}/2|} - \frac{1}{|\boldsymbol{\rho} - \boldsymbol{R}/2|}. \tag{S4}$$

The nuclear reduced mass is $\mu_N = M/2$ and the electron reduced mass is $\mu_e = 2M/(2M + 1)$. Because the two $X^+$ have equal mass, the midpoint of $\boldsymbol{R}$ is their centre of mass and the coordinates are orthogonal with respect to the kinetic energy: there is no mass-polarisation (cross) term $\nabla_R \cdot \nabla_\rho$. The masses enter only through the kinetic energy, and $\widehat{H}$ is a family of operators on a fixed Hilbert space, analytic in $\lambda = 1/M$: the coefficient of $\nabla_R^2$ is $\lambda$, and that of $\nabla_\rho^2$ is $1/2 + 1/4\,\lambda$. Two consequences are used throughout. (i) The isolated ground-state eigenvalue, and the ground state itself, are real-analytic in $M$ (Kato's analytic perturbation theory), so no sharp transition can occur in the ground state at finite $M$. (ii) Ground states at different $M$ can be compared directly (overlaps, fidelities).

## S1.2 Thresholds

The lowest dissociation channel is a hydrogen-like atom $(\mathrm{X\,e})_{1s}$, with reduced mass $M/(M + 1)$, and a distant $X^+$:

$$E_{\mathrm{th}}(M) = -\frac{M}{2(M+1)}. \tag{S5}$$

The same threshold holds for every natural-parity $J$ (S1.10). In the BO approximation, the electronic energy for an electron of reduced mass $\mu_e$ tends to $-\mu_e/2 = -M/(2M + 1)$; with the diagonal adiabatic correction it tends to $-\mu_e/2 + \mu_e^2/(4M)$, which equals eqn (S5) to first order in $1/M$ only. The BO and exact limits differ by $M/[(2M + 1)(2M + 2)] \approx 1/(4M)$, that is, $1.36 \times 10^{-4}$ Eh for $H_2^+$ and 0.083 Eh at $M = 1$. Each model (BO, BO+ad, exact) is therefore counted at its own dissociation limit, and a level of that model is bound when its energy lies below that limit. The critical (threshold) mass $M_c(J, v)$ of a level is the mass at which its energy crosses the dissociation limit.

## S1.3 The BO electronic problem and Coulomb scaling

The clamped-nuclei electronic Hamiltonian for an electron of mass $\mu_e$ is the unit-mass $H_2^+$ Hamiltonian with all lengths divided by $\mu_e$ and all energies multiplied by $\mu_e$. Hence, exactly,

$$E_{\mathrm{el}}(R; M) = \mu_e\, U(\mu_e R), \qquad \phi_0(\boldsymbol{\rho}; R) = \mu_e^{3/2}\, \phi(\mu_e \boldsymbol{\rho}; \mu_e R), \tag{S6}$$

where $U(r)$ is the unit-mass $1s\sigma_g$ potential curve (electronic energy plus $1/R$) and $\phi$ the unit-mass electronic function. A single curve $U(r)$, and a single adiabatic function $D(r)$, therefore serve the whole family. We write $\Delta U(r) = U(r) + 1/2$ and $\Delta D(r) = D(r) - 1/4$ for both functions measured from their asymptotes ($U(\infty) = -1/2$, $D(\infty) = 1/4$).

## S1.4 The zero-energy BO equation as a linear eigenvalue problem

The BO radial equation for a level with angular momentum $J$, with the nuclear kinetic coefficient $1/M$ of eqn (S4), is (eqn (2) of the main text, without the adiabatic term)

$$\left[-\frac{1}{M}\frac{d^2}{dR^2} + \frac{J(J+1)}{MR^2} + \mu_e U(\mu_e R)\right]\chi(R) = E\,\chi(R). \tag{S7}$$

A new level appears when its energy reaches the dissociation limit, $E = -\mu_e/2$ (zero binding energy). At that energy,

$$\left[-\frac{1}{M}\frac{d^2}{dR^2}+\frac{J(J+1)}{MR^2}+\mu_e\,\Delta U(\mu_e R)\right]\chi=0. \tag{S8}$$

With $r=\mu_e R$, $d^2/dR^2=\mu_e^2\,d^2/dr^2$ and $1/R^2=\mu_e^2/r^2$, every kinetic term acquires the factor $\mu_e^2/M$. Multiplying by $M/\mu_e^2$ gives

$$-\chi''+\frac{J(J+1)}{r^2}\chi=\frac{M}{\mu_e}[-\Delta U(r)]\chi, \qquad \frac{M}{\mu_e}=M+\frac{1}{2}\equiv\Lambda. \tag{S9}$$

The operator on the left and the weight $-\Delta U(r)$ on the right do not depend on $M$; the mass enters only as the eigenvalue $\Lambda$. A BO level therefore binds when $\Lambda$ reaches an eigenvalue $\Lambda_c$ of a fixed (Sturmian, or coupling-constant) eigenproblem, and

$$M_c=\Lambda_c-1/2. \tag{S10}$$

The diagonal adiabatic correction, $E_{\text{ad}}(R;M)=\mu_e^2 D(\mu_e R)/M$, is of order $1/M$ in the energy; after the multiplication by $M/\mu_e^2$ it becomes the $M$-independent term $\Delta D(r)$. Including it with a switch $c$ ($c=0$: BO; $c=1$: BO+ad) gives eqn (3) of the main text,

$$-u''+\left[\frac{J(J+1)}{r^2}+c\,\Delta D(r)\right]u=\Lambda[-\Delta U(r)]u. \tag{S11}$$

## S1.5 Constant shifts: first-order perturbation theory for $\Lambda_c$

Write eqn (S11) as $\hat{A}u=\Lambda\hat{B}u$, with $\hat{A}=-d^2/dr^2+J(J+1)/r^2$ and $\hat{B}=-\Delta U>0$ in the region where $u$ lives. For an $M$-independent perturbation $\delta\hat{A}$, ordinary Rayleigh–Schrödinger theory for the generalised eigenproblem gives

$$\delta\Lambda_c=\frac{\langle u|\delta\hat{A}|u\rangle}{\langle u|\hat{B}|u\rangle}, \tag{S12}$$

where $u$ is the zero-energy solution; the integrals converge because $\hat{B}\propto r^{-4}$ at large $r$. Any correction of order $1/M$ to the energy is, after the scaling of S1.4, an $M$-independent $\delta\hat{A}$, so it shifts $\Lambda_c$, and therefore $M_c$, by an $M$-independent amount. For the adiabatic correction, $\delta\hat{A}=\Delta D$, which gives eqn (4) of the main text:

$$M_c^{\text{BO}}-M_c^{\text{BO+ad}}\simeq\frac{\langle u|\,1/4-D|u\rangle}{\langle u|-\Delta U|u\rangle}. \tag{S13}$$

**Why the shift is nearly the same for every level (interpretation).** In the semiclassical (WKB) limit, $u^2\approx 1/p$ with $p=[\Lambda(-\Delta U)]^{1/2}$, so every zero-energy level samples the same weight, and $\delta\Lambda$ tends to a $v$-independent limit; the remaining $v$ dependence is a quantum correction that decreases with $v$. For the adiabatic shift the values are 1.016, 1.021 and 1.030 for $v=$ 1–3, with the WKB limit 1.058. More than 99% of the weight lies at $r<8$ bohr, so the offset measures the corrections inside the molecular region and not at dissociation. The value of about 1 arises because $D(\infty)-D(R)$ (about 0.075 near $R_e$) and the well depth (about 0.10) are comparable in that region. A linearisation of the binding energy, $\Delta M\approx\Delta E/(dE_b/dM)$, does not work at a $J = 0$ threshold, because $E_b \propto (M - M_c)^2$ there and $dE_b/dM = 0$ at $M_c$; the coupling-constant form is the correct linear version of the same idea.

## S1.6 Second-order effective nuclear Hamiltonian and the mass functions

Partition the wave function as $\Psi = \phi_0\chi + (Q \text{ part})$, with $P = |\phi_0\rangle\langle\phi_0|$ acting on $\boldsymbol{\rho}$ at each $\boldsymbol{R}$ and $Q = 1 - P$. Löwdin partitioning gives

$$\hat{H}_{\text{eff}} = P\hat{H}P + P\hat{T}_N Q\left[E - Q\hat{H}Q\right]^{-1} Q\hat{T}_N P. \tag{S14}$$

$P\hat{H}P$ is the BO Hamiltonian plus the diagonal (adiabatic) correction. With $\hat{T}_N = -(1/2\mu_N)\nabla_R^2$,

$$Q\hat{T}_N(\phi_0\chi) = -\frac{1}{\mu_N} Q(\partial_i\phi_0)\,\partial_i\chi + Q(\hat{T}_N\phi_0)\,\chi. \tag{S15}$$

We take the leading resolvent, $[E - Q\hat{H}Q]^{-1} \approx -Q(\hat{H}_{\text{el}} - E_0)^{-1}Q$, whose corrections are relatively $O(1/M)$, and keep the term with two nuclear derivatives:

$$\Delta\hat{H} = \frac{1}{\mu_N^2}\,\partial_i A_{ij}(\boldsymbol{R})\,\partial_j, \qquad A_{ij}(\boldsymbol{R}) = \langle\partial_i\phi_0|Q(\hat{H}_{\text{el}} - E_0)^{-1}Q|\,\partial_j\phi_0\rangle_\rho, \tag{S16}$$

with the derivatives taken at fixed $\boldsymbol{\rho}$. Combined with $\hat{T}_N$,

$$\hat{T}_N + \Delta\hat{H} = -1/2\;\partial_i[(\mu_N\mathbb{1} + 2\boldsymbol{A})^{-1}]_{ij}\,\partial_j + O(M^{-3}), \tag{S17}$$

so the nuclear reduced mass becomes the tensor $\mu_N\mathbb{1} + 2\boldsymbol{A}(\boldsymbol{R})$. Since $\hat{H}_{\text{el}} - E_0 > 0$ on $Q$, $\boldsymbol{A}$ is positive semidefinite: the effective mass always increases, binding increases, and the exact thresholds lie below the BO+ad ones, which is the observed sign of the offsets. Neglected here are the potential-like second-order terms (with $Q\hat{T}_N\phi_0$), the $O(1/M)$ corrections to the resolvent and third-order terms; all are $O(M^{-2})$ in the energy, i.e. $O(1/M)$ in $\Lambda_c$.

**Symmetry.** Cylindrical symmetry gives $A_{ij} = A_\parallel n_i n_j + A_\perp(\delta_{ij} - n_i n_j)$, $\boldsymbol{n} = \boldsymbol{R}/R$. The parallel function $f_\parallel = \partial_R\phi_0|_\rho$ is $\sigma$ and gerade (inversion $\boldsymbol{\rho} \to -\boldsymbol{\rho}$ commutes with $\partial_R$ at fixed $\boldsymbol{\rho}$), so only $\Sigma_g$ states couple; the 2p$\sigma_u$ state, degenerate with 1s$\sigma_g$ as $R \to \infty$, is excluded by symmetry, which keeps the resolvent finite. The perpendicular function comes from rotating the frame, $\partial_x\phi_0|_\rho = \pm(i/R)\hat{L}_y\phi_0$ (electron angular momentum about the midpoint); it is $\Pi_g$, so only $\Pi_g$ states couple. As a check, $\langle f_\parallel|f_\parallel\rangle + 2\langle f_\perp|f_\perp\rangle = D(R)$.

**Scaling and definition of $m_v$, $m_r$.** With eqn (S6), $\partial_R \to \mu_e\,\partial_r$ and $\hat{H}_{\text{el}} - E_0 \to \mu_e(\hat{h} - e_0)$, so $A^{(\mu_e)}(R) = \mu_e A(\mu_e R)$ with $A$ the unit-mass quantity. The radial equation of $\hat{T}_N + \Delta\hat{H} + V$ is

$$-\frac{1}{2}\frac{d}{dR}\frac{1}{\mu_N + 2A_\parallel}\frac{du}{dR} + \frac{J(J+1)}{2(\mu_N + 2A_\perp)R^2}u + Vu = Eu. \tag{S18}$$

Multiplying by $M/\mu_e^2$ and setting $r = \mu_e R$, the radial kinetic prefactor becomes exactly $M/(M + 4\mu_e A_\parallel) = \Lambda/(\Lambda + 4A_\parallel)$, and similarly for the centrifugal term. Hence

$$m_v(r) = 4A_\parallel(r), \qquad m_r(r) = 4A_\perp(r), \qquad \Lambda \to \Lambda + m, \tag{S19}$$

which is eqns (6) and (7) of the main text in the mass-scaled convention.

**Separated-atom limit.** For $R \to \infty$, $\phi_0 \to (a+b)/\sqrt{2}$ with atomic 1s functions on the two centres, and both $f_\parallel$ and $f_\perp$ reduce to the atomic quantity $1/4\,\langle \partial_z 1s|(h-e)^{-1}|\,\partial_z 1s\rangle$. With $p_z = i[h,z]$ (unit mass), $(h-e)^{-1}p_z|1s\rangle = iz|1s\rangle$, so $\langle p_z 1s|(h-e)^{-1}|p_z 1s\rangle = 1/2$, $A_\parallel, A_\perp \to 1/8$ and

$$m_v(\infty) = m_r(\infty) = 1/2. \tag{S20}$$

This limit is exact in our convention: the relative motion of X⁺ and the atom has $2\mu = 2M(M+1)/(2M+1)$, so $\Lambda_\infty = 2\mu/\mu_e = M+1 = \Lambda + 1/2$. The non-adiabatic mass correction restores the ion–atom reduced mass at dissociation.

**Mass convention.** Our BO and BO+ad calculations use the bare nuclear reduced mass $M/2$ for the nuclear motion and $\mu_e$ for the electron, and the mass functions are defined relative to this choice. In a clamped-nuclei convention (electron mass 1, $1/\mu_{\mathrm{vib}} = (2/M)(1 + A/M)$, Kutzelnigg) the asymptote is also $1/2$. With atomic instead of nuclear masses, part of the non-adiabatic offset would move into the BO+ad thresholds; the theory is unchanged, but the split between the adiabatic and non-adiabatic parts depends on this choice (S1.8).

## S1.7 Folding the mass functions with the zero-energy states

**Non-perturbative form (used for all predictions).** With the mass functions included, eqn (S11) with $c = 1$ becomes eqns (8) and (9) of the main text,

$$-\frac{d}{dr}\left(q_v \frac{du}{dr}\right) + \left[q_r \frac{J(J+1)}{r^2} + \Delta D\right] u = -\Lambda\,\Delta U\,u, \qquad q_{v,r}(r) = \frac{\Lambda}{\Lambda + m_{v,r}(r)}, \tag{S21}$$

solved directly for $\Lambda_c$ by shooting.

**First-order form.** With $u$ the BO+ad zero-energy solution,

$$\Delta M_c^{\mathrm{nad}} = M_c^{\mathrm{BO+ad}} - M_c \simeq \frac{\int [m_v(r)\,u'^2 + m_r(r)J(J+1)u^2/r^2]dr}{\Lambda_c \int (-\Delta U)u^2\,dr} \tag{S22}$$

$$= \int m_v\,w_{\mathrm{rad}}\,dr + \int m_r\,w_{\mathrm{cen}}\,dr,$$

which defines the radial and centrifugal threshold weights $w_{\mathrm{rad}}$, $w_{\mathrm{cen}}$. The two forms differ by $O(m^2/\Lambda)$, formally of the same order as the neglected terms, so their spread is an honest error indicator: at most 0.0024 in $M_c$ for the levels of Table S6.

## S1.8 The $\kappa^6$ coefficient and its convention dependence

At large $M$, the order-$\kappa^6 = M^{-3/2}$ term of $E - E_{\mathrm{BO+ad}}$ is the $R$-dependent vibrational mass: second-order perturbation theory gives an operator $(1/M^2)\,\partial_R a(R)\,\partial_R$, which, acting on a vibrational function with $\langle p^2\rangle \propto M^{1/2}$, is of order $M^{-3/2}$; non-adiabatic potential-like terms are $O(M^{-2}) = \kappa^8$. Writing the correction as $M \to M + m_v(R)$ in the radial kinetic energy, first order gives $E - E_{\mathrm{BO+ad}} \approx -(m_v/M)\langle T_{\mathrm{rad}}\rangle$. In the harmonic limit $\langle T_{\mathrm{rad}}\rangle = \omega/4$, and $\sqrt{M}\,\omega/4 = \sqrt{2k}/4 = 0.113452$, with $k = U''(R_e) = 0.1029705$ (unit mass). Hence

$$c_6 = -m_v(R_e)\,\frac{\sqrt{2k}}{4} \quad \Rightarrow \quad m_v(R_e) = 0.3465 \pm 0.0003. \tag{S23}$$

An independent route, $m_v^{\mathrm{eff}}(M) = -M(E - E_{\mathrm{BO+ad}})/\langle T_{\mathrm{rad}}\rangle_{\mathrm{BO+ad}}$ extrapolated in $\kappa^2$, gives 0.34654 ± 0.0001; the computed function gives $m_v(R_e) = 0.346552$. **Convention.** The potential convention ($\mu_e$-scaled BO plus adiabatic, or unscaled $U(R)$ plus an adiabatic correction that includes the electron reduced-mass term) does not affect $c_6$: the two agree to $O(1/M)$ in the potential and differ at $O(M^{-2})$. The kinetic-mass convention does: with a nuclear mass $M + \delta$ in the reduced mass, $m_v$ shifts by $-\delta$ and $c_6 \to c_6 + 0.113452\,\delta$. For the nuclear mass ($\delta = 0$, used here) $c_6 = -0.03931$; for "nucleus plus half an electron" ($\delta = 1/2$) $+0.01742$; for the atomic mass ($\delta = 1$) $+0.07414$. Any quoted $\kappa^6$ coefficient must therefore state the convention.

## S1.9 Unequal nuclear masses

**System and coordinates.** Nuclei A (mass $m_A = M$, the lighter) and B ($m_B = qM$, $q \geq 1$), and an electron; $\boldsymbol{R} = \boldsymbol{R}_B - \boldsymbol{R}_A$ and $\boldsymbol{\rho}$ the electron relative to the nuclear centre of mass. Then

$$\hat{H} = -\frac{1}{2\mu_N}\nabla_R^2 - \frac{1}{2\mu_e}\nabla_\rho^2 + \frac{1}{R} - \frac{1}{r_A} - \frac{1}{r_B}, \quad \mu_N = \frac{qM}{1+q}, \quad \mu_e = \frac{(1+q)M}{(1+q)M+1}, \tag{S24}$$

with no mass-polarisation term. The nuclear centre of mass lies at $+\beta\boldsymbol{R}$ from the midpoint (towards B), with $\beta = (q-1)/[2(1+q)]$ ($\beta = 1/6$ for $q = 2$, $1/10$ for $q = 3/2$). **Thresholds.** The lowest continuum is A⁺ + $(\mathrm{B\,e})_{1s}$, $E_{\mathrm{th}} = -\mu_{Be}/2$ with $\mu_{Be} = qM/(qM+1)$; the second channel, B⁺ + (A e), lies $(\mu_{Be} - \mu_{Ae})/2 \approx (q-1)/(2qM)$ higher.

**Linear map.** The clamped-nuclei problem is that of $H_2^+$ in midpoint coordinates, so $E(R) = \mu_e U(\mu_e R)$ exactly. With $r = \mu_e R$ and multiplication by $M_{\mathrm{eff}}/\mu_e^2$ ($M_{\mathrm{eff}} = 2\mu_N$), the coupling is again linear in $M$, $\Lambda = M_{\mathrm{eff}}/\mu_e = 2qM/(1+q) + 2q/(1+q)^2$, so that

$$M_c = \frac{(1+q)\Lambda_c}{2q} - \frac{1}{1+q}, \tag{S25}$$

which reduces to eqn (S10) at $q = 1$ (eqn (5) of the main text). The BO couplings $\Lambda_c^{\mathrm{BO}}$ are those of the symmetric family; only the map changes, and constant shifts in $\Lambda$ are multiplied by $(1+q)/(2q)$ (3/4 for $q = 2$).

**Adiabatic correction.** At fixed $\boldsymbol{\rho}$, $\partial_R|_\rho = \partial_R|_{\rho_m} + \beta\,\partial_z$, so the diagonal correction of $1s\sigma_g$ acquires a term $\beta^2 t_g$ with $t_g = \langle p^2\rangle_g = -2(U + rU')$ (virial):

$$-u'' + \frac{J(J+1)}{r^2}u + \Lambda\,\Delta U\,u + \left[\Delta D + \beta^2(t_g - 1)\right]u = 0. \tag{S26}$$

This g-only model has the wrong asymptote for $q \neq 1$: its limit is the mean of the two atomic thresholds, and the exact open threshold lies $\beta$ (scaled) below it. We therefore report two conventions, BOad (own asymptote) and BOad_thr (measured from the exact threshold).

**Two-channel model.** The coupling $\beta\,\partial_z$ connects 1s$\sigma_g$ with $\sigma_u$ states, and 2p$\sigma_u$ is degenerate with 1s$\sigma_g$ as $R \to \infty$, so second-order perturbation theory fails exactly where the asymmetric threshold is decided. We treat 2p$\sigma_u$ explicitly, $\Psi = \phi_g\chi_g + \phi_u\chi_u$:

$$-u'' - 2\boldsymbol{a}\,u' - \boldsymbol{a}'u + \left[\frac{J(J+1)}{r^2} + \boldsymbol{K}(r)\right]u + \Lambda\,\mathrm{diag}(U_g, U_u)\,u = \frac{M_{\mathrm{eff}}}{\mu_e^2}E\,u, \tag{S27}$$

with $a_{gu} = \beta\langle g|\,\partial_z|u\rangle = \beta(E_u - E_g)\langle g|z|u\rangle$, $K_{gg} = D_g + \beta^2 t_g$, $K_{uu} = D_u + \beta^2 t_u$ and $K_{gu} = \beta k_{gu}$, all computed from the exact BO wave functions. In the limit, $\boldsymbol{K}(\infty) = \begin{pmatrix} 1/4+\beta^2 & -\beta \\ -\beta & 1/4+\beta^2 \end{pmatrix}$, with eigenvalues $(1/2 \mp \beta)^2$, which are exactly the two atomic reduced-mass shifts: the two-channel model has the exact threshold built in, and at $q = 1$ it decouples into the symmetric BO+ad problem.

**Second-order mass matrices.** Löwdin partitioning with the model space $\{g, u\}$ gives an $R$-dependent $2 \times 2$ mass matrix, $m_{ij} = 4A_{ij}$ with $A_{ij}(e) = \langle f_i|Q(h-e)^{-1}Q|f_j\rangle$ symmetrised over the two reference energies, radial functions $f_i = F_i + \beta Z_i$ ($\Sigma$ blocks) and perpendicular functions $f_i = G_i/R + \beta S_i$ ($\Pi$ blocks), entering as $Q = (1 + \boldsymbol{m}/\Lambda)^{-1}$. By parity, $m_{gg} = m_v + \beta^2 m_{gg2}$ and $m_{gu} = \beta m_{gu1}$; exact checks are $m^{\perp}_{gg2} \equiv 2$ and $m^{\parallel}_{gg2} + m^{\parallel}_{uu2} \equiv 4$, and the atomic limits give the open-channel value $m = 1/2 - 2\beta + 2\beta^2 = 2/(1+q)^2$, exactly the value required by the exact open-channel reduced mass. The u-row pieces diverge as $r \to 0$ (united-atom degeneracy); there the u channel is closed by at least $0.9\Lambda$, and these pieces are frozen below $r = 2$ (variant: 3).

**Models and null models.** Table S1 lists the models; the frozen prediction is 2ch+m, and its band includes the spread of the variants.

**Table S1** Zero-energy models for unequal masses.

| Model | Content | Role |
|---|---|---|
| BO | g channel, $\Lambda\Delta U_g$ | baseline ($\Lambda_c$ identical to the symmetric family) |
| BOad | $+\,\Delta D + \beta^2(t_g - 1)$, own asymptote | conventional BO+ad |
| BOad_thr | BOad measured from the exact threshold | BO+ad, correct threshold, no g/u mixing |
| BOad+m | BOad + g-row masses | symmetric theory transferred |
| 2ch | g and u channels, couplings $\boldsymbol{a}$, $\boldsymbol{K}$, exact threshold | non-perturbative g/u mixing |
| **2ch+m** | 2ch + $2 \times 2$ mass matrices | **parameter-free theory (frozen)** |
| 2ch+m_asym, 2ch+m_rc3 | u-row masses atomic, or frozen below $r = 3$ | error-bar variants |

Null models (frozen with the predictions): NS, the BO+ad threshold minus the exact symmetric-family offset of the same level; NT, the same measured from the exact threshold; NP (persistence), the offset of the previous $v$ of the same family.

## S1.10 Exchange symmetry and thresholds for $J > 0$

A natural-parity state ($P = (-1)^J$) is written in the bipolar form $\Psi_{JM} = \sum_{l_1+l_2=J} |\boldsymbol{a}|^{l_1}|\boldsymbol{b}|^{l_2}[Y_{l_1}(\hat{\boldsymbol{a}}) \otimes Y_{l_2}(\hat{\boldsymbol{b}})]_{JM} f_{l_1 l_2}(r_{01}, r_{02}, r_{12})$, with $\boldsymbol{a} = \boldsymbol{x}_0 - \boldsymbol{x}_2$ and $\boldsymbol{b} = \boldsymbol{x}_1 - \boldsymbol{x}_2$. The exchange $P_{01}$ maps $\boldsymbol{a} \leftrightarrow \boldsymbol{b}$ and $(l_1, l_2) \to (l_2, l_1)$ with phase $(-1)^{l_1+l_2-J} = +1$, so both exchange symmetries exist for every $J$. **Which belongs to the ground manifold.** In the BO

limit $\Psi \approx \phi_{\rm el}(\boldsymbol{\rho};\boldsymbol{R})\chi(R)Y_{JM}(\widehat{\boldsymbol{R}})$; $P_{01}$ sends $\boldsymbol{R} \to -\boldsymbol{R}$ and leaves $\boldsymbol{\rho}$ unchanged, the $1s\sigma_g$ function is even under exchange of the centres, and $Y_{JM}(-\widehat{\boldsymbol{R}}) = (-1)^J Y_{JM}(\widehat{\boldsymbol{R}})$. Hence $P_{01} = (-1)^J$: even $J$ symmetric, odd $J$ antisymmetric. The other symmetry at the same $J$ correlates with $2p\sigma_u$. With nuclear spin, for spin-½ X the symmetric levels pair with the nuclear-spin singlet (para) and the antisymmetric ones with the triplet (ortho); for spinless bosonic X the odd-$J$ levels of the ground manifold do not exist. **Threshold.** The lowest continuum of natural parity $J$, of either exchange symmetry, is X⁺ + $(\mathrm{X\,e})_{1s}$ with relative angular momentum $J$, so $E_{\rm th} = -M/[2(M+1)]$ for every $J$. Unnatural parity would require an excited atom ($n = 2$). For $J \geq 1$ the open channel has a centrifugal barrier and a $-\alpha/(2R^4)$ polarisation tail; the zero-energy solution $u \sim R^{-J}$ is normalisable, so the binding vanishes linearly at $M_c$ and the state does not become a halo.

## S1.11 Threshold laws and halo size

Near a $J = 0$ threshold, the new state is an X⁺ weakly bound in an s wave to the neutral polarisable atom $(\mathrm{X\,e})_{1s}$. The long-range interaction is the polarisation tail $-C_4/R^4$, with no $1/R^2$ term (the atom is in a non-degenerate 1s state) and an exponentially small g/u splitting, so the binding is of short-range type: the scattering length diverges linearly in the coupling, $k_b = (2\mu E_b)^{1/2}$ is linear in $M - M_c$, and

$$E_b \simeq A(M - M_c)^2 \quad (J = 0), \qquad E_b \simeq B(M - M_c) \quad (J > 0), \tag{S28}$$

the second for $J > 0$ because the centrifugal barrier keeps the zero-energy state normalisable. As $M \to M_c$ at $J = 0$, the size approaches the universal zero-range value $\langle r_{XX}\rangle \to 1/(2k_b)$, with $\mu$ the ion–atom reduced mass. A Coulomb-tail threshold (as for the critical nuclear charge of H⁻-like ions) would instead give a linear law at $J = 0$, so the exponent distinguishes the two classes. For a $p$ wave with a $-C_4/R^4$ tail the threshold is marginal, and logarithmic or non-analytic corrections near the $J = 1$ threshold are possible; the (1,1) level is linear only asymptotically (S3.3).

## S1.12 Electron–nuclear entanglement

We partition the ground state between $\boldsymbol{R}$ and $\boldsymbol{\rho}$, both mass-independent Jacobi vectors in which the kinetic energy is separable. For $J = 0$,

$$\Psi = \sum_l g_l\,(R,\rho) \sum_m Y_{lm}\,(\widehat{\boldsymbol{R}})Y^*_{lm}(\widehat{\boldsymbol{\rho}}), \qquad g_l = 2\pi\int du\,\Psi\,P_l(u), \tag{S29}$$

with $u = \cos\angle(\boldsymbol{R},\boldsymbol{\rho})$. The reduced density matrix of $\boldsymbol{R}$ is block-diagonal in $l$; block $l$ has the radial kernel $\int \rho^2 d\rho\, g_l(R,\rho) g_l(R',\rho)$, and each of its eigenvalues has multiplicity $2l+1$. Exchange of the two X⁺ ($\boldsymbol{R} \to -\boldsymbol{R}$) leaves only even $l$. With $w_l$ the weight of partial wave $l$ and $S_l$ the entropy of the normalised radial block $l$, the von Neumann entropy splits exactly (eqns (16) and (17) of the main text):

$$S = S_{\rm orient} + S_{\rm rad}, \qquad S_{\rm orient} = -\sum_l w_l \ln\frac{w_l}{2l+1}, \qquad S_{\rm rad} = \sum_l w_l\,S_l. \tag{S30}$$

$S_{\rm orient}$ is fixed by the partial-wave weights alone; it is the coupled-angular-momentum (frame) entanglement of a $J = 0$ state in a laboratory-frame partition. The orientational coherence of the X–X axis after tracing out the electron is $D(\Theta) = \sum_l w_l\, P_l(\cos\Theta)$, normalised so that $D(0) = 1$. Radial-only measures follow from tracing out the directions (local operations): the logarithmic negativity $E_N$ of $\rho_{\mathrm{rad}} = \sum_l (2l + 1)|g_l\rangle\langle g_l|$ and the classical mutual information $I(|R|;|\rho|)$. The BO product state is $\Psi_{\mathrm{BO}} = [P_{\mathrm{BO}}(R)/(4\pi R^2)]^{1/2}\phi_{1s\sigma_g}(\boldsymbol{\rho};R)$. The leading orders at large $M$ follow from a justified, not proven, argument (second Schmidt coefficient $p_2 \propto \langle\delta R^2\rangle \propto \kappa^2$): $S_{\mathrm{orient}} - S(\infty) \propto \kappa^2$, $I \propto \kappa^2$, $E_N \propto \kappa$ and $S_{\mathrm{rad}} \propto \kappa^2 \ln(1/\kappa)$.

## S1.13 Fidelity susceptibilities

With the variable $\ln M$, $F(M, Me^h) = |\langle\psi(M)|\psi(Me^h)\rangle| = 1 - 1/2\,\chi h^2 + O(h^3)$ and $\chi = \langle\dot\psi|\dot\psi\rangle - |\langle\psi|\dot\psi\rangle|^2$, $\dot\psi = d\psi/d\ln M$. Overlaps are taken in the fixed Jacobi coordinates. $\chi_F$ is the plain susceptibility; $\chi_F^{\mathrm{inv}}$ removes a global dilation, $U_s\psi(\boldsymbol{R},\boldsymbol{\rho}) = s^3\psi(s\boldsymbol{R}, s\boldsymbol{\rho})$ with generator $\widehat{D} = \boldsymbol{x}\cdot\nabla + 3$:

$$\chi_F^{\mathrm{inv}} = \chi_F - \frac{\langle\dot\psi|\widehat{D}\psi\rangle^2}{\parallel \widehat{D}\psi \parallel^2}; \tag{S31}$$

$\chi_{\mathrm{sep}}$ removes separate dilations of $\boldsymbol{R}$ and $\boldsymbol{\rho}$ (generators $\widehat{D}_R = \boldsymbol{R}\cdot\nabla_R + 3/2$, $\widehat{D}_\rho = \boldsymbol{\rho}\cdot\nabla_\rho + 3/2$), $\chi_{\mathrm{sep}} = \chi_F - \boldsymbol{v}^{\mathrm{T}}\boldsymbol{G}^{-1}\boldsymbol{v}$, with $\boldsymbol{v} = (\langle\dot\psi|\widehat{D}\psi\rangle, \langle\dot\psi|\widehat{D}_R\psi\rangle)$ and $\boldsymbol{G}$ the Gram matrix of $(\widehat{D}\psi, \widehat{D}_R\psi)$. In perimetric coordinates $\widehat{D} = s_0\,\partial_0 + s_1\,\partial_1 + s_2\,\partial_2 + 3$ (an exact rescaling of the basis), and

$$\widehat{D}_R = r_{01}\,\partial_{r_{01}} + \frac{r_{01}^2 + r_{02}^2 - r_{12}^2}{4r_{02}}\partial_{r_{02}} + \frac{r_{01}^2 - r_{02}^2 + r_{12}^2}{4r_{12}}\partial_{r_{12}} + \frac{3}{2}. \tag{S32}$$

All three susceptibilities tend to 1/32 as $M \to \infty$: the vibrational width scales as $M^{-1/4}$, and a dilation at fixed $R_e$ cannot absorb it, which gives $1/2\,(d\ln\eta/d\ln M)^2 = 1/32$ with $\eta \propto M^{-1/4}$.

## S1.14 Width of the ground-state crossover

The Lindemann ratio $\mathcal{L} = \Delta R/\langle R\rangle$ of the X–X distance has a maximum near $M \approx 0.2$–$0.3$ and falls as $\kappa$ in the BO limit. We define its local exponent (eqn (13) of the main text)

$$f(M) = -4\,\frac{d\ln\mathcal{L}}{d\ln M}, \tag{S33}$$

which is 0 at the maximum and tends to 1 in the BO limit. Definition (a) of the width is the range in which $f$ passes from 0.1 to 0.9; definition (b) is the range over which $\mathcal{L}$ falls from 90% to 10% of its maximum (the 10% point requires an extrapolation of the $M^{-1/4}$ tail). Both depend on the choice of descriptor and on the level convention.

# S2 Codes and numerical methods

## S2.1 Overview, software and tests

All calculations were done with in-house Python codes (Python 3.12.7, NumPy 1.26.4, SciPy 1.13.1, SymPy 1.13.2, mpmath 1.3.0): about 5,600 lines in the solver packages and about 13,700 lines in the analysis and run scripts. The codes are deterministic: no random numbers enter any result. They are designed for continuous families of mass ratios: all matrices are mass independent, so a mass scan, a threshold search or a derivative with respect to $M$ needs only new diagonalisations.

**Table S2** Codes and their test suites.

| Package | Purpose | Tests |
|---|---|---|
| threebody | variational three-body solver, perimetric Laguerre basis, any masses and charges, $J = 0$–4 natural parity (extended to $J = 5$–7 by the map scripts), multi-block bases, thresholds, observables | 29 (test_angular, test_validation, test_threshold, test_hamiltonian, test_observables; full suite with slow tests: 38) |
| bo_h2plus | exact BO and BO+ad reference: electronic curves 1s$\sigma_g$ and 2p$\sigma_u$, adiabatic correction, nuclear levels, BO critical masses, unequal-mass couplings | 26 (test_bo, test_unequal) |
| massfunc | non-adiabatic mass functions $m_v(R)$, $m_r(R)$ (Sternheimer) and their fold | 4 |
| unequal_masses | two-channel model, $2 \times 2$ mass matrices, exact unequal-mass runs | 8 |
| threshold_map_L7hp | double-double assembly of the matrices for $J = 7$ | 9 |
| harmonic_model | exactly solvable harmonic models (toolkit calibration) | 19 |

## S2.2 The three-body solver: coordinates, basis and matrix elements

**Perimetric coordinates.** For total $J = 0$ the wave function depends only on the three distances. We use

$$s_k = r_{kj} + r_{kl} - r_{jl}, \qquad r_{ij} = 1/2\,(s_i + s_j), \qquad d\tau \propto r_{01} r_{02} r_{12}\, ds_0\, ds_1\, ds_2, \qquad \text{(S34)}$$

in which the triangle inequalities become $0 \le s_k < \infty$ independently. **Basis (Pekeris type).**

$$F_{\boldsymbol{a}} = \prod_{k=0}^{2} L_{a_k}(\alpha_k s_k)\, e^{-\alpha_k s_k/2}, \qquad w_0 a_0 + w_1 a_1 + w_2 a_2 \le N, \qquad \text{(S35)}$$

with Laguerre polynomials $L_n$, weights $w$ (default (1,1,1), total degree; (1,1,2) for heavy X) and the size parameter $N$. The basis spans exactly the Hylleraas space $\{r_{01}^i r_{02}^j r_{12}^k\} \times \exp(-c_{01} r_{01} - c_{02} r_{02} - c_{12} r_{12})$ with $\alpha_0 = c_{01} + c_{02}$, etc., but is far better conditioned (scaled overlap condition number about $2 \times 10^6$ at $N = 28$). Exchange of the identical particles is $s_0 \leftrightarrow s_1$ (with $\alpha_0 = \alpha_1$); symmetry-adapted functions give separate symmetric and antisymmetric blocks. There are only two (symmetric) or three (general) non-linear parameters, chosen deterministically by Nelder–Mead minimisation of the chosen root at a modest size $N_{\text{opt}}$ (default 12), started from a hydrogenic rule; the convergence sequence then uses these fixed scales at increasing $N$, which gives nested spaces and strictly monotonic variational energies (Hylleraas–Undheim–MacDonald).

**Hamiltonian.** For $f(r_{01}, r_{02}, r_{12})$ and particle $i$ with partners $j$, $k$,

$$\nabla_i^2 f = f_{,ij,ij} + \frac{2}{r_{ij}} f_{,ij} + f_{,ik,ik} + \frac{2}{r_{ik}} f_{,ik} + 2\cos\theta_i \, f_{,ij,ik}, \qquad \cos\theta_i = \frac{r_{ij}^2 + r_{ik}^2 - r_{jk}^2}{2 r_{ij} r_{ik}}. \tag{S36}$$

Matrix elements use the symmetric, integrated-by-parts form $\langle a|\hat{T}|b\rangle = \sum_i (2m_i)^{-1} \int [a_{,ij} b_{,ij} + a_{,ik} b_{,ik} + \cos\theta_i (a_{,ij} b_{,ik} + a_{,ik} b_{,ij})] d\tau$, in which $\cos\theta_i \, r_{01} r_{02} r_{12}$ is a polynomial. With $\partial/\partial r_{ij} = \partial_{s_i} + \partial_{s_j} - \partial_{s_k}$, every integrand (overlap, kinetic, Coulomb, $r_{ij}^n$ with $n \geq -1$) is a polynomial in $s$ times products of 1D basis functions or their first derivatives. The Hamiltonian matrix is linear in the inverse masses and in the couplings, $\boldsymbol{H} = \sum_i m_i^{-1} \boldsymbol{T}_i + \sum_{i<j} Z_{ij} \boldsymbol{V}_{ij}$, with mass-independent $\boldsymbol{T}_i$, $\boldsymbol{V}_{ij}$; for the family, $\boldsymbol{H}(M) = (\boldsymbol{T}_0 + \boldsymbol{T}_1)/M + \boldsymbol{T}_2 + \boldsymbol{V}_{01} - \boldsymbol{V}_{02} - \boldsymbol{V}_{12}$, and $d\boldsymbol{H}/dM$ is exact.

**Exact 1D integrals.** All integrals reduce to products of $\int (\partial^{d_a} f_a)(\partial^{d_b} f_b) s^p \, ds$, $d_a, d_b \in \{0,1\}$, obtained from the integer Jacobi matrix of the Laguerre polynomials ($\int f_a f_b s^p ds = \alpha^{-(p+1)} (\boldsymbol{J}^p)_{ab}$) and the derivative matrix ($df_n/ds = \alpha \sum_k D_{kn} f_k$, $D_{kn} = -1$ for $k < n$ and $-1/2$ for $k = n$). Integrals between blocks with different scales are done by Gauss–Laguerre quadrature, which is exact for these integrands.

**Eigenproblem.** The generalised eigenproblem $\boldsymbol{Hc} = E\boldsymbol{Sc}$ is solved by canonical orthogonalisation: $\boldsymbol{S}$ is scaled to unit diagonal, and directions with eigenvalue below a cutoff (tol) times the largest are dropped. Single-scale bases need no truncation; multi-block bases use tol $10^{-12}$ (S2.4).

**Verification of the operators** (test_hamiltonian.py). (i) The distance-form Laplacian, eqn (S36), agrees with a brute-force SymPy Cartesian Laplacian in nine coordinates for a generic non-symmetric function to $10^{-11}$ (relative). (ii) The engine's $\boldsymbol{S}$, $\boldsymbol{V}_{ij}$ and $\boldsymbol{T}_i$ (first-derivative form) agree with direct Gauss–Laguerre quadrature of the second-order form for a non-symmetric basis with three different scales to about $10^{-10}$. (iii) Exact limits: with the electron–electron interaction switched off and an infinitely heavy nucleus, $E = -Z^2$; the mass-scaling identity $E(\lambda m) = \lambda E(m)$ holds to $10^{-11}$; the symmetry-adapted code agrees with the unsymmetrised code after relabelling the particles to $10^{-11}$.

**Convergence and error estimates.** The convergence table keeps the scales fixed, so the spaces are nested and every root decreases monotonically. The extrapolation assumes a geometric tail, $d_n = E_{n-1} - E_n$, $\rho = d_n/d_{n-1}$, $E_{\text{ext}} = E_n - d_n\rho/(1-\rho)$, with an error estimate equal to the larger of the tail and the change of $E_{\text{ext}}$ from the previous step. This is a heuristic estimate, not a bound. Changes below about $10^{-13}$–$10^{-11}$ Eh are at the rounding floor set by the conditioning of $\boldsymbol{S}$. **Counting bound states.** Each variational root is an upper bound to the corresponding exact eigenvalue, so a root below $E_{\text{th}}$ proves, up to rounding, that at least that many bound states of that symmetry exist; a root above threshold proves nothing.

## S2.3 Natural-parity states with $J > 0$

The bipolar form of S1.10 has $J + 1$ components; each radial function is expanded in the same perimetric Laguerre products (one or more blocks of scales). The components are coded as Cartesian irreducible tensors (for $J = 1$: $a_k$, $b_k$; for general $J$, the rank-$J$ symmetric traceless tensors obtained by harmonic projection, stored with one entry per index multiset and multinomial weights: 10 entries for $J = 3$, 15 for $J = 4$). All operators are scalars, so each matrix element summed over the tensor index is rotation invariant; it is reduced to the three distances by evaluating it in a canonical frame with SymPy in exact rational arithmetic, times the measure $r_{01}r_{02}r_{12}$. The kinetic energy is used in the symmetric first-derivative form, so that every integrand is again a polynomial in $s$ times first derivatives of 1D Laguerre functions, and the exact 1D machinery of S2.2 is reused unchanged. The completeness of the $l_1 + l_2 = J$ form for natural parity is taken from the literature on two-electron atoms (Bhatia and Temkin; Drake).

**Term lists.** The SymPy term lists are generated once and cached: about 5 s for $J = 1$, 40 s for $J = 2$, 15 min for $J = 4$, 4086 s for $J = 5$ and 9643 s for $J = 6$ (on a loaded machine); $J = 7$ took several hours. **Tests for $J = 3,4$.** Weighted contraction equals the full $3^J$ contraction from an independent detracer formula; the union of the symmetric and antisymmetric spectra equals the unsymmetrised spectrum; a pointwise check of the kinetic integrand and a second-order-form quadrature check ($J = 3$). **Validation for $J \geq 5$.** The difference $\delta(J) = E_{\text{exact}}(0,J) - E_{\text{BO+ad}}(0,J)$ for $H_2^+$ is almost exactly linear in $J(J+1)$ for $J = 0$–4 (quadratic-fit residual $2 \times 10^{-11}$); extrapolated to $J = 5$ it gives $-6.793 \times 10^{-7}$ Eh, against $-6.245 \times 10^{-7}$ from the new $J = 5$ lists ($N = 20$), a difference of $5.5 \times 10^{-8}$ against a rotational energy of $3.9 \times 10^{-3}$ Eh. For $J = 6$ the same test passes at tol $10^{-14}$ (difference $4 \times 10^{-8}$).

## S2.4 Multi-block bases for near-threshold states, and the threshold criterion

A state at its threshold is an atom plus a distant ion (S1.11), which a single compact basis cannot describe. We therefore use a union of blocks, each with its own three scales and exchange symmetry imposed as $\phi = F + PF$ (so a block may have $\alpha_0 \neq \alpha_1$):

- **Compact blocks** (total degree $N$): the ground-state optimum and the root-$v$ optimum, plus an intermediate set for $v \geq 2$ (for $J > 0$ in the map: root 0 plus a copy with $0.6\,\alpha_X$ for $v = 0$; root-0 and root-$v$ optima, and their geometric mean, for $v \geq 1$). Scales are optimised at $N_{\text{opt}} = 8$–12 at the BO+ad threshold mass (×1.03 for $v \geq 2$).
- **Diffuse blocks** describing the atom and its polarisation in $s_0$, $s_2$ ($a_0 + a_2 \leq 4$) times the ion–atom coordinate $s_1 \approx 2R$ (up to 30–60 functions), with scales $(\mu_a, \alpha_s, \mu_a)$ and $\mu_a = M_{\text{ref}}/(M_{\text{ref}} + 1)$, and a mirror block; a nested ladder of $\alpha_s$ from $10^{-2}$ to $3 \times 10^{-5}$ ($J = 0$), five sets for $J = 1$, three for $J = 2,3$ and two for $J \geq 4$ (the centrifugal barrier makes the diffuse part less important).

**Crossing criterion.** For a fixed basis, the matrices are mass independent, and the crossing $M^*$ of root $v$ with $E_{\text{th}}(M)$ (Brent's method on $E_v(M) - E_{\text{th}}(M)$) is a rigorous upper bound to $M_c$ up to rounding, because the variational root lies above the exact level. A certified (rounding-safe) bound requires the root to lie at least $10^{-11}$ Eh below threshold; the observed rounding floor is about $10^{-14}$–$10^{-13}$ Eh, and build-to-build noise at genuine crossings is at most $1.6 \times 10^{-13}$ Eh. The margin is needed because a fixed-basis root crosses threshold with a very small slope (for example $10^{-8}$ Eh per unit of $M$ for $v = 1$, $2 \times 10^{-9}$ for $v = 2$), so a margin-free crossing is a bound only up to rounding. A variational calculation cannot prove that a state is unbound; the central values and the lower side therefore come from extrapolation (threshold-law fits for $J = 0$, $J = 1$ and $J = 2$; geometric extrapolation in the compact size for the other levels).

**Spurious roots and guards.** A too small orthogonalisation cutoff can produce a spurious root from a nearly singular overlap: with tol $10^{-14}$ a spurious root appeared at $v = 2$ ($M = 29.649$, $7 \times 10^{-7}$ Eh below the converged value), and for an $L = 5$ test basis $\lambda_{\min}/\lambda_{\max} = -2.3 \times 10^{-12}$, with a spurious root at $E = -2.41$. All double-precision results therefore use tol $10^{-12}$ (tol $10^{-10}$ as a sensitivity check). The production scripts (i) log $\lambda_{\min}/\lambda_{\max}$ of every basis; (ii) reject any root with $E_0 - E_{\text{th}} < -0.15$ Eh; (iii) require that an extended basis lowers $E_0$; and (iv) check that a crossing is genuine by the slope of the gap: the gap must be positive at $M^*(1 - 10^{-3})$ and negative at $M^*(1 + 10^{-3})$. Canonical orthogonalisation also means that a nominally larger basis is not exactly a superset; this causes rounding-level non-nesting (for example a compact $N = 20$ crossing $1.4 \times 10^{-5}$ higher than $N = 16$ for $v = 3$, about $10^{-13}$ Eh).

## S2.5 Extended-precision assembly for $J = 6$ and 7

At $J = 7$ each matrix element is a sum of about 50 to 1000 angular terms with powers up to $s^{16}$. Summed in double precision, the elementwise rounding errors (about $10^{-15}$ of the largest element) exceed the smallest relative eigenvalues of $\boldsymbol{S}$, which acquires a spurious negative direction ($\lambda_{\min}/\lambda_{\max} = -2 \times 10^{-14}$). Canonical orthogonalisation at tol $10^{-12}$ then removes physical directions, and at tol $10^{-14}$ it fails: the validation against $H_2^+$ at $J = 7$ gave an energy $1.5 \times 10^{-5}$ Eh too high. The precision loss is in the assembly, not in the 1D integrals (exact to $4 \times 10^{-16}$) or in the eigensolver. The fix (HPCoupledBasis) overrides only the assembly: same-scale 1D factors come from exact integer matrices and exact powers of $\alpha$; mixed-scale factors from Gauss–Laguerre quadrature in mpmath (80 digits); the terms are accumulated in vectorised double-double arithmetic (about 32 digits) and the result rounded to double. Orthogonalisation and eigensolver are unchanged. With this assembly $\lambda_{\min}/\lambda_{\max} = -4 \times 10^{-16}$, the variational energies are monotonic, the $J = 7$ validation passes, and tol $10^{-14}$ is safe with the guards of S2.4. A control at $J = 6$ reproduces double precision to $2 \times 10^{-10}$. A two-pass orthogonalisation in double precision is not a cure (it fails the validation at tol $10^{-14}$) and was used only as a cross-check (S3.5). The same assembly was later used to recompute the two $J = 6$ levels, whose double-precision sequences were not monotonic (Table S5, footnote b).

## S2.6 The BO and BO+ad reference

**Electronic problem** (prolate spheroidal separation). The angular equation is expanded in normalised Legendre polynomials (even $l$ for 1s$\sigma_g$, odd for 2p$\sigma_u$), giving a symmetric tridiagonal eigenproblem for the separation constant. The radial equation uses the Jaffé expansion $X = (\xi + 1)^{\sigma} e^{-p\xi} \sum_n g_n t^n$, $t = (\xi - 1)/(\xi + 1)$, $\sigma = R/p - 1$, with the three-term recurrence (derived symbolically)

$$
\begin{aligned}
&(n+1)^2 g_{n+1} + [\lambda \\
&- (2n^2 + 4np + 2n + p^2 + 2p + 1 - a/p\, n - a - a/2p)] g_n \\
&+ (n-1-\sigma)^2 g_{n-1} = 0, \quad a = 2ZR.
\end{aligned} \tag{S37}
$$

The energy $E = -2p^2/R^2$ follows from matching the radial and angular separation constants (Brent's method), in 1–3 ms per point; $E_{\mathrm{el}}$ is accurate to about $10^{-12}$ Eh for $0.01 \leq R \leq 200$ bohr. **Adiabatic correction.** $D(R) = \langle \partial_R \psi | \partial_R \psi \rangle_\rho + \langle L_x^2 + L_y^2 \rangle / R^2$, with the derivative at fixed $\boldsymbol{\rho}$ from the analytic chain rule in spheroidal coordinates and Gauss–Legendre quadrature in $(\xi, \eta)$; accurate to about $10^{-9}$ relative. $U$ and $D$ are tabulated with piecewise Chebyshev interpolation on 15 geometric segments over [0.01, 200] bohr, with the asymptotic form $-1/2 - 9/(4R^4) - 15/(2R^6)$ beyond. **Nuclear problem.** Sinc-DVR (Colbert–Miller radial form) in the mapped variable $x$, $R = R_s \sinh x$, with a step adapted to the largest local wavenumber and a box enlarged automatically until the DVR finds as many levels as an independent node count and $R_{\max} \geq 25/k$ for the least bound level; each result is recomputed on a second grid to give a per-level error. **Level count and BO critical masses.** The number of nodes of the threshold-energy solution (Numerov from a 12-term Frobenius series, with the exact zero-energy $-C_4/R^4$ solution beyond the grid for $J = 0$) is bisected in $\log M$ at two step sizes and Richardson-extrapolated. **Zero-energy coupling solver.** Eqns (S11) and (S21) are solved by shooting with the exact $-C_4/r^4$ zero-energy tail for matching; it reproduces every BO critical mass of the DVR/node-count route to $\leq 10^{-9}$ (0.9649154241, 11.0814565935, 31.1677742733 and 60.8466743844 for $v = 0$–3).

## S2.7 Mass functions

$A_\parallel$ and $A_\perp$ are obtained without a sum over physical states, from the Hylleraas functional $A = -\min_\psi \{ \langle \psi | h - e_0 | \psi \rangle + 2 \langle \psi | Qf \rangle \}$, whose minimiser solves the Sternheimer equation $(h - e_0)\psi_1 = -Qf$; in a finite basis the result is a lower bound converging from below. The basis is prolate spheroidal: $\Sigma_g$, $L_i(x) e^{-x/2} P_l(\eta)$ with even $l$; $\Pi_g$, $\sqrt{\xi^2 - 1}\, L_i(x) e^{-x/2} P_l^1(\eta) \cos\varphi$, with $x = 2\alpha(\xi - 1)$ and $\alpha = R(-e_0/2)^{1/2}$. All overlap, kinetic, Coulomb and centrifugal integrals are exact (Gauss quadrature of polynomial × $e^{-x}$), and every matrix is a sum of Kronecker products of 1D matrices. The right-hand sides come from the exact BO wave function, with $f_\parallel = \partial_R \phi|_{\xi\eta} - [\xi(\xi^2 - 1)\phi_\xi + \eta(1 - \eta^2)\phi_\eta] / [R(\xi^2 - \eta^2)]$ (five-point difference in $R$) and, for $\Pi_g$, $g = \sqrt{(\xi^2 - 1)(1 - \eta^2)}(\eta\phi_\xi - \xi\phi_\eta)/(\xi^2 - \eta^2)$. The production grid has 2597 points from 0.02 to 60 bohr; basis convergence is $3 \times 10^{-11}$ ($m_v$) and $10^{-12}$ ($m_r$). Checks: the norms reproduce $D(R)$ to $10^{-4}$ relative (limited by the completeness for $f$, not for $A$); $m - 1/2 \propto R^{-3}$ at large $R$ (ratio 8.1 between 20 and 40 bohr). For unequal masses the same code gives the $\Sigma_g$, $\Sigma_u$, $\Pi_g$ and $\Pi_u$ blocks of the $2 \times 2$ mass matrices (convergence $\leq 2 \times 10^{-12}$ for $r \geq 0.1$).

## S2.8 Unequal-mass calculations

The model problems of Table S1 are solved by outward integration of the regular solutions from $r = 10^{-3}$, inward integration of closed (decaying) and open (tail) solutions from $r_m = \max(40, 30/\sqrt{2\beta})$, QR renormalisation with sign tracking, and a matching determinant at $r = 6$. For one channel the code reproduces the symmetric zero-energy solver to $10^{-10}$ and the symmetric non-perturbative offset (0.5281 for $v = 1$). The exact calculations use the unsymmetrised version of the multi-block solver (no exchange symmetry), with the same crossing criterion and guards. Validation at $q = 1$ requires the closed channel to receive its own diffuse functions (a mirror basis), because it is degenerate with the open one (S3.9). $HD^+$ $v = 0$ and 1 at the physical masses were compared with published values before the freeze.

## S2.9 Entanglement, fidelity and density codes

**Entanglement.** Composite Gauss–Legendre quadrature in $R$ on $[\langle R\rangle - 10\Delta R, \langle R\rangle + 14\Delta R]$, in $\rho$ on $[0, 14\langle r_{Xe}\rangle]$ (panels uniform in $\sqrt{\rho}$) and in $\theta \in [0, \pi/2]$ (128 nodes, with the coalescence kinks at the endpoints), $l_{\max} = 64$ (fine grid: about 1.5 times more points and $l_{\max} = 96$). Spectra by Nyström discretisation (SVD of $\sqrt{a_i} g_l(R_i, \rho_j)\sqrt{b_j}$). The partial-wave weights decay as $l^{-6}$ (electron–X cusps); the tail beyond $l_{\max}$ is estimated and is below $10^{-5}$. **Fidelity.** In a fixed basis $\partial \boldsymbol{H}/\partial \ln M = -(\boldsymbol{T}_0 + \boldsymbol{T}_1)/M$ exactly, and $\dot{\psi}$ is the solution of the bordered linear system $\begin{pmatrix} \boldsymbol{H} - E\boldsymbol{S} & \boldsymbol{Sc} \\ \boldsymbol{c}^{\mathrm{T}}\boldsymbol{S} & 0 \end{pmatrix}\begin{pmatrix} \boldsymbol{x} \\ \lambda \end{pmatrix} = \begin{pmatrix} -(\boldsymbol{H}' - E'\boldsymbol{S})\boldsymbol{c} \\ 0 \end{pmatrix}$, $\chi_F = \boldsymbol{x}^{\mathrm{T}}\boldsymbol{Sx}$. All integrands except two are polynomial × exponential and integrated exactly; the two singular pieces of $\| \widehat{D}_R\psi \|^2$ are done in polar coordinates in which the $1/r$ factor cancels the Jacobian. **Density of X about the centre of mass.** With $\boldsymbol{s} = \boldsymbol{R}/2 - \boldsymbol{r}/(2M+1)$ the vector from the centre of mass to $X_0$, $\rho_X(s) = 8\int d^3r\, |\Psi(\boldsymbol{R} = 2s\hat{z} + 2\boldsymbol{r}/(2M+1), \boldsymbol{r})|^2$; the remaining 2D integral is done by Gauss–Legendre quadrature, and the topology indicator is the curvature $a_2$ in $\rho(s)/\rho(0) = 1 + a_2(s/\ell)^2 + a_4(s/\ell)^4$.

## S2.10 Hardware and use of generative AI

Calculations ran on a laptop (Apple silicon, 10 cores) and on a Mac Studio (Apple M1 Ultra, 20 cores, 128 GB). Typical costs: a mass point of the ground-state scan, 2–4 min; a complete $J = 0$ threshold, a few hours; a level of the map, 1–20 h (the largest check bases have up to about 8000 functions, about 6 GB); the (7,0) level with the double-double assembly, about one day. The development and use of the codes with the assistance of a generative AI coding agent are described in the Methods section of the main text; every result was produced by the deterministic codes described here, which are archived with the paper together with their tests.

# S3 Calculation details and additional results

## S3.1 Validation

**Table S3** Ground-state validation of the three-body solver (total-degree basis, scales optimised for the ground state at $N = 12$ and then fixed; energies in Eh). $H_2^+$ at the CODATA 2022 proton mass.

| System | $N$ (size) | Variational $E$ | Extrapolated | Reference |
|---|---|---|---|---|
| He, $m_{\text{nuc}} = \infty$ | 28 (2360) | −2.90372437702875 | −2.90372437703255(4) | −2.903724377034 |
| $H^-$, $m_{\text{nuc}} = \infty$ | 28 (2360) | −0.52775101646188 | −0.52775101652007(6) | −0.527751016544 |
| $Ps^-$ ($M = 1$) | 28 (2360) | −0.26200507020127 | −0.26200507022359(2) | −0.262005070233 |
| $H_2^+$, total degree | 32 (3417) | −0.59713906306287 | −0.5971390630883(3) | |
| $H_2^+$, weights (1,1,2) | 40 (3311) | −0.59713906308051 | −0.59713906308051(1) | |

| System | $N$ (size) | Variational $E$ | Extrapolated | Reference |
|---|---|---|---|---|
| $Mu_2^+$, $M = 206.7682827$ | 28 (2360) | −0.58512609817815 | converged | |

References: He and $H^-$, Nakashima and Nakatsuji, *J. Chem. Phys.*, 2007, **127**, 224104; $Ps^-$, Drake and Grigorescu, *J. Phys. B*, 2005, **38**, 3377 (both rounded to 12 decimals). No literature value exists at exactly our $H_2^+$ and $Mu_2^+$ masses; see the comparisons with Tang *et al.* and King *et al.* below. Uncertainties in parentheses refer to the last digits shown. In every case the variational value lies above the reference. The extrapolated values lie $1.6 \times 10^{-12}$ (He), $2.4 \times 10^{-11}$ ($H^-$) and $9.4 \times 10^{-12}$ Eh ($Ps^-$) above the references, so for these light systems the geometric-tail estimates in parentheses, which are heuristic, understate the remaining error by one to two orders of magnitude; the sequences are monotonic and geometric, and the virial ratio equals 2 to $10^{-9}$. For $H_2^+$ the four lowest $J = 0$ levels ($N = 40$, weights (1,1,2)) are −0.59713906308051, −0.58715567909958, −0.5777519044 and −0.5689084987 Eh, with estimated errors of $3 \times 10^{-13}$, $2 \times 10^{-12}$, $2 \times 10^{-11}$ and $6 \times 10^{-10}$. **Rotational levels.** The He 2 ¹P and 2 ³P levels agree with Drake and Bondy to $10^{-10}$ Eh; the $H_2^+$ levels $(v, J) = (0,0)$, (1,0), (0,1), (1,1), (0,2) and (1,2) agree with Tang *et al.* to $10^{-12}$–$10^{-13}$ Eh at their proton mass, 1836.15267247; (0,3) agrees to $1.2 \times 10^{-11}$ Eh; (0,4) lies $3.1 \times 10^{-9}$ Eh below a Dunham interpolation of the $J = 0$–3 levels (no published value). **A resolved discrepancy.** A frequently quoted $H_2^+$ value, −0.597139063123 Eh, lies $4 \times 10^{-11}$ below ours; it belongs to the older proton mass 1836.152701, at which we obtain −0.5971390631232.

**BO reference for $H_2^+$.** $U(R = 2) = -0.602634214495265$; $R_e = 1.99719332$ bohr, $E_{\min} = -0.6026346191$, $k = U''(R_e) = 0.1029705$; at the CODATA 2022 mass, $E_{\mathrm{BO}}(0,0) = -0.597233978551$ and $E_{\mathrm{BO+ad}}(0,0) = -0.597138541555$ Eh; 20 bound $J = 0$ levels; spacings 2191.144 cm⁻¹ (BO) and 2191.300 cm⁻¹ (BO+ad); $E(J = 1) - E(J = 0) = 58.234$ cm⁻¹. **Comparison with published values.** $U(2)$ agrees with the reference value −0.6026342144949 Eh (Madsen and Peek, reproduced by Scott *et al.*) to $4 \times 10^{-13}$ Eh; $R_e$ and $E_{\min}$ agree with Scott *et al.* (1.997193319970 bohr, −0.6026346191065 Eh) to $3 \times 10^{-10}$ bohr and $4 \times 10^{-13}$ Eh; the $1s\sigma_{\mathrm{g}}$ (10 points) and $2p\sigma_{\mathrm{u}}$ (9 points) energies of Scott *et al.* for $R = 0.5$–20, given to 9 digits, agree to a relative $10^{-8}$ or better; the 7-decimal table of Wind ($R = 0.05$–20) agrees within 2 units of the last digit at every legible entry. Internal checks (35 tests): independent ODE shooting agrees with the Jaffé recurrence to $10^{-13}$; doubling the basis changes $E_{\mathrm{el}}$ by less than $2 \times 10^{-11}$; the mass-scaling identity holds to below $10^{-12}$; the adiabatic correction agrees with an independent finite-difference evaluation to $10^{-10}$ and reproduces $D \to 1/4$.

**Comparison with King, Longford and Cox.** For this comparison all energies were recomputed with their masses, which are CODATA 2010 values ($m_p = 1836.15267245$, $m_\mu = 206.7682843$, $m_d = 3670.4829652$, $m_t = 5496.9215267$). All 43 rows of their Table III (42 finite-mass rows, 37 distinct mass ratios with six rows at $M = 1$, plus infinite-mass $H^-$) agree within one unit of their last certain digit (worst case 0.64 unit, for (6,6,1)); their marker $g$ agrees to $\leq 1.4 \times 10^{-8}$. For their Table II (2856-term values), $Mu_2^+$, $H_2^+$ (theirs −0.597139063079, ours −0.5971390630793) and $\mu D_2^+$ agree to all 12 printed digits; our energy is lower for $D_2^+$ (by $2.7 \times 10^{-10}$) and for $H^-$, $Ps^-$ and $Mu^-$ (by $1$–$4 \times 10^{-11}$), a better upper bound. Their "literature" column is reproduced to $\leq 10^{-11}$ with the older mass constants listed in their Table I, except $\mu D_2^+$ ($4 \times 10^{-11}$) and $\mu H^-$ ($6.4 \times 10^{-9}$, unresolved). We locate the minimum of their marker $g(M) = (E_0 - E_{\mathrm{th}})/E_{\mathrm{th}}$ without fitting, by solving $dg/dM = 0$ with the exact Hellmann–Feynman derivative: $M_{\min} = 0.4924222(1)$, $g_{\min} = 0.0451139164$ (their fit: $0.491 \pm 0.003$). Refitting their fit form to our exact data places the minimum at 0.4903–0.4904 and moves it with the number of terms (7 terms: 0.5187; 9: 0.4888; 11: 0.4903; 13: 0.49238), so the bias comes from the truncated form and not from their energies. Errata noticed in their paper: the ratio of $(\tau, \tau, e)$ is printed 3477.500 instead of 3477.15; the energy of (4.7,4.7,1) lacks its minus sign; the printed coefficients of their eqn (9) do not reproduce their data.

## S3.2 Vibrational thresholds, $v = 1$–4

**Basis.** Compact total-degree blocks ($N = 16$; ground-state optimum, root-$v$ optimum, and an intermediate set for $v \geq 2$) plus two diffuse blocks (atom × ion–atom coordinate, $a_0 + a_2 \leq 4$, $a_1 \leq 40$ and $\leq 30$) with the $\alpha_s$ ladder; 2130 functions ($v = 1$) and 2655 ($v = 2,3$); tol $10^{-12}$. At each $M$ the lowest root over the ladder is taken.

**Convergence** (three masses above the crossing for each check): compact $N$ 16→20, $\leq 1.3 \times 10^{-13}$ Eh; diffuse $a_0 + a_2 \leq 4 \to 6$, $\leq 2 \times 10^{-13}$; diffuse sizes 40/30→60/45, $\leq 1.1 \times 10^{-12}$; tol $10^{-12} \to 10^{-10}$, $\leq 6 \times 10^{-12}$ (smaller space, higher energy, as expected).

**Threshold-law fits.** Data: 12 masses per level from $M^* + 0.001$ to $M^* + 1$, with $E_b$ from $2 \times 10^{-10}$ to $2 \times 10^{-4}$ Eh. Free power laws $E_b = Ax^p$ ($x = M - M_c \leq 0.05$) give $p =$ 2.0012(2), 2.0044(8), 2.0057(7) and 2.011(2) for $v = 1$–4; $\sqrt{E_b} = c_1 x + c_2 x^2 + c_3 x^3$ fits within a relative rms of $10^{-7}$ ($v = 1$) to $5 \times 10^{-6}$ ($v = 3$) for $x \leq 0.1$, with $c_1 = 0.013737$, 0.004131 and 0.001666 ($v = 1$–3; $7.67732 \times 10^{-4}$ for $v = 4$). The quoted $M_c$ are the third-order fits on $x \leq 0.1$; the error bars are the full spread over second- and third-order fits on windows $x \leq 0.05$ to 0.3, plus the free power-law fits. A fixed-basis cross-check, crossings decreasing monotonically towards the fitted value as the diffuse block is made more diffuse, is consistent (Table S4).

**Table S4** Fixed-basis crossing masses $M^*$ against the most diffuse scale $\alpha_s$.

| $v$ | $\alpha_s = 10^{-2}$ | $3 \times 10^{-3}$ | $10^{-3}$ | $3 \times 10^{-4}$ | $10^{-4}$ | $M_c$ (fit) |
|---|---|---|---|---|---|---|
| 1 | 9.52830 | 9.52644 | 9.52591 | 9.52572 | 9.52566 | 9.525634(4) |
| 2 | 29.64449 | 29.64094 | 29.63991 | 29.63955 | 29.63944 | 29.63939(1) |
| 3 | 59.32764 | 59.32139 | 59.31957 | 59.31893 | 59.31875 | 59.31866(2) |

**Bounds.** Margin-free crossings 9.525652, 29.639433 and 59.318722; certified bounds 9.52586, 29.64016, 59.32056 and 98.26962 (Table 1 of the main text). For $v = 4$ the margin-free crossings (98.26522 to 98.26564) scatter by $3 \times 10^{-4}$, because the root crosses threshold with a slope of only $3 \times 10^{-11}$–$2 \times 10^{-10}$ Eh per unit of $M$; the certified bound lies where $E_b$ reaches the $10^{-11}$ Eh margin, $M - M_c \approx 0.004$. **Halo.** For $v = 1$, $\langle r_{XX} \rangle \approx$ 11 000 bohr at $M - M_c \approx 0.001$, and $2k_b \langle r_{XX} \rangle \to 1$; one data point of $v = 3$ closest to threshold is not converged and is excluded.

## S3.3 Rotational levels, the (1,1) level and $J = 3,4$

**Method.** Fixed bases with mass-independent matrices: compact sets of increasing size (two scale sets) plus nested diffuse sets (atom × ion–atom relative motion, with a mirror set); crossings by Brent's method; a check stage with the largest basis at tol $10^{-10}$ and with the $10^{-11}$ Eh margin; binding energies at $M^*(1 + x)$, $x = 5 \times 10^{-4}$ to 0.2. **Results.** $M_c(1,0) =$

$4.2082827(5)$ (bound $4.2082830$); $M_c(2{,}0) = 11.4894217(3)$ $(11.4894218)$; $M_c(1{,}1) = 17.094282(1)$: crossings with compact sizes $N = 16$, 18, 20 and five diffuse sets (up to 7214 functions, 5166 kept) agree to $10^{-9}$ at 17.094281912, certified bound 17.094281939; $M_c(3{,}0) = 21.376199(1)$ (extrapolated 21.37619905(82), best crossing 21.376199865, bound 21.376203385); $M_c(4{,}0) = 33.914349(2)$ (33.91434859(17), best crossing 33.914350279, bound 33.914359616). **Threshold laws.** $E_b/(M - M_c)$ is $2.50 \times 10^{-3}$ ($J = 2$), $1.712 \times 10^{-3}$ ($J = 3$) and $1.211 \times 10^{-3}$ ($J = 4$) Eh per unit of $M$, nearly constant for $x$ from 0.0005 to 0.2. For (1,1) it rises from $3.9 \times 10^{-4}$ to $6.3 \times 10^{-4}$ for $x = 0.0005$ to 0.1, so a quadratic-fit root (17.0970) is not reliable, and the threshold rests on the converged crossings. **Exploratory symmetric $J = 1$ level** (2p$\sigma_u$-like, bound by the long-range polarisation well): $M_c = 813.54(7)$ (extrapolated 813.5374(64), best crossing 813.6015, rounding-safe bound 813.6485, compact-size ratio 0.585), with a binding slope of only $7.9 \times 10^{-9}$ Eh per unit of $M$; BO 815.5587; no BO+ad value (adiabatic correction for 2p$\sigma_u$ not implemented). This value is determined to about 0.1 in $M$ and no better.

## S3.4 The threshold map (14 levels)

**Targets.** All levels of the 1s$\sigma_g$ manifold with a BO+ad threshold below 101 that had not been computed before (from the zero-energy BO map for $J = 0$–10, all roots up to $M = 105$); no 2p$\sigma_u$ level binds in this range (its first BO threshold is 330.6), and the first $J = 8$ BO threshold lies above 105. **Method.** As in S3.3, generalised to any $(J, v)$; compact sequence $N_D - 4$ to $N_D + 2$, with $N_D$ the largest even size for which the largest check basis has at most 8000 functions; root-order check at every crossing (the gaps of roots $v - 1$ and $v + 1$ are stored; root $v - 1$ must be bound); geometric extrapolation in the compact size.

**Table S5** The threshold map: exact (extrapolated), certified upper bound, BO and BO+ad threshold masses.

| $(J, v)$ | $M_c$ | Certified bound | $M_c^{\mathrm{BO}}$ | $M_c^{\mathrm{BO+ad}}$ | BO $-M_c$ | BO+ad $-M_c$ |
|---|---|---|---|---|---|---|
| (2,1) | 27.69724706(1) | 27.697247072 | 29.019661 | 28.066887 | 1.3224 | 0.3696 |
| (1,2) | 39.30862402(1) | 39.308624131 | 40.755428 | 39.766531 | 1.4468 | 0.4579 |
| (3,1) | 41.07434726(1) | 41.074347285 | 42.355695 | 41.408480 | 1.2813 | 0.3341 |
| (5,0) | 49.1153755(49) | 49.115380349 | 50.318074 | 49.353402 | 1.2027 | 0.2380 |
| (2,2) | 52.58916447(2) | 52.589164528 | 53.968672 | 53.002721 | 1.3795 | 0.4136 |
| (4,1) | 57.1649498(5)[a] | 57.164949247 | 58.421973 | 57.475700 | 1.2570 | 0.3108 |
| (6,0)[b] | 66.983717(1) | 66.983798633 | 68.181554 | 67.215424 | 1.1978 | 0.2317 |
| (3,2) | 69.0138886(2) | 69.013888822 | 70.347388 | 69.393851 | 1.3335 | 0.3800 |
| (1,3) | 71.19123222(1) | 71.191232553 | 72.664208 | 71.661196 | 1.4730 | 0.4700 |
| (5,1) | 75.9486530(8) | 75.948654001 | 77.189949 | 76.242851 | 1.2413 | 0.2942 |
| (2,3) | 86.7873305(3) | 86.787330745 | 88.206149 | 87.225441 | 1.4188 | 0.4381 |
| (7,0)[c] | 87.521468(1) | 87.521471609 | 88.715954 | 87.748556 | 1.1945 | 0.2271 |
| (4,2) | 88.3541941(3) | 88.354193850 | 89.656384 | 88.708986 | 1.3022 | 0.3548 |
| (6,1)[b] | 97.416732(2) | 97.416776689 | 98.647176 | 97.698595 | 1.2304 | 0.2819 |

[a] The extrapolated value lies $7 \times 10^{-7}$ above the best crossing (57.164949070) and the certified bound; the extrapolation error is underestimated by about a factor of 2, irrelevant at the 0.03 level. [b] $J = 6$: recomputed with the double-double assembly of S3.5 (tol $10^{-14}$); the certified column keeps the double-precision values. In double precision the compact-size sequence was not monotonic (best crossings 66.983757757 and 97.416774043). In the

rerun, the crossings decrease monotonically: compact $N = 6$, 8, 10: 66.98408545, 66.98373151, 66.98371877 ((6,0)), and $N = 6$–12: 97.44357292, 97.42011510, 97.41744657, 97.41692694 ((6,1)). The quoted values extrapolate the sequences with two diffuse sets, $N = 8$, 10, 12: 66.98372871, 66.98371861, 66.98371717 (geometric ratio 0.14, tail $2.4 \times 10^{-7}$) and 97.41684723, 97.41673702, 97.41673227 (ratio 0.04, tail $2.1 \times 10^{-7}$). The error bars also cover the change on adding the second diffuse set ($+2.7 \times 10^{-7}$ and $-4.2 \times 10^{-7}$ at $N = 10$). For these largest bases the conditioning flag of S3.5 fires at the level of rounding ($\lambda_{\min}/\lambda_{\max} = -1.2 \times 10^{-16}$ to $-3.6 \times 10^{-16}$), while the crossing-slope guard is passed, so no spurious root is present; we accept these rows. The best crossings that pass both guards, 66.98371834 and 97.41673744 (one diffuse set at $N = 10$), lie above the quoted values, as they must. [c] Double-double assembly, certified bound at tol $10^{-14}$ (S3.5). Uncertainties in parentheses refer to the last digits. Errors are $10^{-9}$–$3 \times 10^{-8}$ for (2,1), (1,2), (3,1), (2,2), (1,3), (3,2) and (2,3), and $3 \times 10^{-7}$–$5 \times 10^{-6}$ for (4,1), (4,2), (5,0) and (5,1), whose compact-size sequences converge more slowly (ratios 0.76–0.79 against 0.09–0.23 for $J \leq 3$).

**The two $J = 6$ levels.** In double precision their compact-size sequences were not monotonic. The double-double rerun (footnote b of Table S5) gives monotone sequences and the values $M_c(6{,}0) = 66.983717(1)$ and $M_c(6{,}1) = 97.416732(2)$, which pass the pre-specified test with misses of $-0.0012$ and $-0.0025$. **Order and counts.** The order of appearance of the 23 levels is the same in the exact, BO and BO+ad calculations (Table S7).

## S3.5 The (7,0) level

The validation of the double-precision $J = 7$ basis failed (S2.5), and the term lists were verified to be uncorrupted by a fresh regeneration. With the double-double assembly (tol $10^{-14}$, single-pass canonical orthogonalisation), every accepted row passed the crossing-slope guard (gap $+1.5 \times 10^{-5}$ to $+4.7 \times 10^{-5}$ Eh at $M^*(1 - 10^{-3})$ and $-4.7 \times 10^{-5}$ at $M^*(1 + 10^{-3})$). Accepted crossings: compact $N = 6$, 8, 10, 12: 87.5224238, 87.5215196, 87.5214751, 87.5214699 (monotonic); $N = 10$ plus one and two diffuse sets: 87.5214717, 87.5214716; $N = 12$ plus two diffuse sets: 87.5214689 (best bound). Subtracting the geometric compact-size correction ($6.7 \times 10^{-7}$, ratio 0.115) gives $M_c(7{,}0) = 87.521468 \pm 1 \times 10^{-6}$, where the error combines the compact correction, the last diffuse increment ($10^{-7}$) and the two-pass cross-check ($10^{-7}$). The certified bound (margin $10^{-11}$ Eh, $N = 10$ plus two diffuse sets) is 87.5214716; at tol $10^{-12}$ the crossing is 87.5218653, a looser but valid bound. **Excluded basis.** One check basis ($N = 8$ plus two diffuse sets) produced a spurious bound level at $M = 86.25$, although it is a subset of $N = 10$ plus two diffuse sets ($\lambda_{\min}/\lambda_{\max} = -3.1 \times 10^{-16}$); the cause is the single-pass orthogonalisation at tol $10^{-14}$, and with a second pass the same basis gives a normal crossing, 87.5215171. **Cross-check.** A complete rerun with two-pass orthogonalisation reproduces every accepted crossing to $\leq 10^{-7}$ for $N \geq 10$. It is not used for the results because it fails the validation at tol $10^{-14}$. The run-5 levels computed at tol $10^{-12}$ are not affected: two-pass orthogonalisation shifts them by at most $4.3 \times 10^{-10}$.

## S3.6 Comparison with earlier estimates

**Table S6** Our threshold masses compared with the earlier estimates of Frolov and co-workers. Sources: FY89, Frolov and Yeremin, *J. Phys. B*, 1989, **22**, 1263; BF92, Bishop and Frolov, *J. Phys. B*, 1992, **25**, 3049; F92, Frolov, *J. Phys. B*, 1992, **25**, 3059; FSB94, Frolov, Smith and Bishop, *Phys. Rev. A*, 1994, **49**, 1686; FW11, Frolov and Wardlaw, *Eur. Phys. J. D*, 2011, **63**, 339.

| Level | This work | Earlier estimates (source) | Relation |
|---|---|---|---|
| (1,0), $J = 1$ antisym. | 4.2082827(5) | 4.219 (FY89); about 4.20 (BF92); 4.2133 (FSB94) | 0.25% and 0.12% above; 4.20 a rounded estimate within its bracket (bound at 4.25, unbound at 4.0) |
| (2,0), $J = 2$ sym. | 11.4894217(3) | 11.485 (BF92); 11.489 (unbound) to 11.490 (bound) (FSB94) | ours lies inside both brackets; 11.485 is a rounded estimate (bound at 11.5, unbound at 11.4), 0.04% below |
| (0,1) | 9.525634(4) | 10.204 (FY89); 10.045 (F92) | 7.1% and 5.5% above |
| (0,2) | 29.63939(1) | 30.70 (F92) | 3.6% above |
| (0,3) | 59.31866(2) | 61.135 (F92) | 3.1% above |
| (1,1) | 17.094282(1) | 17.128 (FY89); 17.125 (F92); 17.0944 (FW11) | 0.20%, 0.18% and 0.0007% above |
| (2,1), (1,2), (2,2) | 27.6972471, 39.3086240, 52.5891645 | 27.715, 39.375, 53.295 (F92) | 0.06%, 0.17% and 1.3% above |
| (3,0) | 21.376199(1) | 19.63 (F92, empirical formula) | 8% below |
| (4,0) | 33.914349(2) | 28.14 (F92, empirical formula) | 17% below |
| (5,0), (6,0), (7,0), (3,1), (4,1), (5,1), (6,1), (3,2), (4,2), (1,3), (2,3), (0,4) | Tables 1, 2 and S5 | none | first determination (together with (3,0) and (4,0): 14 levels) |

The earlier estimates come from variational calculations in exponential bases of 600 to 1500 functions, at three to five significant figures and without error bars. Every level they find bound lies above our threshold; points listed as unbound at $M = 10.025$, 30.5 and 61.10 are in fact bound. The empirical formula $1/M_c = \sum_n a_n\,(v)(L + 3v)^{-n}$ behaves as $1/L$ at large $L$ and fails to extrapolate; our values give $M_c/L^2 = 4.21$, 2.87, 2.38 and 2.12 for $J = 1$–4.

## S3.7 Level counts

**Table S7** Number of bound levels of the $1s\sigma_g$ manifold (natural parity, including the ground state).

| $M$ | 2 | 5 | 10 | 20 | 30 | 40 | 50 | 60 | 70 | 80 | 90 | 100 |
|---|---|---|---|---|---|---|---|---|---|---|---|---|
| Exact | 1 | 2 | 3 | 5 | 8 | 10 | 12 | 15 | 17 | 19 | 22 | 24 |
| BO | 1 | 1 | 2 | 5 | 7 | 9 | 11 | 14 | 16 | 19 | 22 | 24 |
| BO+ad | 1 | 2 | 2 | 5 | 7 | 10 | 12 | 15 | 17 | 19 | 22 | 24 |

**Table S8** BO critical masses of the $J = 0$ levels ($m_{\text{eff}} = (2M + 1)/4$).

| $v$ | 0 | 1 | 2 | 3 | 4 | 5 | 6 | 7 | 8 | 9 |
|---|---|---|---|---|---|---|---|---|---|---|
| $M_{\iota}^{]}$ | 0.964915424 | 11.0814565936 | 31.16777427 | 60.8466744 | 99.7973934 | 147.663680 | 204.060272 | 268.640823 | 341.228978 | 421.94099 |

$\sqrt{m_{\text{eff}}}$ at successive thresholds is spaced by about 1.55, as WKB predicts. The $v = 0$ entry is a BO artefact: the exact ground state is bound at all $M$, and the BO level at $M = 1$ ($1.09 \times 10^{-5}$ Eh below $-1/3$, $\langle R\rangle \approx 154$ bohr) is physically meaningless.

## S3.8 Pre-specified predictions: protocol and log

Each prediction used the same code path: the BO+ad zero-energy problem of the level, the computed mass functions, and the non-perturbative fold of eqn (S21), with no fitted parameter. The prediction file was made read-only and its SHA-256 hash recorded; the band and the pass criterion ($|\mathrm{exact} - \mathrm{pred.}| \leq$ band + exact error) were written in the same file. The hashes were verified after the exact calculations.

**Table S9** Freezes and what was known at each.

| Freeze | Levels; band | Hash (first 8) | Before the freeze | After the freeze |
|---|---|---|---|---|
| 1, 14 Nov 2026, 17:22 | (0,4); ±0.0095 | 11b44609 | no exact calculation of the level | short code tests at two masses (15 Nov, 14:07–14:13); a two-point estimate from them implies about 98.27 |
| 2, 15 Nov, 20:17 | (1,1), (3,0), (4,0); ±0.0295 | 551b8f28 | scale optimisation of the (1,1) basis, 20:11–20:13, no crossing (it showed only that the level is bound at its BO+ad mass) | small-basis pipeline tests: approximate crossings 17.0944, 21.3762, 33.9144 |
| 3, 16 Nov, 09:35 | 14 levels of the map; ±0.0295 | 6e494c54 | the earlier exact thresholds, the BO and BO+ad map, Frolov's estimates for (2,1), (1,2), (2,2); no exact calculation of any target; a non-frozen run of the same prediction code 9 min before | short tests, e.g. a small-basis crossing of (2,1) at 27.6973 |
| 4, 16 Nov, 09:48 | 7 unequal-mass levels; ±0.03 (0.0368 for (1,0)), lenient 0.10 | 7ac806bf | $HD^+$ validation energies at the physical masses; an unconverged small-basis code check at $M = 8$ and 12, no crossing search | — |

The frozen files of freeze 1 also contained empirical variants (fits to the exact offsets of $v =$ 1–3), which were labelled as such; the primary prediction is the parameter-free one. The freezes were held by the author and not time-stamped by a third party, so we call the predictions pre-specified rather than blind. **Calibration set.** The five levels known before any prediction, $v =$ 1–3, $J = 1$ and $J = 2$, set the bands: ±0.0095 is the largest $J = 0$ residual ($v = 1$) and ±0.0295 the largest $J > 0$ residual ($J = 1$). **Residuals.** The residuals of $v =$ 1–4 are $+0.0095$, $-0.0062$, $-0.0046$ and $-0.0028$ in the offset (a smooth monotonic trend); all $J > 0$ residuals have the same sign, and 13 of the 14 residuals of the map are negative (exact below prediction). **Persistence.** For $v = 4$, "the offset of $v = 4$ equals that of $v = 3$" predicts 98.2615 (error +0.0040), and the mean of the $v = 2,3$ offsets predicts 98.2569 (+0.0086); both lie within the band. For (1,1), $J = 3$ and $J = 4$, repeating the offset of a neighbouring sector gives errors of 0.027 ($J = 3$ from $J = 2$), 0.015 ($J = 4$ from $J = 3$), 0.067 and 0.108 ((1,1) from $J = 1$ or from $v = 1$), 4.6 to 8.4 times the theory errors; these null models were chosen after the results were known.

## S3.9 Unequal masses

**Table S10** Unequal masses: exact thresholds, models and null models (threshold mass $M$ of the lighter nucleus).

| $q$, $(J, v)$ | Exact $M_c$ | Bound | BO | BO+ad | Theory (2ch+m) | Exact − theory | Exact − null: NS / NT / NP |
|---|---|---|---|---|---|---|---|

| $q$, $(J,v)$ | Exact $M_c$ | Bound | BO | BO+ad | Theory (2ch+m) | Exact − theory | Exact − null: NS / NT / NP |
|---|---|---|---|---|---|---|---|
| 2, (0,1) | 11.573761(8) | 11.57393265 | 8.3528 | 7.5496 | 11.7133 | −0.1396 | +4.43 / −2.91 / – |
| 2, (0,2) | 29.005219(29) | 29.00580969 | 23.4175 | 22.6098 | 29.0886 | −0.0834 | +6.78 / −4.81 / +2.37 |
| 2, (0,3) | 53.41741(10) | 53.41904003 | 45.6767 | 44.8621 | 53.4741 | −0.0567 | +8.93 / −7.42 / +2.16 |
| 2, (0,4) | 84.3454(4) | 84.34973515 | 74.8897 | 74.0691 | 84.3858 | −0.0404 | +10.65 / −11.17 / +1.72 |
| 2, (1,0) | 5.405970 | 5.40597031 | 4.1948 | 3.4264 | 5.5819 | −0.1760 | +2.25 / −0.83 / – |
| 2, (2,0) | 11.029758(1) | 11.02975806 | 9.5965 | 8.8329 | 11.1478 | −0.1180 | +2.41 / −0.51 / – |
| 1.5, (0,1) | 11.675521(7) | 11.67568620 | 9.2512 | 8.3868 | 11.8117 | −0.1362 | +3.74 / −2.33 / – |

BO+ad is counted at its own asymptote. Along the $J = 0$ series at $q = 2$ the exact thresholds lie 4.02, 6.40, 8.56 and 10.28 above the BO+ad ones, so the theory misses 3.5%, 1.3%, 0.7% and 0.4% of this offset; it predicts the level-to-level changes ($-2.315$, $-2.133$, $-1.705$) within 0.06, 0.03 and 0.02 of the exact ones ($-2.371$, $-2.160$, $-1.721$). Other frozen predictions at $q = 1.5$ (not computed exactly): (0,2) 30.610, (0,3) 57.310, (0,4) 91.429, (1,0) 5.408, (2,0) 11.453.

**Validation and noise checks.** (i) At $q = 1$, the production basis gave 9.5873 against the symmetric exact 9.525634 ($+0.0617$): a diffuse ladder was missing in the closed channel, which at $q = 1$ is degenerate with the open one. A mirror basis (the closed channel receives the open-channel sets) gives 9.52565573 (compact sizes 14, 16, 18: 9.52565574, 9.52565573, 9.52565550), and a larger basis (14 000 functions) 9.52564891, that is $+1.5 \times 10^{-5}$ from the exact value, about twice the last enlargement step ($6.8 \times 10^{-6}$). Polarisation alone, without the mirror, does not help (9.5873, 9.5871). (ii) Enlargement tests at $q = 2$ and 1.5 move the $J = 0$ thresholds by less than $1.2 \times 10^{-6}$, within the quoted errors ($J = 0$ only). (iii) A spurious near-threshold crossing appeared in an early run before the slope guard existed; those stages were discarded and redone. (iv) Build-to-build noise of the gap at genuine crossings is at most $1.6 \times 10^{-13}$ Eh. (v) The hashes of the prediction files match, and the first exact run is later than the freeze. **Candidate causes of the miss** (none tested): the treatment of the second dissociation channel, $B^+$ + (A e), which lies $\approx (q-1)/(2qM)$ above the open one; g/u mixing beyond the two-channel truncation (2p$\pi_u$ and higher $\Sigma$ states, flagged as uncalibrated in the frozen file); and $O(1/M)$ terms odd in the mass asymmetry, which vanish at $q = 1$. The miss at $q = 1.5$ is almost equal to that at $q = 2$, although $\beta^2$ is 0.010 against 0.028, so it does not scale as $\beta^2$.

## S3.10 Muonic molecular ions

**Table S11** Level existence in pp$\mu$, dd$\mu$ and tt$\mu$ predicted from our thresholds.

| Ion | $M$ | Predicted bound $(J,v)$ | Absent (predicted and published) | Agreement | Closest threshold: level, $(M-M_c)/M_c$ |
|---|---|---|---|---|---|
| pp$\mu$ | 8.8802 | (0,0), (1,0) | (0,1), (0,2), (1,1), (2,0), (2,1), (3,0), (4,0) | 8/8 | (0,1), −6.8% |
| d | 17.7 | (0,0), (0,1), (1,0), (1,1), (2,0) | (0,2), (2,1), (3,0), (4,0) | 8/8 | (1,1), +3.8% |

| Ion | $M$ | Predicted bound $(J, v)$ | Absent (predicted and published) | Agreement | Closest threshold: level, $(M - M_c)/M_c$ |
|---|---|---|---|---|---|
| d $\mu$ | 517 | | | | |
| tt$\mu$ | 26.5849 | (0,0), (0,1), (1,0), (1,1), (2,0), (3,0) | (0,2), (2,1), (4,0) | 8/8 | (2,1), −4.0% |

$M = m_{\text{nucleus}}/m_\mu$ with the masses of Frolov. Agreement: predictions that match the published spectra, excluding (0,0), which is always bound. Closest threshold: the level whose threshold mass lies nearest to the $M$ of the ion, with the relative distance; a positive value means that the level is bound, but only just ((1,1) in dd$\mu$, with $M_c = 17.094$; consistent with the very weak binding of that level, $E_b = 3.5 \times 10^{-4}$ muon atomic units), a negative value that it is just absent ((2,1) in tt$\mu$, with $M_c = 27.697$, so that the sixth bound state of tt$\mu$ is (3,0), at $M_c = 21.376$, and not (2,1); (0,1) in pp$\mu$, with $M_c = 9.526$). Presence of a level rests on the certified upper bound and the monotonic growth of the binding energy with $M$; absence rests on the extrapolated $M_c$. Published counts: 2, 5 and 6 bound states (Frolov and Wardlaw). **Binding energies** at Frolov's mass ratios ($M = 17.751673$ and $26.584936$; three compact sizes identical to $10^{-12}$): dd$\mu$ (1,1) $3.510170 \times 10^{-4}$, dd$\mu$ (2,0) $1.5372612 \times 10^{-2}$, tt$\mu$ (2,0) $3.0694483 \times 10^{-2}$, tt$\mu$ (1,1) $8.034498 \times 10^{-3}$ (45.21 eV), in muon atomic units, in agreement with the published values to all digits given (relative differences $1\text{–}3 \times 10^{-7}$).

## S3.11 Mass scans

**Grid.** 30 masses from 0.05 to $10^4$ (0.05, 0.1, 0.2, …, 1, 1.2, 1.5, 2, 3, 4.5, 6, 9.5256, 12, 15, 20, 29.639, 50, 100, 206.768, 500, 1000, 1836.15, 4000, $10^4$), plus 0.62–0.68 and 1.05–1.15 to locate two crossovers; the physical-mass rows were recomputed at CODATA 2022 (S4). **Exact** $J = 0$**.** Scales optimised for root 0 by Nelder–Mead ($N_{\text{opt}} = 12$ for $M < 150$; 20 with weights (1,1,2) above), warm-started from the neighbouring mass, then fixed; nested sequences $N = 16\text{–}28$ ($M < 150$, size 2360), 28–40 ($150 \le M < 3000$, 3311), 32–44 ($M = 4000$, 4324), 36–48 ($M = 10^4$, 5525). Estimated errors: $\le 3 \times 10^{-11}$ Eh for $M \le 1$, $\le 10^{-12}$ for $3 \le M \le 4000$, $2.2 \times 10^{-9}$ at $M = 10^4$ (5% of $E - E_{\text{BO+ad}}$ there). **Descriptors.** $\langle R \rangle$, $\langle R^2 \rangle$, $\Delta R/\langle R \rangle$, $\langle 1/R \rangle$, $\langle 1/R^2 \rangle$, the virial ratio, the Hellmann–Feynman X kinetic energy $K_X = -dE/d\ln M$ (exact for the variational eigenvalue at fixed basis) and $d^2E/d(\ln M)^2$ (central differences of the Hellmann–Feynman derivative, $h = 0.02$ and 0.01, Richardson); each also at the next smaller basis as convergence evidence (Lindemann ratio to $\le 3 \times 10^{-9}$ for $M \le 1$ and $< 5 \times 10^{-10}$ for $3 \le M \le 4000$). **Rotational levels.** Natural parity, $J = 1$ (antisymmetric) and $J = 2$ (symmetric), computed only above their thresholds; at $M = 10^4$ with bases up to $N = 46$ (14 400 functions): $E(J = 0) = -0.60032852464759(80)$, $E(J = 1) = -0.6002789726735(14)$, $E(J = 2) = -0.60017994134773(50)$.

**Table S12** Fitted exponents in $\kappa$ at large $M$ (fits biased by the next order; local slope at the largest mass in the third column).

| Quantity | Fitted exponent | Local slope | BO expansion |
|---|---|---|---|
| Lindemann ratio | 0.980(4) | 0.986 | 1 |
| $\langle R \rangle - R_e$ | 2.036(8) | 2.025 | 2 |
| $-dE/d\ln M$ | 2.11(2) | 2.08 | 2 |
| $d^2E/d(\ln M)^2$ | 2.20(4) | 2.15 | 2 |

| Quantity | Fitted exponent | Local slope | BO expansion |
|---|---|---|---|
| Vibrational spacing | 1.905(23) | 1.94 | 2 |
| Rotational spacing $d_1$ | 3.952(13) | 3.97 | 4 |
| $E - E_{\mathrm{BO}}$ | 3.995(2) | | 4 |
| $E - E_{\mathrm{BO+ad}}$ | 6.07(1) | 6.056 | 6 |

The quoted ± are not statistical errors, because the bias from the next order dominates. **$\kappa^4$ and $\kappa^6$ coefficients.** $M(E - E_{\mathrm{BO}}) \to 0.1746$ (two-term fit $a\kappa^4 + b\kappa^6$: $a = 0.1750$), equal to $D(R_e) = 0.1748$. Fixed-exponent fits of $M^{3/2}(E - E_{\mathrm{BO+ad}}) = c_6 + c_8\kappa^2(+c_{10}\kappa^4)$ give $c_6 = -0.039278$, −0.039303, −0.039311 and −0.039315 (three terms; windows $M \geq 100$, 206.8, 500, 1000) and −0.039335, −0.039327, −0.039320 (two terms; $M \geq 500$, 1000, 1836), hence $c_6 = -0.03931 \pm 0.00003$ and $c_8 \approx -0.074 \pm 0.002$.

**Table S13** Rotational spacings ($d_1 = E(J=1) - E(J=0)$, $d_2 = E(J=2) - E(J=0)$; exact).

| $M$ | $M\,d_1$ | $d_2/d_1$ | $d_1/d_1^{\mathrm{BO+ad}}$ |
|---|---|---|---|
| 20 | 0.322 | 2.475 | 0.9969 |
| 50 | 0.395 | 2.749 | 0.9987 |
| 100 | 0.430 | 2.864 | 0.9993 |
| 206.77 | 0.454 | 2.932 | 0.99966 |
| 1836.15 | 0.487 | 2.992 | 0.99996 |
| $10^4$ | 0.4955 | 2.99853(10) | – |

BO reproduces $d_2/d_1$ to $\leq 10^{-4}$ for $M \geq 100$; the BO+ad spacing errors are about $+0.15/M$ (vibration) and $+0.075/M$ ($d_1$); plain BO spacing errors change sign ($d_1$ between $M = 100$ and 207, vibration between 500 and 1000). For $H_2^+$, the exact vibrational spacing is 2191.10 cm$^{-1}$ and $d_1 = 58.232$ cm$^{-1}$; $Md_1/(2\langle R^{-2}\rangle) = 0.62$ at $M = 4.5$ and 0.999 at 1836.

## S3.12 The ground-state crossover

**Width.** $f(M)$ of eqn (S33) passes 0.1, 0.5 and 0.9 at $M = 0.455$, 1.52 and 95 (width 5.3 in $\ln M$, 2.3 decades); the Lindemann ratio falls from 90% to 10% of its maximum between $M = 1.66$ and about $3 \times 10^4$ (4.3 decades; the 10% point is an extrapolation of the $M^{-1/4}$ tail). The ground-state markers 0.2–1.11 span 1.71 in $\ln M$, 0.32 of width (a) and 0.17 of width (b); with the fidelity marker (1.39), 0.36 of width (a); the markers proposed in the literature (0.4–0.98) span 0.17 of width (a); the thresholds (4.21–98.27) span 1.16 widths. All levels move by a few percent with the grid.

**Descriptors.** The Lindemann ratio is almost flat (0.52–0.55) for $M \leq 1$, with a maximum of 0.5507 near $M \approx 0.2$–0.3 (below the two-body value $1/\sqrt{3} = 0.577$), 0.367 at $M = 9.53$ and 0.112 at 1836. The X density about the centre of mass changes from a maximum to a hole at $M = 0.635 \pm 0.01$ (the sign change was confirmed by an independent adaptive quadrature, flip between 0.64 and 0.66); for Ps$^-$ ($M = 1$), $n(0)/n_{\max} = 0.92$. The maximum of the X kinetic energy (zero of $d^2E/d(\ln M)^2$) lies at $M = 1.11$ (BO+ad: 1.10). BO+ad overestimates the Lindemann ratio by up to 66% and the second derivative by up to 60% below $M \approx 3$, and plain BO has no bound state below $M = 0.965$. **Literature markers.** Mátyus *et al.*: density of the identical particles about the centre of mass, change in $0.4 < M < 0.8$ (coarse scan, no error analysis). Kais and Shi: minimum of the charge-stability boundary at their parameter $\xi_m = 0.352$, converted by us to $M = \xi_m/(1-\xi_m) = 0.543$ (not recomputed). Salas *et al.*: sign change of a second derivative of the marginal distribution near a mass ratio of 1 (their $R = m_s/m_d$ is the inverse of our $M$); 0.98 is our linear interpolation between two tabulated points of a one-term trial-function calculation (0.984–0.987 in $M$, $1/M$ or $\ln M$) and could shift by tens of percent.

**Fidelity susceptibilities.** $\chi_F$ decreases monotonically from 1.220 ($M = 0.05$) to 0.0331 ($10^4$); at $M = 1$, $\chi_F = 0.4698$ against $\chi_F^{\rm inv} = 0.00998$ (98% dilation). $\chi_F^{\rm inv}$ rises most steeply at $M = 1.3856$, has a broad maximum of 0.026527 at $M = 6.8945$ and a minimum of 0.025850 at $M = 25.07$ (the peak is 2.6% above the minimum, half-width about 0.4 in $\ln M$). $\chi_{\rm sep}$ increases monotonically, with inflections at 0.2442, 1.428, 9.78 and 42.9. All three tend to 1/32 (fitted limits 0.031246, 0.031242, 0.031246). Degree-6 polynomial fits in $\ln M$ on dense windows (0.17–48.5) leave residuals of $10^{-9}$–$2 \times 10^{-7}$, at the basis error: no sharp feature. **Validation:** at $M = 1$, perturbative $\chi_F = 0.469825574773$ against a Richardson finite difference of fixed-basis overlaps, 0.469825574844; the Coulomb mass-scaling identity (an exact dilation) gives $\chi_F^{\rm inv} = \chi_{\rm sep} = 5.5 \times 10^{-9}$ (exact 0); a brute-force quadrature in Jacobi coordinates reproduces the $\widehat{D}_R$ integrals to about $10^{-6}$.

**Solvable models (toolkit calibration).** Model A (all harmonic, structureless) and model B (harmonic nuclear coordinate plus $g/R^2$, with a bond; the exactly solvable $1/r^2$ harmonium of Müller-Herold) separate exactly, so their $\boldsymbol{R}|\boldsymbol{\rho}$ entanglement is identically zero (a null test). $\chi_F^{\rm inv}$ has a plateau of 3/64 in model A, larger than 1/32 in model B, so a plateau is not a signature of structure; $\chi_{\rm sep}$ is 0 for model A and 1/32 for model B. About the centre of mass, model B develops a density hole only above $M_c = 0.633$ ($g = k = 1$); our closed form reproduces Müller-Herold's critical relation, as corrected by Ludeña *et al.*, to $1.6 \times 10^{-14}$. The Lindemann ratio is constant (0.422) in model A and $\propto M^{-1/4}$ in model B. These models are not Coulomb systems and carry no quantitative statement about the family.

## S3.13 Electron–nuclear entanglement

**Table S14** Electron–nuclear entanglement of the exact ground state (nats).

| $M$ | $S$ | $S_{\rm orient}$ | $S_{\rm rad}$ |
|---|---|---|---|
| 0.05 | 1.49524 | 1.02376 | 0.47148 |
| 0.4 (near the maximum of $S$, at 0.414) | 1.60394 | 1.10712 | 0.49681 |
| 1 ($Ps^-$) | 1.51623 | 1.05104 | 0.46520 |
| 206.768 ($Mu_2^+$) | 0.38342 | 0.31388 | 0.06954 |
| 1836.15 ($H_2^+$) | 0.29877 | 0.27045 | 0.02832 |
| $10^4$ | 0.27211 | 0.25831 | 0.01379 |
| $\infty$ | 0.249338 | 0.249338 | 0 |

**Extrema.** Maxima of $I(|R|;|\rho|)$ at 0.357, $S_{\rm rad}$ 0.369, $S_0$ 0.387, $S$ 0.414, $S_{\rm orient}$ 0.433 and $S_{\rm lin}$ 0.445; minima of $w_0$ at 0.460 and $D(90°)$ at 0.475 (spline in $\ln M$, about ±0.01). **BO limit.** $w_0 = 0.95865$, $w_2 = 0.03926$, remainder in $l \geq 4$; $D(90°) \to 0.93966$. **Approach to the limit** (fits on $M \geq 500$): $S_{\rm orient} - S(\infty) \propto \kappa^{2.03}$, $I \propto \kappa^{2.00}$, $p_2 \propto \kappa^{1.96}$, $E_N \propto \kappa^{1.04}$, and $S_{\rm rad} = \kappa^2[0.391\ln(1/\kappa) + 0.479]$ for $M \geq 206.77$ (residuals $\leq 10^{-5}$). **BO product state.** $S = 2.42$ against 1.17 (exact) at $M = 3$ and 0.758 against 0.645 at $M = 20$; at large $M$, $S_{\rm BO} - S = 2.1 \times 10^{-3}$ ($H_2^+$), $1.0 \times 10^{-3}$ (4000) and $4.0 \times 10^{-4}$ ($10^4$), roughly $\propto 1/M$ (three points;

tentative). **Convergence.** $S$ to about $10^{-5}$ absolute (quadrature $1.2 \times 10^{-5}$; basis $N \to N - 4$ at most $2.5 \times 10^{-7}$; $l$ tail at most $10^{-5}$); $E_N$ only to about $5 \times 10^{-3}$ for $M < 1$. **Validation:** separable Gaussian and exponential functions give $S = 10^{-26}$; a coupled Gaussian with an analytic entropy (0.856460302494) is reproduced to $5 \times 10^{-15}$; Parseval sums agree with the direct norm to the $l$ tail ($\leq 4 \times 10^{-7}$). **Body-frame comparison.** Tracing out the orientation instead of keeping it as a subsystem gives, for $H_2^+$, $S_{\mathrm{vN}} = 0.047954$ bits (0.033239 nats) and $S_L = 0.010466$, against 0.0479102 bits and 0.0104556 of Sanz-Vicario *et al.* (0.1%); converged to $2 \times 10^{-8}$ bits between two grids; unchanged at the CODATA 2022 mass.

# S4 Physical constants

**Table S15** Mass ratios used in this work.

| Constant | CODATA 2022 (this work) | CODATA 2010 (King *et al.* comparison only) |
|---|---|---|
| $m_p/m_e$ | 1836.152673426 | 1836.15267245 |
| $m_\mu/m_e$ | 206.7682827 | 206.7682843 |
| $m_d/m_e$ | 3670.482967655 | 3670.4829652 |
| $m_t/m_e$ | 5496.92153551 | 5496.9215267 |
| $m_\alpha/m_e$ | 7294.29954171 | |
| $E_h/(hc)$ | 219 474.63136314 cm$^{-1}$ | |

The CODATA 2022 values were taken from the NIST database (P. J. Mohr, D. B. Newell, B. N. Taylor and E. Tiesinga, *Rev. Mod. Phys.*, 2025, **97**, 025002); the CODATA 2010 values are those of P. J. Mohr, B. N. Taylor and D. B. Newell, *Rev. Mod. Phys.*, 2012, **84**, 1527. Literature comparisons use the masses of the work compared with: King *et al.*, CODATA 2010; Tang *et al.*, $m_p = 1836.15267247$; Frolov's muonic ions, his mass ratios. Threshold masses, offsets, predictions and fits are expressed in the mass ratio and do not depend on these constants.

**Table S16** Physical-mass values at CODATA 2022 (Eh).

| Quantity | $H_2^+$ | $Mu_2^+$ |
|---|---|---|
| $E_0$ (exact) | −0.5971390630805(3) | −0.585126098178(1) |
| $E_1$ (exact, $v = 1$) | −0.587155679099 | −0.558828428995 |
| $E_{\mathrm{BO}}$ | −0.597233978551 | −0.585961513660 |
| $E_{\mathrm{BO+ad}}$ | −0.597138541555 | −0.585111176441 |
| $E(J = 1)$ | −0.5968737388108 | −0.5829284525435 |
| $E(J = 2)$ | −0.5963452054792 | −0.5786835774492 |

The CODATA 2022 masses are within $4 \times 10^{-9}$ (proton) and $3 \times 10^{-7}$ (muon) of the CODATA 2018 masses of the mass-scan grid, and the CODATA 2022 energies reproduce the 2018 grid values to the last digits ($H_2^+$ −0.597139063081; $Mu_2^+$ shifted by $dE/dM\,\Delta M = 1.4 \times 10^{-11}$), an independent rerun check. At the CODATA 2010 masses the ground-state energies are −0.5971390630793 ($H_2^+$) and −0.585126098252 ($Mu_2^+$).

## S5 Data and code availability

The code, the numerical results (including the frozen pre-specified predictions with their SHA-256 hashes) and the scripts that regenerate every figure and table are openly available at https://github.com/peverati/Ethreshold_BOoffset and archived on Zenodo (https://doi.org/10.5281/zenodo.24398482), under MIT (code) and CC BY 4.0 (data) licences.